\documentclass[authoryear]{elsarticle}

\usepackage{booktabs}
\usepackage{multirow}
\usepackage{graphicx}  
\usepackage[ruled]{algorithm2e}
\usepackage[font=small,format=plain,labelfont=bf,up,textfont=it,up]{caption}

\usepackage{subfigure}
\usepackage{geometry}
\usepackage{amsmath}
\usepackage{amsfonts}
\usepackage{amsthm}
\usepackage{appendix}

\usepackage{float}

\usepackage{pgf,tikz}
\usetikzlibrary{shapes,arrows,calc,automata}

\usepackage{fancyhdr}
\newtheorem{remark}{Remark}

\title{\textbf{Finite-Volume Simulation of Capillary-Dominated Flow \\
       in Matrix-Fracture Systems using Interface Conditions}}
\date{\today}

\author[stanford]{Ammar H. Alali\corref{cor1}}
\ead{ammara@stanford.edu}
\author[LBNL]{Fran{\c{c}}ois P. Hamon\corref{cor2}}
\author[chevron]{Bradley T. Mallison}
\author[stanford]{Hamdi A. Tchelepi}

\cortext[cor1]{Corresponding author, now with Saudi Aramco}
\cortext[cor2]{Now with Total Exploration \& Production}
\address[stanford]{Energy Resources Engineering, Stanford University, USA}
\address[LBNL]{Center for Computational Sciences and Engineering, Lawrence Berkeley National Laboratory, Berkeley, USA}
\address[chevron]{Chevron ETC, 6001 Bollinger Canyon Rd, San Ramon, USA}

\usepackage{soul}

\begin{document}

\begin{abstract}
In numerical simulations of multiphase flow and transport in
fractured porous media, the estimation of the hydrocarbon
recovery requires accurately predicting the capillary-driven
imbibition rate of the wetting phase initially present in
the fracture into the low-permeability matrix. In the fully
implicit finite-volume scheme, this entails a robust methodology
that captures the capillary flux at the interface between the
matrix and the fracture even when very coarse cells
are used to discretize the matrix. Here, we investigate the
application of discrete interface conditions at the
matrix-fracture interface to improve the accuracy of the
flux computation without relying on extreme grid refinement.
In particular, we study the interaction of the upwinding
scheme with the discrete interface conditions. Considering
first capillary-dominated spontaneous imbibition and then
forced imbibition with viscous, buoyancy, and capillary
forces, we illustrate the importance of the interface
conditions to accurately capture the matrix-fracture flux
and correctly represent the flow dynamics in the problem.
\end{abstract}

\begin{keyword} Fractured porous media 
\sep 
Two-phase flow and transport 
\sep 
Capillarity-dominated flow 
\sep  
Implicit finite-volume schemes 
\sep 
Interface conditions.
\end{keyword}

\maketitle

\section{\label{section_introduction}Introduction}

Reservoir simulation is an essential tool for a 
comprehensive understanding of both petroleum recovery and
groundwater processes. The modeling of multiphase flow and
transport in porous media entails the difficult
task of solving the governing partial differential equations (PDEs).
In heterogeneous porous media, a major modeling challenge stems
from the fact that the saturation-dependent flux coefficients
representing the relative permeabilities and capillary pressures
are spatially discontinuous between different rock types, leading
to complex capillary-driven flow dynamics characterized by
discontinuous pressure and saturation profiles.

These discontinuities arise because capillary pressure depends on
saturation but also on intrinsic rock properties exhibiting large
and abrupt spatial variations, such as the pore throat size
and wettability. These contrasts in intrinsic rock properties and
saturation functions are particularly severe in fractured porous
media. For these problems, the wetting phase propagating into the
fracture quickly invades the lower-permeability matrix driven by
a large capillary pressure gradient, while the non-wetting phase
is displaced into the fractures and can be produced. This
counter-current imbibition process at the matrix-fracture interface
is a key oil recovery mechanism that must be resolved
accurately to obtain a reliable prediction of the flow dynamics.

Therefore, it is essential to design with an accurate and
specialized numerical scheme and solution strategy able
to capture abrupt spatial jumps in the relative permeability
and capillary pressure functions. The accuracy of standard flux 
approximations applied to fractured systems is often limited
by the presence of large cells in the matrix next 
to the fracture, since this configuration often leads to large errors
in the flux computation. This issue is particularly severe for 
Discrete Fracture Model (DFM) simulations
\citep{karimi2003efficient}
and dual-permeability models. Therefore, we particularly seek
robust numerical schemes that produce physically-consistent
predictions of the capillary-driven imbibition rate in the context
of fractured media on coarse grids.

Following the work of \cite{van1995effect}, interface
conditions have been used in the finite-volume method to improve
the accuracy of the flux in the presence of capillary
heterogeneity \citep{enchery2006numerical,cances2009finite,
evje2012numerical,brenner2013finite,hamon2016capillary,
ahmed2018cell}. In this approach, the flux is computed by solving
a local nonlinear system that enforces a capillary equilibrium
at the interface. Along with the interface conditions, 
the introduction of additional interface-based degrees of freedom
makes it possible to resolve complex interfacial flow
dynamics that would otherwise require extreme spatial refinement. 
Interface conditions have initially been applied to ensure
physically consistent simulations of capillary trapping taking
place during drainage in the context of DNAPL trapping
\citep{niessner2005interface,papafotiou2010numerical}, and
more recently, to represent capillary imbibition in
matrix-fracture systems
\citep{brenner2017immiscible,brenner2018hybrid,aghili2019two}.
For completeness,
we mention here the alternative numerical approaches to handle
capillary approaches in the mixed finite-element method
\citep{hoteit2005multicomponent} and in discontinuous Galerkin
schemes
\citep{evje2012numerical,arbogast2013discontinuous}.

\textcolor{black}{In this work, we focus on a widely used
fully implicit finite-volume method with cell-centered variables
to demonstrate that interface conditions can be used in
existing reservoir simulators to improve the accuracy
of oil recovery predictions. We show that, in the context
of high-contrast matrix-fracture systems with capillary
heterogeneity, a straightforward modification to the standard flux
computations to better account for the local capillary equilibrium
at the interface results in a more accurate imbibition
rate.}

To do that, we extend the analysis performed in
\cite{hamon2016capillary} to demonstrate the
usefulness of the discrete interface conditions for the simulation
of capillary-dominated counter-current imbibition. We perform
a detailed analysis of the impact of the upwinding  of the
saturation-dependent coefficients of the flux on the accuracy
of a fully implicit finite-volume scheme based on the interface
conditions.
We consider both the standard Phase-Potential Upwinding (PPU)
and the Implicit Hybrid Upwinding (IHU). The latter was initially
developed to improve the nonlinear solver convergence for
two-phase transport with buoyancy but has been extended to
coupled flow and transport with three phases and with capillarity
\citep{hamon2016analysis,hamon2016implicit,lee2016c,hamon2016capillary,lee2018hybrid,moncorge2019consistent}.

In the first numerical example, we simulate capillary-driven
spontaneous imbibition.
\textcolor{black}{We demonstrate that the use of interface conditions improves the
accuracy of the interfacial flux on practical (coarse) grids, 
which results in a more accurate prediction of the imbibition 
rate.} 
We also note that IHU-C, combining the IHU scheme
with interface conditions, is more accurate than the
PPU-based schemes.
Then, we consider the capillary interactions with viscous
and buoyancy forces in a forced imbibition case, and focus on the
ability of the schemes to capture non-wetting phase trapping
in the matrix due to the capillary end-effect. The
conclusions are supported by a truncation error analysis.

In the remainder of the paper, we first review the governing
equations and the interface conditions in
Section~\ref{section_mathematical_model}. Then,
we present the fully implicit finite-volume scheme in
Section~\ref{section_fully_implicit_finite_volume_scheme}, with 
an emphasis on the approximation of the saturation-dependent 
coefficients in PPU or IHU. After analyzing
the truncation error of the schemes for the homogeneous case in 
Section~\ref{section_truncation_error_analysis}, we present
numerical examples in Section~\ref{section_numerical_examples}.

\section{\label{section_mathematical_model}Mathematical model}

\subsection{\label{section_governing_equations}Governing equations}

We consider two-phase flow
in an incompressible porous medium. The incompressible wetting
and non-wetting phases are denoted by the subscripts $w$ and
$n$, respectively. The mass conservation laws read
\begin{equation}
\phi\frac{\partial S_\ell}{\partial t} + \nabla \cdot \boldsymbol{u}_\ell = q_\ell  \quad \forall \ell \in \{w, n\},
\label{mass_conservation}
\end{equation}
where $\phi(\boldsymbol{x})$ is the porosity of the medium,
$S_\ell(\boldsymbol{x},t)$ is the saturation of the phase
$\ell$, and $t$ is the time. The source/sink term $q_\ell$ is
positive for injection and negative for production. The phase
velocity, $\boldsymbol{u}_\ell$, is defined using
the multiphase extension of Darcy's law as
\begin{equation} \label{phase_velocity}
\boldsymbol{u}_\ell := -k\lambda_\ell(\nabla p_\ell - \rho_\ell g \nabla z) \quad \forall \ell \in \{w, n\}.
\end{equation}
Here, $k(\boldsymbol{x})$ is the scalar absolute permeability
of the medium, $\lambda_\ell$ is the phase mobility defined as
$k_{r\ell}/\mu_\ell$, where $k_{r\ell}(\boldsymbol{x},S_\ell)$
is the phase relative permeability, $\mu_\ell$ is the phase
viscosity, and $\rho_{\ell}$ is the phase density. We assume
that the relative permeabilities are strictly increasing
functions of their own saturation. In this work, we set
\textcolor{black}{$\rho_{n} = 800 \, \text{kg}.\text{m}^{-3}$} and 
$ \rho_{w} = 1000 \, \text{kg}.\text{m}^{-3}$. We assume
the same viscosity $\mu_w = \mu_{\textit{n}} = 1 \, \text{cP}$ for both phases.
The gravitational acceleration is denoted by $g$, and the
depth by $z$ (positive going downward).
\textcolor{black}{Even though studying buoyancy effects
is not the main focus of this work, the gravity term is included
in (\ref{phase_velocity}) for completeness.}
The saturations of both phases are constrained
by the following equation:
\begin{equation}
S_{n} + S_w = 1.
\label{saturation_constraint}
\end{equation}
To account for capillary heterogeneity, we divide the domain
$\Omega$ into non-overlapping subdomains $\Omega_{\alpha}$
referred to as rock regions such that
$\cup_{\alpha} \, \Omega_{\alpha} = \Omega$.
That is, capillary pressure and relative permeabilities are
functions of saturation that can be spatially discontinuous
between rock regions. 
In each rock region, the capillary
pressure constraint relates the two phase pressures with
\begin{equation}
P^{(\alpha)}_{c}(S_{w}) = p_{n} - p_w.
\end{equation}
We assume that capillary pressure is a strictly decreasing
function of the wetting-phase saturation.
Capillary pressure functions that do not satisfy this property
require a specific treatment in the numerical scheme
(see \cite{brenner2017immiscible}) and will be considered
in future work.

\subsection{\label{section_pressure_equations}Pressure and transport equations}

In this work, we consider two types of upwinding methods in
the flux approximation required by the finite-volume scheme.
Although the Phase-Potential Upwinding (PPU) methodology is
applied directly to (\ref{mass_conservation}), the Implicit
Hybrid Upwinding (IHU) approach is constructed as the
approximation of the flux in fractional flow form. This
formulation is obtained by decomposing the governing
equations into a flow problem and a transport problem for
one of the two phases. To obtain this decomposition, we first
sum the phase velocities to define the total velocity as
\begin{equation}
\boldsymbol{u}_T := \boldsymbol{u}_w + \boldsymbol{u}_{n}= -k \lambda_T \nabla p + k ( \lambda_w \rho_w + \lambda_{n} \rho_{n}) g \nabla z + k \lambda_w \nabla P^{(\alpha)}_{c}, 
\label{total_velocity}
\end{equation}  
where $\lambda_T$ is the total mobility, defined as the sum
of the phase mobilities. The total velocity is a function
of space as well as of the reference pressure chosen to be
$p = p_{n}$ and the reference saturation denoted by 
$S = S_{w}$. Summing (\ref{mass_conservation}) and then
using the saturation constraint results in the
elliptic equation governing the temporal
evolution of the pressure field:
\begin{equation}
\nabla \cdot \boldsymbol{u}_T = q_w + q_{n}. 
\label{flow_equation}
\end{equation}  
Next, the highly nonlinear parabolic transport equation
is obtained by using  (\ref{total_velocity}) to formally
eliminate the pressure variable from (\ref{mass_conservation}),
yielding
\begin{equation}
\textcolor{black}{
\phi\frac{\partial S}{\partial t} + \nabla \cdot 
\Bigg(
\underbrace{\frac{\lambda_w}{\lambda_T}\boldsymbol{u}_T}_{\substack{\text{viscous} \\ \text{term}}}  + 
\underbrace{k \frac{\lambda_w \lambda_{n}}{\lambda_T}(\rho_w - \rho_{n}) 
g\nabla z}_{\substack{\text{buoyancy} \\ \text{term}}}
-
\underbrace{k D^{(\alpha)}(S) 
\nabla S}_{\substack{\text{capillary} \\ \text{term}}}
\Bigg)  =  
q_w,}
\label{transport_equation}
\end{equation}
where we have linearized the capillary pressure term of (\ref{phase_velocity})
to introduce a nonlinear capillary diffusion coefficient defined as
 \begin{equation}
 D^{(\alpha)}(S) :=  - \frac{ \lambda_w \lambda_{n}}{\lambda_T} \frac{\partial P^{(\alpha)}_{c}(S)}{\partial S}.
\label{nonlinear_capillary_diffusion_coefficient}
\end{equation}
The capillary diffusion coefficient is
used to construct a differentiable and bounded capillary
flux in the IHU approach. The function
$S \mapsto D^{(\alpha)}(S)$ depends on space and saturation, 
is nonnegative, and is equal to zero only at the saturation
endpoints. The maximum value of the capillary diffusion coefficient
is denoted by $D^{(\alpha)}_{\textit{max}}$.
We consider two rock regions, the matrix and the 
fracture, but the methodology is applicable to any
number of regions. We refer to the matrix
region with the superscript $\alpha = m$, and to the
fracture region with the superscript $\alpha = f$.
\textcolor{black}{
In the matrix, we rely on the capillary pressure model of
\cite{skjaeveland1998capillary}. Capillary pressure is
written as a function of saturation as}  
\begin{equation}
P_{c}^{(m)}(S) := p_{e}^{(m)} S^{-1/\theta^{(m)}} - p_{e}^{(m)}(1 - S)^{-1/\theta^{(m)}},
\label{brooks_corey_pc}
\end{equation} 
where $p_{e}^{(m)}$ is the entry pressure in the matrix, and
$\theta^{(m)}$ determines the saturation exponent. 
In this work, we set $\theta^{(m)} = 4$.
\textcolor{black}{This type of capillary pressure function --
used to fit experimental curves in \cite{masalmeh2005improved}
and described in the context of numerical studies in
\cite{tavassoli2005analysis,schmid2013universal} --
is representative of a capillary imbibition process in a
mixed-wet system. The wetting-phase saturation
$S^{\star}$ that satisfies $P_c(S^{\star}) = 0$ is used
to distinguish two imbibition regimes -- namely,
a spontaneous imbibition regime in which the capillary pressure
function is positive ($S < S^{\star}$), and a forced imbibition
regime in which the capillary pressure function is negative
($S > S^{\star}$).}

\begin{figure}[H]
\begin{center}
\subfigure[]{
\begin{tikzpicture}
\node[anchor=south west,inner sep=0] at (0,0){
\includegraphics[width=0.29\textwidth]{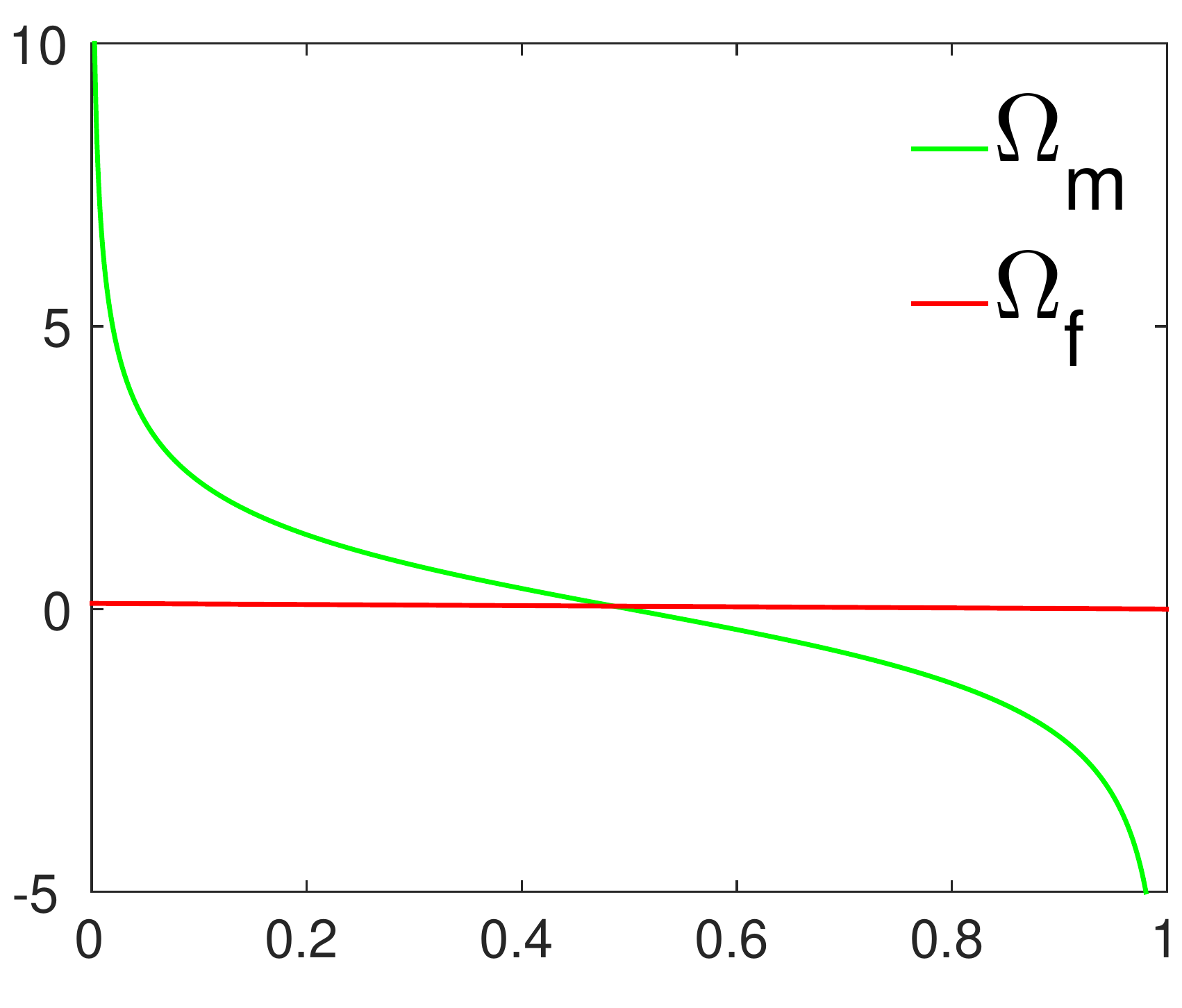}
};
\path (2.4,0.1) node (e) {$S$};
\node (ib_1) at (0.05,2.15) {\small $P_c$};
\end{tikzpicture}
\label{fig:scal_data_capillary_pressure}
}
\hspace{-0.35cm}
\subfigure[]{
\begin{tikzpicture}
\node[anchor=south west,inner sep=0] at (0,0){
\includegraphics[width=0.29\textwidth]{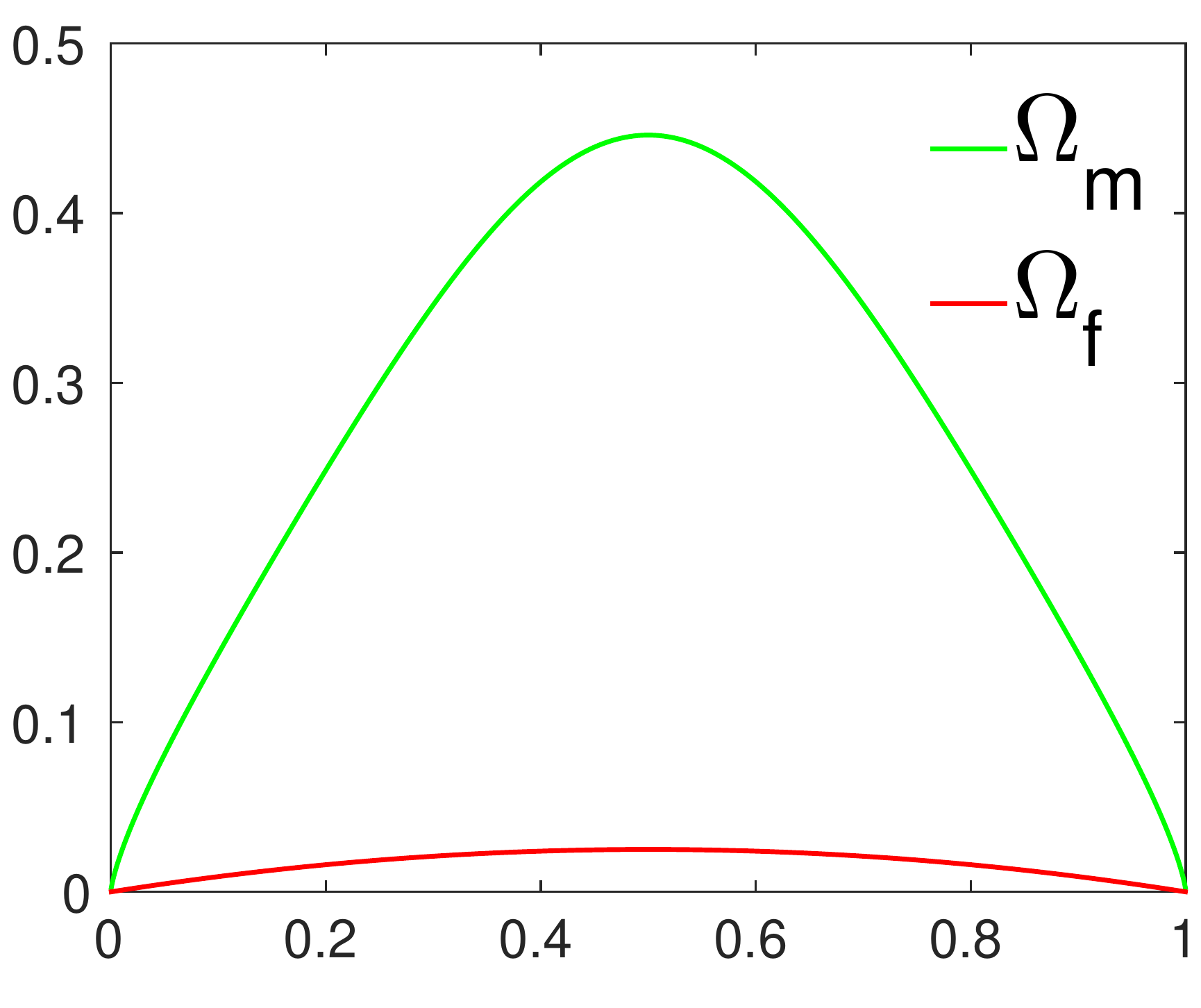}
};
\path (2.45,0.1) node (e) {\small $S$};
\node (ib_1) at (0.0,2.1) {\small $D$};
\end{tikzpicture}
\label{fig:scal_data_diffusion_coefficient_quadratic}
}
\hspace{-0.35cm}
\subfigure[]{
\begin{tikzpicture}
\node[anchor=south west,inner sep=0] at (0,0){
\includegraphics[width=0.3\textwidth]{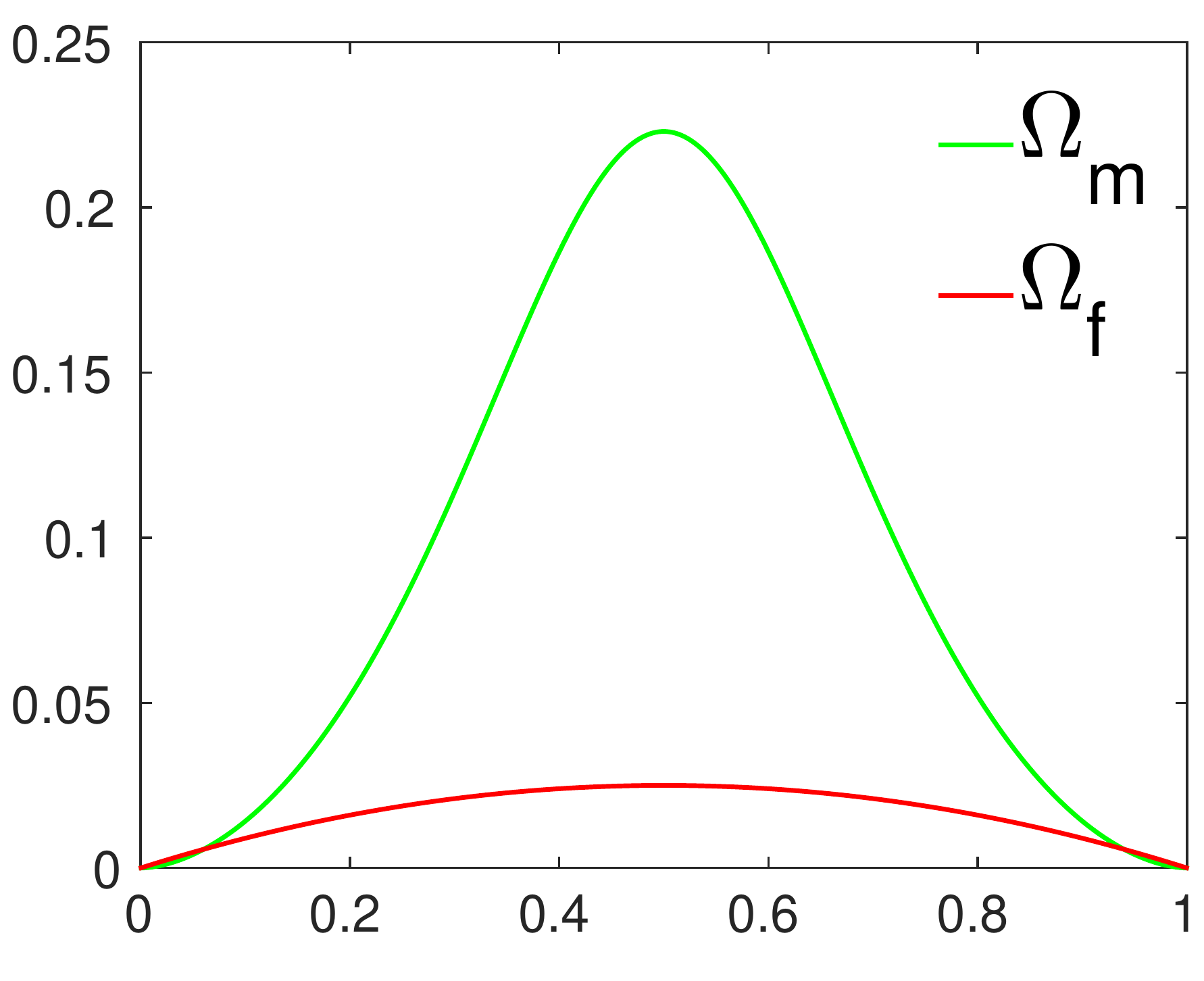}
};
\path (2.6,0.1) node (e) {\small $S$};
\node (ib_1) at (0.1,2.1) {\small $D$};
\end{tikzpicture}
\label{fig:scal_data_diffusion_coefficient_cubic}
}
\vspace{-0.4cm}
\caption{\label{fig:scal_data}Capillary pressures in
\subref{fig:scal_data_capillary_pressure} and
capillary diffusion coefficients in
\subref{fig:scal_data_diffusion_coefficient_quadratic}
and
\subref{fig:scal_data_diffusion_coefficient_cubic}. The parameters of the capillary pressure functions are
$p^{(m)}_e = 3 \, \text{psi}$, $p^{(f)}_{\textit{max}} = 0.1 \, \text{psi}$, and $ \theta^{(m)} = 4$.
The relative permeabilities in the fracture are linear,
whereas those of the matrix are quadratic in
\subref{fig:scal_data_diffusion_coefficient_quadratic}
and cubic in \subref{fig:scal_data_diffusion_coefficient_cubic}.
The capillary pressures are shown in psi, and the capillary
diffusion coefficients are in psi/cP.}
\label{input}
\end{center}
\end{figure}

We highlight that the capillary
pressure curve obtained in \textcolor{black}{(\ref{brooks_corey_pc})}
is unbounded when $S$ is close to zero or one. Therefore, to avoid
numerical problems with the standard Phase-Potential Upwinding (PPU)
scheme with this analytical capillary pressure, we modify the 
formulation given by (\ref{brooks_corey_pc}) in a preprocessing step. 
This procedure aims at enforcing upper and lower bounds on 
the capillary pressure in the matrix while preserving its monotonicity
and differentiability with respect to saturation. The details
are given in \ref{appendix_capillary_pressure_interpolation}.
In the fracture, we use the following linear capillary pressure model:
\begin{equation}
P_{c}^{(f)}(S) := 
p_{\textit{max}}^{(f)} (1-S),
\end{equation}
where the maximum capillary pressure in the
fracture is a constant denoted by $p_{\textit{max}}^{(f)} > 0$.
The relative permeabilities in
the fracture are linear. Figure~\ref{input} shows
the capillary pressure curves used in this work as well as
the corresponding capillary diffusion coefficients. 

\subsection{Interface conditions for local capillary equilibrium}

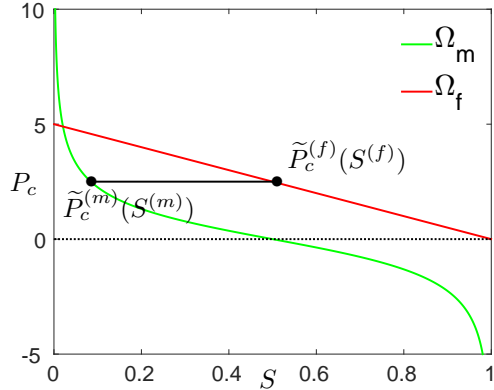
\begin{figure}[H]
\begin{center}
\begin{tikzpicture}
\node[anchor=south west,inner sep=0] at (0,0){
\includegraphics[width=0.4\textwidth]{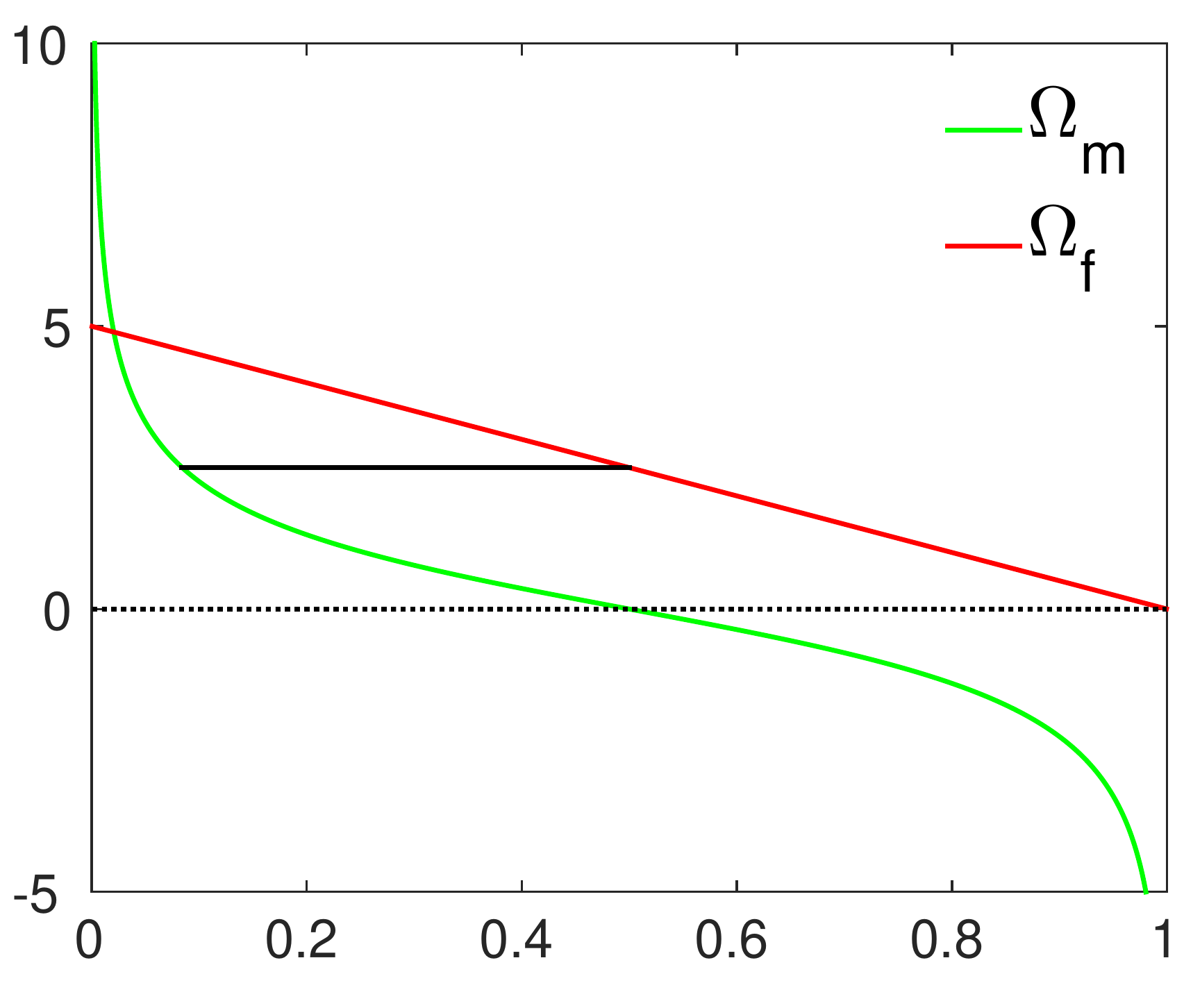}
};
\path (3.35,0.3) node (e) {$S$};
\node (ib_1) at (0.1,2.9) {$P_c$};

\node (ib_1) at (3.47,2.925) {$\bullet$};
\node (ib_1) at (1.0,2.925) {$\bullet$};

\node (ib_1) at (1.5,2.6) {$\widetilde{P}^{(m)}_c(S^{(m)})$};
\node (ib_1) at (4.4,3.25) {$\widetilde{P}^{(f)}_c(S^{(f)})$};

\end{tikzpicture}
\vspace{-0.3cm}
\caption{\label{fig:fig_extended_pressure_condition}Illustration of
  the local equilibrium for a capillary pressure
  discontinuity between two rock regions.}
\end{center}
\end{figure}

In heterogeneous porous media, interface conditions can be
used to enforce a local capillary equilibrium at the interfaces
between rock regions with different capillary pressure functions
\citep{van1995effect}. Once discretized in the numerical scheme,
these conditions provide a methodology to better capture the
capillary jump in the computation of the flux. Here, we consider
these conditions in the context of matrix-fracture systems.
Considering an interface $\Gamma_{m f}$ between the matrix,
$\Omega_m$, and the fracture, $\Omega_f$, the two conditions
imposed at the discontinuity are given below.
\paragraph{Mass flow rate conservation condition at the discontinuity}
The first condition ensures that the mass
flow rate of each phase is conserved at the
interface between rock regions $\Omega_m$ and $\Omega_f$. It reads
\begin{equation}
\sum_{\alpha \in \{m,f\}} \boldsymbol{u}_\ell \cdot \boldsymbol{n}_\alpha =  0 \quad \text{along}  \quad \Gamma_{m f} \quad \text{for} \quad \ell \in \{w, n\},
\label{mass_conservation_at_discontinuity}
\end{equation}  
where $\boldsymbol{n}_{\alpha}$ is the outward normal to the
interface with respect to each rock region.
\paragraph{Extended pressure condition at the discontinuity}
We first define the extended pressure as
\begin{equation}
 \widetilde{P}^{(\beta)}_c (S^{(\beta)}) := \min \big( P_{c,\textit{max}}^{(\alpha)}, \max \big( P^{(\beta)}_c (S^{(\beta)}), P_{c,\textit{min}}^{(\alpha)} \big)\big), \quad \forall \, S^{(\beta)} \in [0, 1], \, \textcolor{black}{\beta \in \{m,f\}, \beta \neq \alpha,}
 \label{extended_pressure}
\end{equation}
where $P^{(\alpha)}_{c,\textit{min}}$ and $P^{(\alpha)}_{c,\textit{max}}$
denote, respectively, the minimum and the maximum of the capillary
pressure function on the saturation interval. 
Assuming that the capillary pressure curves have been
bounded using the procedure described in
\ref{appendix_capillary_pressure_interpolation} and
therefore have finite endpoints, the quantities
$P^{(\alpha)}_{c,\textit{min}}$ and $P^{(\alpha)}_{c,\textit{max}}$
are well defined. Using that, the extended pressure condition
enforces a local capillary equilibrium at the interface with
the requirement that
\begin{equation}
\widetilde{P}^{(m)}_c (S^{(m)}) = \widetilde{P}^{(f)}_c (S^{(f)}) \quad \text{along} \quad \Gamma_{m f}.
\label{extended_pressure_condition_at_discontinuity}
\end{equation} 
\textcolor{black}{The left-hand side and right-hand side of 
(\ref{extended_pressure_condition_at_discontinuity}) reduce to,
respectively,}
\begin{equation}
\widetilde{P}^{(m)}_c (S^{(m)}) = \min \big( p_{\textit{max}}^{(f)}, \max \big( P^{(m)}_c (S^{(m)}), 0 \big)\big), \qquad \text{and} \qquad \widetilde{P}^{(f)}_c (S^{(f)}) = P^{(f)}_c (S^{(f)}).
 \label{extended_pressure_reduced}
\end{equation}
The extended pressure condition states that although 
the capillary pressure functions are different 
in $\Omega_f$ and in $\Omega_m$, the phase pressures
are still continuous at the interface whenever 
$S^{(m)} \in [(P^{(m)}_c)^{-1}\big(p^{(f)}_{\textit{max}}\big),$ $(P^{(m)}_c)^{-1}\big(0\big)]$. 
\textcolor{black}{This
condition limits the possible saturation pairs
satisfying this local capillary equilibrium
if $S^{(m)} \in
[(P^{(m)}_c)^{-1}\big(p^{(f)}_{\textit{max}}\big),$ 
$(P^{(m)}_c)^{-1}\big(0\big)]$, and otherwise
enforces a physically consistent discontinuity
in the saturation profile at the interface.} Figure
\ref{fig:fig_extended_pressure_condition} shows 
example of the capillary equilibrium in a
matrix-fracture system.

\section{\label{section_fully_implicit_finite_volume_scheme}Fully
implicit finite-volume scheme}

The domain is discretized into $N$ cells, with $N^{(m)}$
(respectively, $N^{(f)}$) cells in the matrix
(respectively, in the fracture). \textcolor{black}{A fully implicit finite-volume
discretization of (\ref{mass_conservation})
results in the following equation}:
\begin{equation}
V_i\phi_i \frac{S_{\ell,i}^{n+1} - S_{\ell,i}^n}{\Delta t} +  \sum_{j \in \textit{adj}(i)} F_{\ell,ij}^{n+1} = V_iq_{\ell,i}, \quad \forall i \in \{1, \dots,  N\} \quad \forall \ell \in \{w, n\},
\label{discrete_transport_equation}
\end{equation}  
where $F_{\ell,ij}^{n+1}$ is the numerical flux of phase $\ell$
for interface $(ij)$ between two cells $i$ and $j$,
$adj(i)$ is the set of neighbors of cell $i$,
$V_i$ is the bulk volume of the cell $i$, and
$\Delta t$ is the time step size.

We compute the numerical flux with a two-point flux
approximation (TPFA) and focus on one-dimensional uniform grids
for simplicity. We decompose the numerical flux into a static rock
and geometric transmissibility coefficient independent of the
primary variables, $T_{ij}$, and a dynamic phase-dependent part
that is a function of the primary variables. The static term,
$T_{ij}$, is precomputed using a harmonic average of one-sided
transmissibilities, denoted by $\widehat{T}_{ij}$, as
\begin{equation}
 T_{ij} := T_{ji} := (\widehat{T}^{-1}_{ij} + \widehat{T}^{-1}_{ji})^{-1} := \bigg({\displaystyle\frac{d_{i,(ij)}}{k_i \mathcal{A}_{ij}} + \displaystyle\frac{d_{j,(ji)}}{k_j \mathcal{A}_{ji}}}\bigg)^{-1},
\label{transmissibility}
\end{equation} 
where $k_i$ (respectively, $k_j$) is the absolute permeability
in cell $i$ (respectively, $j$), $d_{i,(ij)}$ (respectively,
$d_{j,(ji)}$) is the distance from the center of cell $i$
(respectively, $j$) to the interface, and
$\mathcal{A}_{ij} = \mathcal{A}_{ji}$ is the area
of the interface. For the simple grids considered in this work, we
have $d_{i,(ij)} = d_{j,(ji)}$. 
The dynamic term at the interface is computed using a first-order
upwinding scheme relying on cell-averaged quantities.
The two upwinding schemes used in this work are discussed next. 

\subsection{\label{subsection_phase_potential_upwinding}
Phase-Potential Upwinding}
In Phase-Potential Upwinding \citep[PPU,][]{aziz1979petroleum},
the numerical flux is constructed as an
approximation of the interfacial integral of the phase velocity
written with Darcy's law: 
\begin{equation}
 F_{\ell,ij}^{n+1} := T_{ij}\lambda_{\ell,ij}^{\textit{P}} \Delta \Phi_{\ell,ij},
\label{ppu_flux}
\end{equation}
where the superscript $P$ denotes the PPU scheme.
The phase mobility is approximated at interface ($ij$)  
using the upstream saturation with respect to the discrete
phase-potential difference as
\begin{equation}
\lambda_{\ell,ij}^{P} := \left\{
\begin{array}{ll}
      \lambda_\ell (S_i) & \text{if} \quad \Delta \Phi_{\ell,ij} \geq 0 \\
      \lambda_\ell (S_j) & \text{otherwise,}
\end{array} 
\right. 
\end{equation}
where the discrete phase-potential difference involves a
viscous, gravitational, and capillary contribution:
\begin{align}
\textcolor{black}{\Delta \Phi_{w,ij}} &\textcolor{black}{:= \Phi_{w,i} - \Phi_{w,j} := \Delta p_{ij} - \rho_w g \Delta z_{ij} - (\Delta P_{c})_{ij},} \\
\textcolor{black}{\Delta \Phi_{n,ij}} &\textcolor{black}{:= \Phi_{n,i} - \Phi_{n,j} := \Delta p_{ij} - \rho_{n} g \Delta z_{ij}}.
\label{discrete_phase_potential_difference}\end{align}
In fractional flow form, the PPU numerical flux is decomposed
into a viscous flux, $V_{\ell,ij}^{n+1}$, a buoyancy flux,
$G_{\ell,ij}^{n+1}$, and a capillary flux, $C_{\ell,ij}^{n+1}$.
Specifically, for $m \neq \ell$,
\begin{align}
\textcolor{black}{F_{\ell,ij}^{n+1}} 
&\textcolor{black}{:=  V_{\ell,ij}^{n+1} 
+G_{\ell,ij}^{n+1} +  C_{\ell,ij}^{n+1},}
 \nonumber
 \\
&\textcolor{black}{:=  \frac{\lambda_{\ell,ij}^{P}}{\lambda_{w,ij}^{P}+\lambda_{n,ij}^{P}} \textcolor{black}{\overline{u}_{T,ij}} + T_{ij}\frac{\lambda_{\ell,ij}^{P}\lambda_{m,ij}^{P}}{\lambda_{w,ij}^{P}+\lambda_{n,ij}^{P}} (\rho_m - \rho_{\ell})g \Delta z_{ij} +  T_{ij}\frac{\lambda_{\ell,ij}^{P}\lambda_{m,ij}^{P}}{\lambda_{w,ij}^{P}+\lambda_{n,ij}^{P}} (\Delta P_{c,m} - \Delta P_{c,\ell})_{ij},}
\label{ppu_transport_flux}
\end{align}
\textcolor{black}{where $\Delta P_{c,w} 
:= \Delta P_c$ and $\Delta P_{c,n} := 0$.}
This methodology yields a monotone numerical flux that
is not differentiable in the presence of counter-current
flow due to buoyancy or capillarity
\citep{sammon1988analysis,brenier1991upstream,kwok2008convergence}.
Previous authors identified this non-differentiability
as the cause of nonlinear convergence difficulties for
Newton solvers when large time steps are taken
\citep{wang2013trust,li2014nonlinearity}.
This limitation has motivated the design of a robust Implicit
Hybrid Upwinding detailed in the next section. 

In heterogeneous
media, the standard PPU scheme presented above is employed
for all interfaces without special treatment for the interfaces
between two rock regions with different capillary
pressure functions. Therefore, this formulation does not strictly
enforce the interface conditions at the interfaces, and is
referred to as standard PPU in the remainder of this paper.
But, in addition to this standard PPU scheme,
we also consider the combination of PPU with discrete
interface conditions to better resolve the capillary jumps.
This extended formulation is discussed in 
Section~\ref{subsection_discrete_interface_conditions}
and is referred to as PPU-C.

\subsection{\label{subsection_implicit_hybrid_upwinding}Implicit Hybrid Upwinding}

In the Implicit Hybrid Upwinding (IHU) strategy, the viscous,
buoyancy, and capillary fluxes are upwinded separately to
achieve a differentiable and bounded flux for the transport
problem and therefore improve nonlinear convergence. This
robust approach was originally developed for two-phase
transport with buoyancy \citep{eymard1989hybrid,lee2015hybrid},
and was later extended to three-phase flow and transport with
capillary pressure \citep{hamon2016implicit,lee2016c,
hamon2016capillary,lee2018hybrid,moncorge2019consistent}
and to a multi-point upwinding strategy reducing grid orientation
effects \citep{hamon2018fully}. The scheme presented in
\cite{hamon2016capillary} was shown to lead to a more robust
Newton convergence than with the standard PPU approach for
challenging simulations with strong buoyancy and capillary
effects. It is briefly reviewed below for the two-phase case.

The IHU numerical  flux is constructed as the approximation of
the interfacial integral of the phase velocity written in
fractional flow form. The upwinding for the viscous flux is
based on the direction of the total flux at
interface ($ij$) denoted by $\overline{u}_{T,ij}$ as
\begin{equation}
 V_{\ell,ij}^{n+1} :=  \frac{\lambda_{\ell,ij}^{V} }{\lambda_{w,ij}^{V}+\lambda_{n,ij}^{V}} \textcolor{black}{\overline{u}_{T,ij}},
\end{equation}
where
\begin{equation}
\lambda_{\ell,ij}^{V}(S_i,S_j) := \left\{
\begin{array}{ll}
      \lambda_\ell (S_i) & \text{if} \quad \textcolor{black}{\overline{u}_{T,ij}} > 0 \\
      \lambda_\ell (S_j) & \text{otherwise}. 
      \label{ihu_viscous_mobility} \\
\end{array} 
\right. 
\end{equation} 
For simplicity, we evaluate the total flux, $\overline{u}_{T,ij}$,
using PPU in our 1D examples. However, in multiple dimensions, a
smooth weighted averaging for the total flux is advantageous
in some cases \citep{hamon2016implicit,hamon2016capillary}.

\textcolor{black}{Although only mild buoyancy effects are considered
in the numerical examples, we describe the discretization of the IHU
gravity flux since the improved nonlinear behavior obtained with IHU
can offset the additional cost induced by the interface conditions
(see Remark~\ref{remark_computational_cost}).}
\textcolor{black}{The IHU gravity flux is upwinded based on the density
difference between phases at interface ($ij$) as}
 \begin{equation}
 \textcolor{black}{G_{\ell,ij}^{n+1}  := T_{ij}\frac{\lambda_{\ell,ij}^{G}\lambda_{m,ij}^{G}}{\lambda_{w,ij}^{G}+\lambda_{n,ij}^{G}}  (\rho_{m} - \rho_{\ell})g \Delta z_{ij} \quad \text{for} \, \ell \neq m,}
\end{equation} 
where
\begin{equation}
\lambda_{\ell,ij}^{G}(S_i,S_j) := \left\{
\begin{array}{ll}
      \lambda_\ell (S_i) & \text{if} \quad \textcolor{black}{(\rho_{m} - \rho_{\ell})g \Delta z_{ij} > 0} \\
      \lambda_\ell (S_j) & \text{otherwise}. 
      \label{ihu_buoyancy_mobility} \\
\end{array} 
\right. 
\end{equation}
In the IHU numerical scheme, the discretization of the 
capillary flux relies on the nonlinear capillary diffusion
coefficient defined in (\ref{nonlinear_capillary_diffusion_coefficient}). 
This term is evaluated at interface $(ij)$ using the maximum
value of the diffusion coefficient between the saturations in
cells $i$ and $j$ as follows:
\begin{equation}
C_{\ell,ij}^{n+1} := T_{ij}D^{(\alpha)}_{ij}\Delta S_{\ell,ij},
\end{equation}
where
\begin{equation}
D^{(\alpha)}_{ij}(S_i,S_j) := \max\limits_{S \in [\min(S_{i},S_{j}),\max(S_{i},S_{j})]} D^{(\alpha)}(S).
\end{equation}
We emphasize that $D^{(\alpha)}_{ij}$ is not a fixed quantity but
depends on $S_i$ and $S_j$. The capillary flux is differentiable and monotone with respect to saturation. By
construction, this flux term is bounded in the entire saturation
range because $D^{(\alpha)}_{ij}(S_i,S_j)$ is bounded and vanishes
at both saturation endpoints. This is in contrast to the PPU
capillary flux that is unbounded when the capillary curve has
an asymptote.

\begin{remark}
\textcolor{black}{As explained above, the IHU scheme
yields a 
bounded capillary flux and therefore does not 
require any modification to the capillary pressure
function given in (\ref{brooks_corey_pc}). 
However, the PPU scheme produces an unbounded 
capillary flux, which creates significant
numerical difficulties in the simulations. To 
overcome this issue, we modify
(\ref{brooks_corey_pc}) using the methodology
detailed in \ref{appendix_capillary_pressure_interpolation}. Although this modification only aims
at improving the behavior of the PPU scheme, we 
use it for both PPU and IHU to perform a fair 
comparison between the two schemes.}
\end{remark}

We have assumed in this section that both
cells are in the same rock region 
denoted by
$\Omega_\alpha$. When the cells belong 
to two rock regions with
different capillary pressure functions, we employ 
a specific treatment that enforces an equilibrium
at the interface to compute the capillary flux
in heterogeneous porous media.
The incorporation of the discrete interface 
condition in the scheme, referred to as IHU-C, 
is presented in 
Section~\ref{subsection_discrete_interface_conditions}.

\subsection{\label{subsection_discrete_interface_conditions}Discrete interface conditions}

In this section, we review the discretization
of the interface conditions that can be used to 
enforce a local capillary equilibrium at the
interface between the fracture and the matrix. 
The formulation follows the work of
\cite{van1995effect} and the discretization 
employed in
\cite{cances2009finite,hamon2016capillary}. 
We introduce two degrees of freedom located at
each interface between two rock regions with 
different capillary pressure functions. 
Specifically, $S^{(m)}_{ij}$ and $S^{(f)}_{ij}$ 
are the saturations located at the
interface on the side of the matrix and on the 
side of the fracture, respectively.
The numerical flux is then computed by finding the
saturation pair, $(S^{(m)}_{ij},S^{(f)}_{ij})$,
that satisfies the following two-by-two
nonlinear system for a fixed pair of
cell-centered saturations $(S_i,S_j)$ 
and a fixed total flux $\textcolor{black}{\overline{u}_{T,ij}}$:
\begin{equation}
 \left\{
\begin{array}{ll}
      F_{w,ij}^{n+1} (\textcolor{black}{\overline{u}_{T,ij}},S_i,S_{ij}^{(\alpha)})= - F_{w,ji}^{n+1}(\textcolor{black}{-\overline{u}_{T,ij}},S_j,S_{ij}^{(\beta)})\\[5pt]
     \widetilde{P}^{(\alpha)}_c (S_{ij}^{(\alpha)}) = \widetilde{P}^{(\beta)}_c (S_{ij}^{(\beta)}). 
     \label{local_nonlinear_system}\\
\end{array} 
\right. 
\end{equation}
The first constraint in (\ref{local_nonlinear_system})
guarantees discrete mass flow rate conservation
of the wetting phase at the interface. Due to the fixed total
flux, equation (\ref{local_nonlinear_system}) also yields
conservation of mass flow rate for the non-wetting
phase since we have
$F^{n+1}_{\textit{n},ij} = \textcolor{black}{\overline{u}_{T,ij}} - F^{n+1}_{w,ij}$.
\textcolor{black}{The second constraint enforces the 
equality of the extended capillary pressures at the 
interface}. In the mass conservation
constraint, the fluxes are computed as 
 \begin{align}
 F_{\ell,ij}^{n+1} (\textcolor{black}{\overline{u}_{T,ij}},S_i,S_{ij}^{(\alpha)}) := V_{\ell,ij}^{n+1} (\textcolor{black}{\overline{u}_{T,ij}},S_i,S_{ij}^{(\alpha)}) +  G_{\ell,ij}^{n+1} (S_i,S_{ij}^{(\alpha)})+  C_{\ell,ij}^{n+1} (S_i,S_{ij}^{(\alpha)}),
 \label{flux_local_nonlinear_system}
\end{align} 
where the third argument of the flux is evaluated with
the interface saturation
in the same rock region.

We assumed in Section~\ref{section_mathematical_model} that
capillary pressure is a strictly decreasing function of
saturation. As a result, the interface conditions yield
a unique solution pair whenever the flux approximation used to
compute (\ref{flux_local_nonlinear_system}) is monotone with
respect to saturation. Therefore, as in the homogeneous case, we
can consider two different methods to discretize this flux.
\textcolor{black}{In the scheme referred to as PPU-C, we use PPU to evaluate 
the fluxes in the first
line of (\ref{local_nonlinear_system}). Instead, in the
IHU-C scheme, IHU is applied to discretize these fluxes.}
These fluxes are computed with the methodologies described
in Sections~\ref{subsection_phase_potential_upwinding} and
\ref{subsection_implicit_hybrid_upwinding}, except that 
each flux is based  on a one-sided transmissibility,
$\widehat{T}_{ij}$, and a half depth difference, $\widehat{\Delta z}_{ij}$. 

We summarize the numerical procedure employed
to solve (\ref{local_nonlinear_system}) in
Algorithm~\ref{alg:local_nonlinear_solver}.
\textcolor{black}{In our current investigation, we can choose
$\Omega_{\alpha}$ to be the matrix since $P_{c,\textit{max}}^{(m)}
> p^{(f)}_\textit{max}$ and
$P_{c,\textit{min}}^{(m)} < p^{(f)}_{\textit{min}} = 0$. 
Then, we use the extended pressure condition
stated in the second line of (\ref{local_nonlinear_system})
to write
\begin{equation}
  S^{(f)}_{ij} = h(S^{(m)}_{ij}) := 
(P^{(f)}_c)^{-1} \big(
  \widetilde{P}^{(m)}_c(S^{(m)}_{ij}) \big),
\end{equation}
as a function of $S^{(m)}_{ij}$. We employ
Newton's method to solve the scalar mass conservation equation 
at the interface using $S^{(m)}_{ij}$ as primary variable.}
The general case can be handled by switching the saturation
variable so that the function $h$ in the local nonlinear system
remains well defined whenever one of the phase pressures is
discontinuous at the interface \citep{alali2018analysis}. 

\textcolor{black}{The derivatives computed in the last two lines of
Algorithm~\ref{alg:local_nonlinear_solver} are then used when the flux
terms are assembled into the global Jacobian matrix. The implicit
function theorem is used to compute the derivatives of the
interfacial saturation, $S^{(\alpha)}_{ij}$, with respect to
the cell-centered variables, $\tau_i$ and $\tau_j$
(\ref{appendix_local_solver_derivatives}). Noting that the
local nonlinear systems are only solved up to a user-defined
tolerance, we have observed
that the performance of the global nonlinear solver is
better when the left flux (respectively, right flux) is
used to compute the derivatives of the variables in the left cell
(respectively, right cell).}

\begin{remark}
Our formulation yields a
unique solution when capillary pressure is a strictly
decreasing saturation function. Extending the formulation
to capillary pressure functions that are not strictly
decreasing requires modifying (\ref{local_nonlinear_system}) as in
\cite{brenner2013finite,brenner2017immiscible} and will
be considered in future work.
\end{remark}

\textcolor{black}{  
\begin{remark}
\label{remark_computational_cost}    
Interface conditions are used to improve the accuracy of the scheme
at the interfaces between different rock types on coarse grids.
As shown in Section~\ref{section_numerical_examples}, achieving
the same accuracy without interface conditions would
require to significantly refine the mesh in the neighborhood of
these interfaces, which, in 3D, becomes extremely expensive.
Instead, although the interface conditions require additional
computations, they only involve local, single-variable solves
at the mesh interfaces where capillary pressure is discontinuous
in space. These local systems are eliminated during the assembly
and therefore do not need to be introduced in the global Jacobian
system.
In addition, we propose to combine interface conditions with a
hybrid discretization of the flux given in
Section~\ref{subsection_implicit_hybrid_upwinding}.
This hybrid flux discretization yields
a robust scheme that is particularly well suited for problems with
strong buoyancy and capillary forces. As shown in
\cite{BRENNER2020109357} on
challenging, three-dimensional problems, the improved global nonlinear
behavior achieved with the hybrid scheme can offset the
additional cost induced by the local interface conditions.
\end{remark}
}

\begin{algorithm}[H]
  \SetAlgoLined
  \caption{\label{alg:local_nonlinear_solver}Solution algorithm for the local nonlinear system (\ref{local_nonlinear_system}).}
  \BlankLine
            
 \KwData{Cell-centered saturations $S_i$ and $S_j$, and total
 flux $\textcolor{black}{\overline{u}_{T,ij}}$.} \vspace{0.1cm}
 \KwResult{Fluxes $F^{n+1}_{\ell,ij}$ and $F^{n+1}_{\ell,ji}$, along with the flux derivatives with respect to the cell-centered primary variables in $i$ and $j$.}
 
 \vspace{0.3cm}
 
 \textbf{Initial guess:} \\ \vspace{0.1cm}
 $S^{(\alpha)}_{ij} \longleftarrow \frac{1}{2}(S_i+S_j)$

 \vspace{0.3cm}
 
 \textbf{Newton loop:} \\ \vspace{0.1cm}
 \For{$\nu = 1, \dots, n^{\textit{loc}}_{\textit{max}}$}{
    
   \vspace{0.2cm}
    
   \textit{\textbf{A)}} Evaluate the residual and its derivative
   \\
   \textcolor{black}{$R_{ij} \longleftarrow F_{\ell,ij}^{n+1} \big(\textcolor{black}{\overline{u}_{T,ij}},S_i,S_{ij}^{(\alpha)} \big) + F_{\ell,ji}^{n+1} \big(-\textcolor{black}{\overline{u}_{T,ij}},S_j,h(S_{ij}^{(\alpha)}) \big)$ \\ \vspace{0.1cm}}
   $J_{ij} \longleftarrow \displaystyle \frac{\partial R_{ij} }{ \partial S_{ij}^{(\alpha)}}$
   
   \vspace{0.3cm}
   
   \textit{\textbf{B)}} Check convergence  
   \\
   \uIf{$|R_{ij} / (\phi_i V_i + \phi_j V_j)| < 10^{-9}$}
   {
    break
   }
   
   \vspace{0.2cm}      
   
   \textit{\textbf{C)}} Update the primary saturation \\
   $S^{(\alpha)}_{ij} \longleftarrow S^{(\alpha)}_{ij} - R_{ij} / J_{ij}$
   
 }
 
 \vspace{0.3cm}
 
 \textbf{Compute the derivatives of the fluxes:} \\ \vspace{0.1cm}
  $\textcolor{black}{\displaystyle \frac{\partial }{ \partial \tau_i} \big[F_{\ell,ij}^{n+1} \big(\overline{u}_{T,ij},S_i,S_{ij}^{(\alpha)} \big) \big]  \longleftarrow
  \frac{\partial F^{n+1}_{\ell,ij} }{ \partial \overline{u}_{T,ij} }  \frac{ \partial \overline{u}_{T,ij} }{ \partial \tau_i }
  + \frac{ \partial F^{n+1}_{\ell,ij} }{ \partial \tau_i }
  + \frac{ \partial F^{n+1}_{\ell,ij} }{ \partial S^{(\alpha)}_{ij} } \frac{ \partial S^{(\alpha)}_{ij} }{\partial \tau_i}}$ \\ 
  
  $\textcolor{black}{\displaystyle \frac{\partial }{ \partial \tau_j} \big[F_{\ell,ji}^{n+1} \big(-\overline{u}_{T,ij},S_j,h(S_{ij}^{(\alpha)}) \big) \big] \longleftarrow
  -\frac{\partial F^{n+1}_{\ell,ji} }{ \partial \overline{u}_{T,ji} } \frac{ \partial \overline{u}_{T,ij} }{ \partial \tau_j }
  + \frac{ \partial F^{n+1}_{\ell,ji} }{ \partial \tau_j }
  + \frac{ \partial F^{n+1}_{\ell,ji} }{ \partial S^{(\beta)}_{ij} } \frac{ \partial h }{\partial S} \frac{\partial S^{(\alpha)}_{ij}}{\partial \tau_j}}$
\end{algorithm}

\section{\label{section_truncation_error_analysis}Numerical flux properties}

We first compare the PPU and IHU numerical
fluxes with viscous and capillary forces in
homogeneous one-dimensional media. We consider 
the fluid properties of the matrix, i.e.,
quadratic relative permeabilities and the 
capillary pressure of
Fig.~\ref{fig:scal_data_capillary_pressure}
defined by $p^{(m)}_e = 3 \, \text{psi}$ and
$\theta^{(m)} = 4$. \textcolor{black}{In this section,
we assume that $\overline{u}_T$ is constant in space and
time.} The behavior of the schemes for 
heterogeneous test cases will
be studied in 
Section~\ref{section_numerical_examples}.

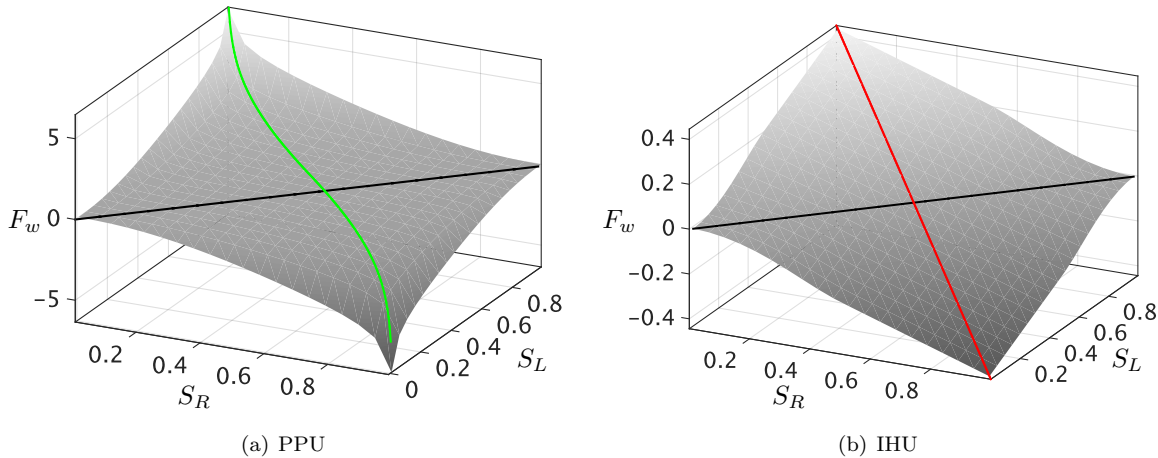
\begin{figure}[H]
\begin{center}
\subfigure[PPU]{
\begin{tikzpicture}
\node[anchor=south west,inner sep=0] at (0,0){
\includegraphics[width=0.447\textwidth]{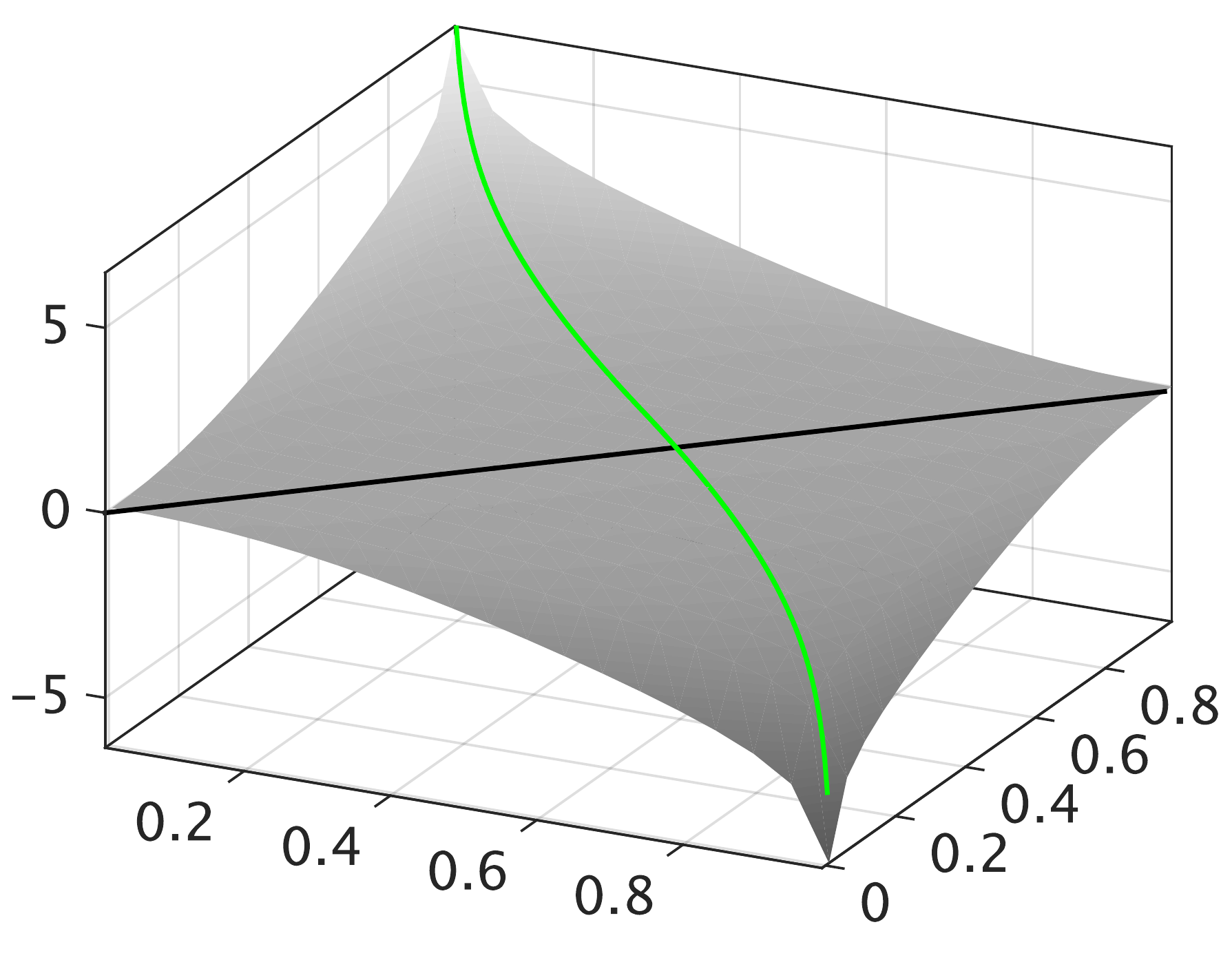}
};
\node (ib_1) at (2.2,0.3) {$S_R$};
\node (ib_1) at (6.7,0.8) {$S_L$};
\node (ib_1) at (-0.05,2.65) {$F_w$};
\end{tikzpicture}
\label{fig:surface_plot_ppu_ut_0}
}
\subfigure[IHU]{
\begin{tikzpicture}
\node[anchor=south west,inner sep=0] at (0,0){
\includegraphics[width=0.447\textwidth]{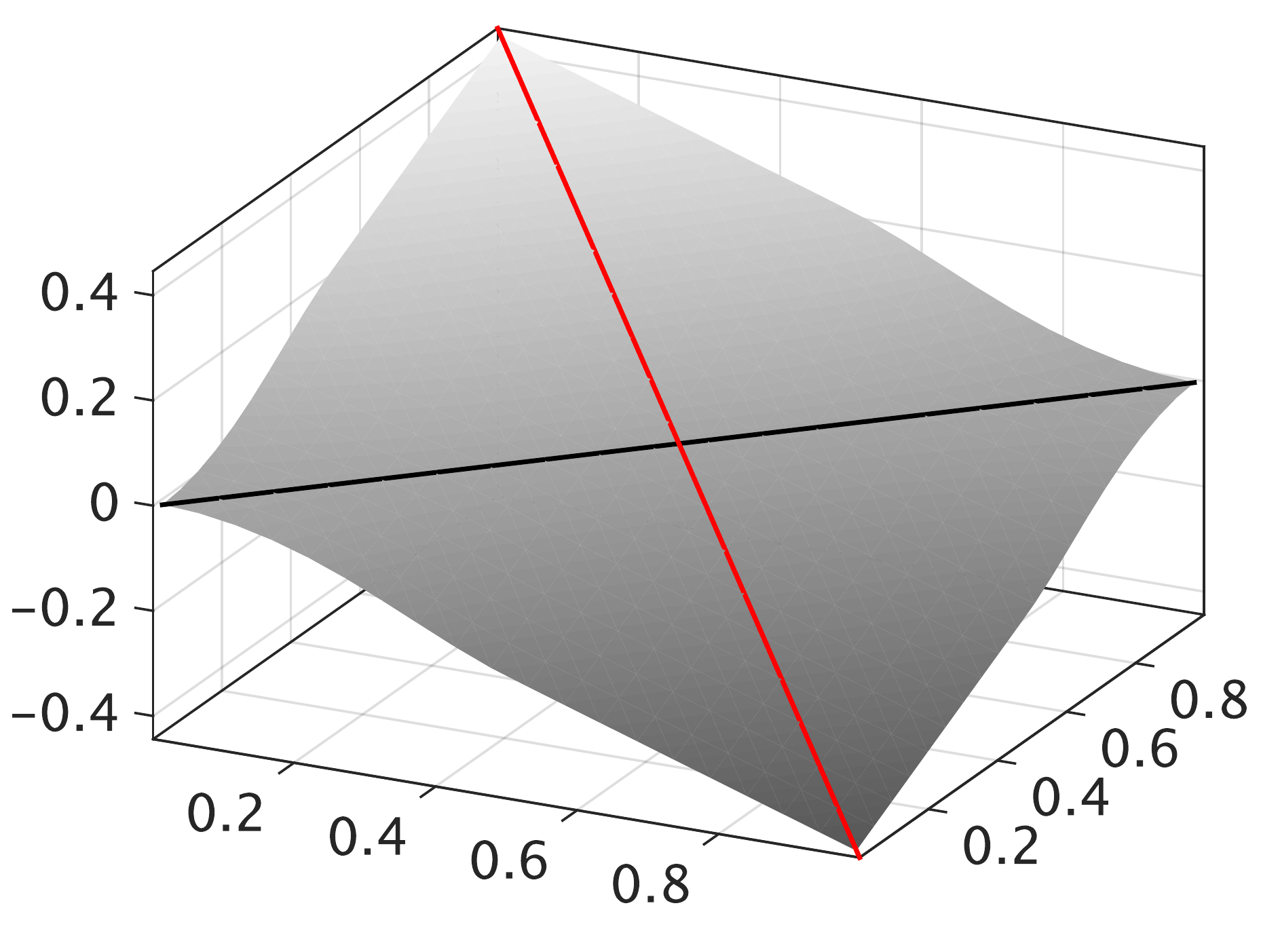}
};
\node (ib_1) at (2.2,0.3) {$S_R$};
\node (ib_1) at (6.7,0.8) {$S_L$};
\node (ib_1) at (-0.05,2.65) {$F_w$};
\end{tikzpicture}
\label{fig:surface_plot_ihu_ut_0}
}
\vspace{-0.4cm}
\caption{\label{fig:surface_plots_ut_0}
Wetting-phase numerical flux as a function
of a left saturation, $S_L$, and a right saturation, $S_R$, using
PPU in \subref{fig:surface_plot_ppu_ut_0} and IHU in
\subref{fig:surface_plot_ihu_ut_0}. We consider the pure capillary
case and neglect buoyancy and viscous forces. We use $\overline{u}_T = 0$,
$T_{LR} = 1$, quadratic relative permeabilities and the
capillary pressure function of
Fig.~\ref{fig:scal_data_capillary_pressure} defined by
$p^{(m)}_e = 3 \, \text{psi}$ and $\theta^{(m)} = 4$.
The solid black line shows the saturations for which
the flow direction changes.} 
\end{center}
\end{figure}

We first discuss the pure capillary case 
($\overline{u}_T = 0$) shown in
Fig.~\ref{fig:surface_plots_ut_0}. For 
saturations close to the endpoints, 0 and 1, we 
observe that the PPU scheme
overestimates the absolute magnitude of the capillary force
since the numerical flux tends to infinity. This difference
stems from the fact that the PPU capillary flux is constructed by
approximating directly the gradient of capillary pressure.
Conversely, the IHU numerical flux is based on a linearized
form involving the capillary diffusion coefficient
(\ref{nonlinear_capillary_diffusion_coefficient})
and remains bounded in the entire saturation range. We show in the
spontaneous imbibition test of
Section~\ref{subsection_spontaneous_imbibition}
that IHU produces more accurate predictions of
the imbibition rate on coarse grids than with PPU.

\begin{figure}[H]
\begin{center}
  \subfigure[PPU]{
\begin{tikzpicture}
\node[anchor=south west,inner sep=0] at (0,0){
  \includegraphics[width=0.447\textwidth]{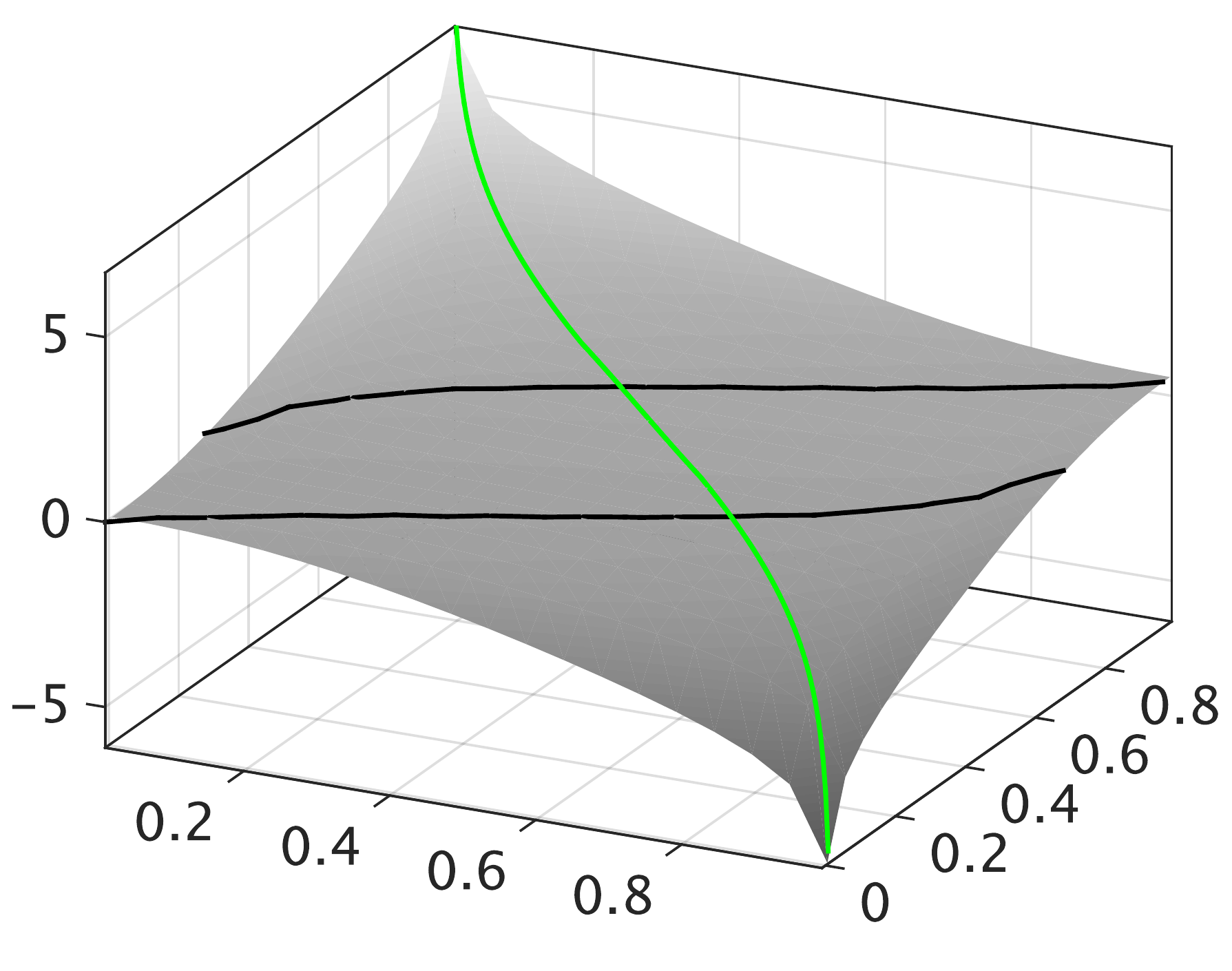}
};
\node (ib_1) at (2.2,0.3) {$S_R$};
\node (ib_1) at (6.7,0.8) {$S_L$};
\node (ib_1) at (-0.05,2.65) {$F_w$};

\end{tikzpicture}
\label{fig:surface_plot_ppu_ut_0.5}
}
\subfigure[IHU]{
\begin{tikzpicture}
\node[anchor=south west,inner sep=0] at (0,0){
  \includegraphics[width=0.447\textwidth]{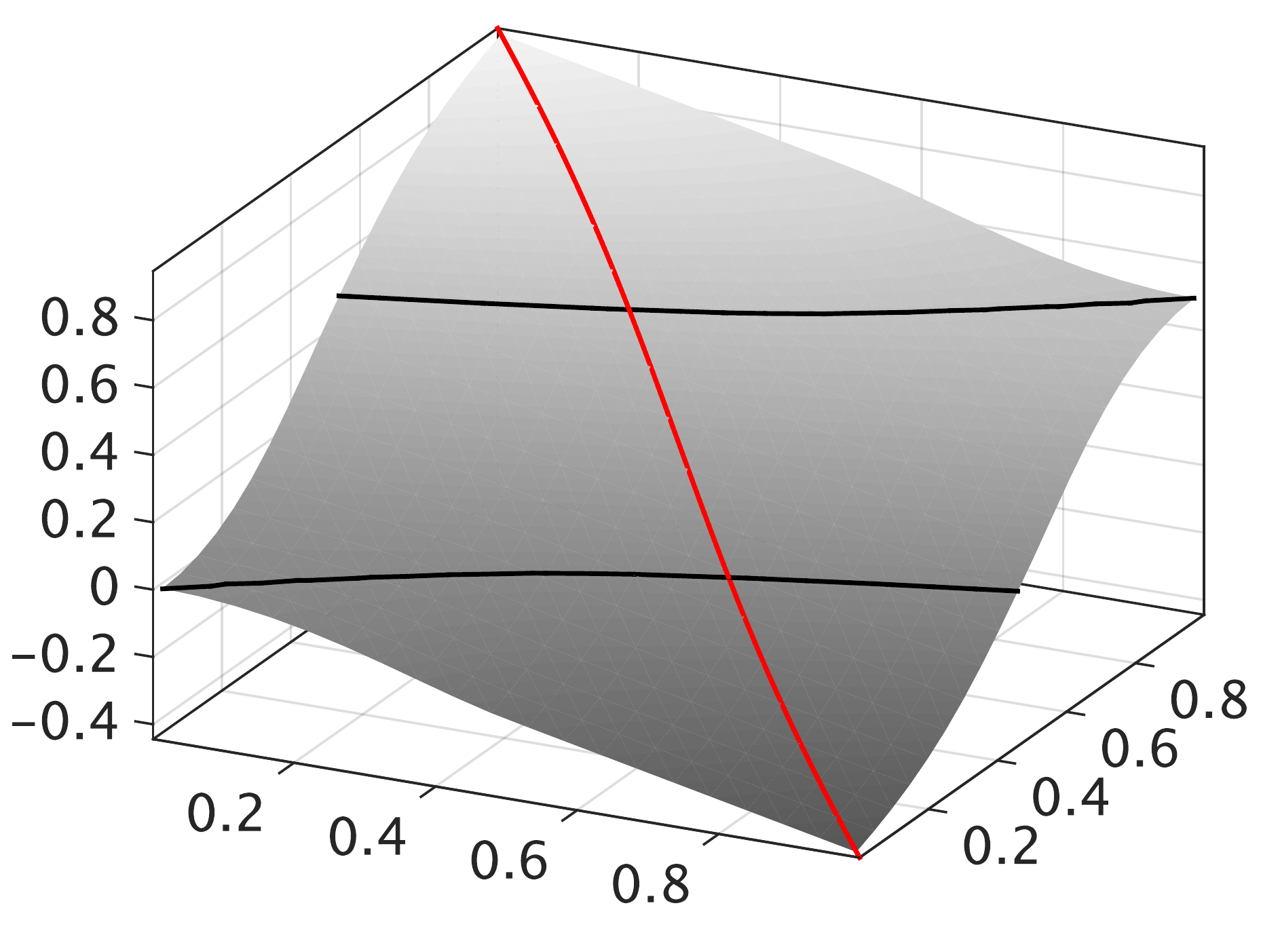}
};
\node (ib_1) at (2.2,0.3) {$S_R$};
\node (ib_1) at (6.7,0.8) {$S_L$};
\node (ib_1) at (-0.05,2.65) {$F_w$};
\end{tikzpicture}
\label{fig:surface_plot_ihu_ut_0.5}
}

\vspace{-0.4cm}
\caption{\label{fig:surface_plots_ut_0.5}
Wetting-phase numerical flux as a function
of a left saturation, $S_L$, and a right saturation, $S_R$, using
PPU in \subref{fig:surface_plot_ppu_ut_0.5} and IHU in
\subref{fig:surface_plot_ihu_ut_0.5}. We consider the
viscous-capillary
case and neglect buoyancy. We use $\overline{u}_T = 1/2$, $T_{LR} = 1$,
quadratic relative permeabilities and the capillary pressure
function of Fig.~\ref{fig:scal_data_capillary_pressure}
defined by $p^{(m)}_e = 3 \, \text{psi}$ and $\theta^{(m)} = 4$.
The solid black lines
show the saturations for which the flow direction for one of the
phases changes. 
}
\end{center}
\end{figure}

\begin{figure}[H]
  \begin{center}
      
\subfigure[]{
\begin{tikzpicture}
\node[anchor=south west,inner sep=0] at (0,0){
\includegraphics[width=0.34\textwidth]{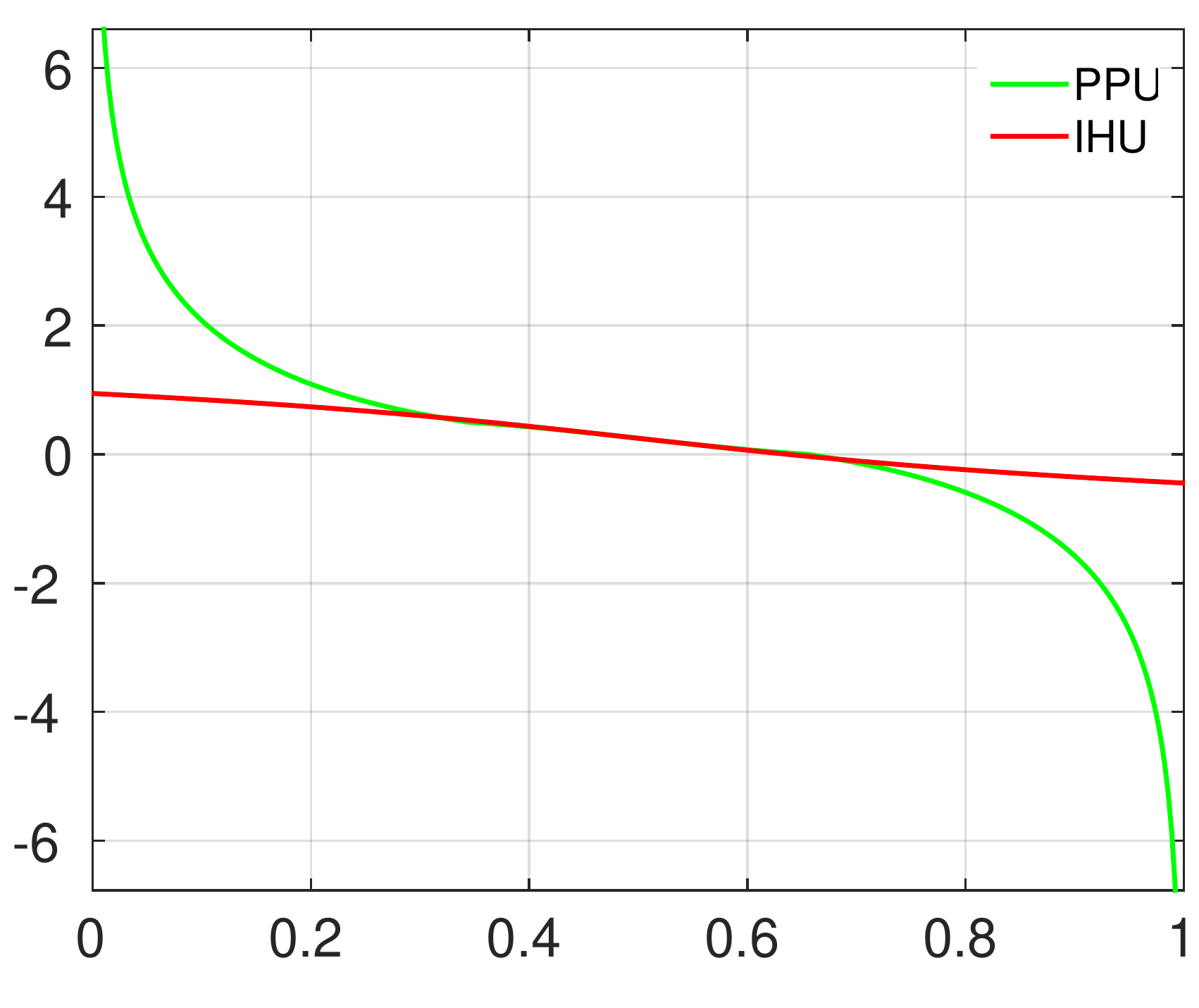}
};
\node (ib_1) at (2.85,-0.05) {$S$};
\node (ib_1) at (-0.2,2.45) {$F_w$};
\end{tikzpicture}
\label{fig:slice_ut_0.5}
}
\hspace{1.4cm}
\subfigure[]{
\begin{tikzpicture}
\node[anchor=south west,inner sep=0] at (0,-0.05){
\includegraphics[width=0.3505\textwidth]{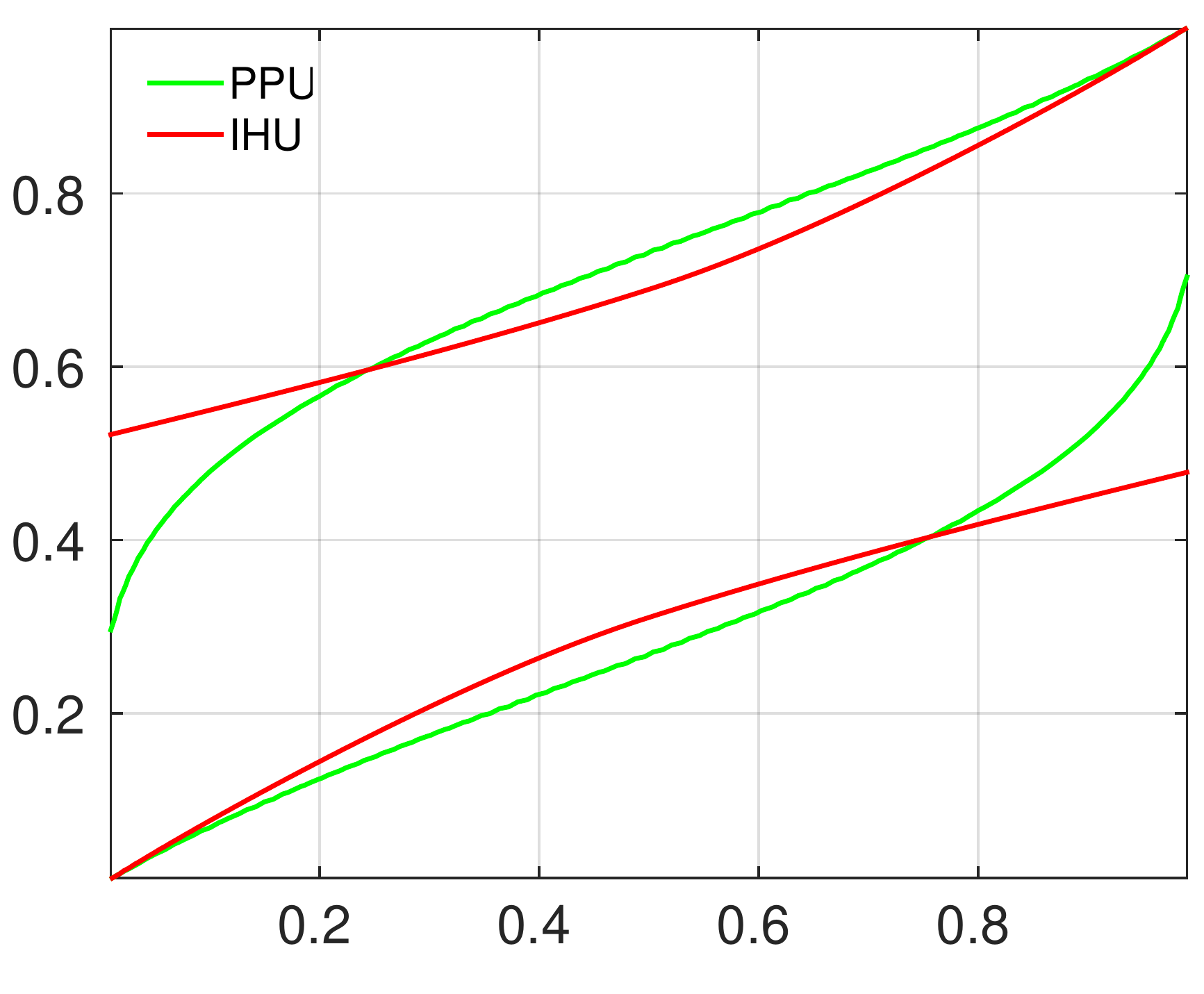}
};
\node (ib_1) at (3.05,-0.05) {$S_R$};
\node (ib_1) at (-0.2,2.5) {$S_L$};

\node (ib_1) at (1.78,3.75) {\scriptsize $w$: $L$ to $R$};
\node (ib_1) at (1.705,3.5) {\scriptsize $\textit{n}$: $R$ to $L$};

\node (ib_1) at (3.85,1.35) {\scriptsize $w$: $R$ to $L$};
\node (ib_1) at (3.775,1.1) {\scriptsize $\textit{n}$: $L$ to $R$};

\node (ib_1) at (2.85,2.65) {\scriptsize $w$: $L$ to $R$};
\node (ib_1) at (2.775,2.4) {\scriptsize $\textit{n}$: $L$ to $R$};

\end{tikzpicture}
\label{fig:contours_ut_0.5}
}
\vspace{-0.4cm}
\caption{\label{fig:contours_slice_ut_0.5}
In \subref{fig:slice_ut_0.5}, wetting-phase numerical fluxes obtained
with IHU and PPU along the line defined by $S_L + S_R = 1$ in
Fig.~\ref{fig:surface_plots_ut_0.5}. In \subref{fig:contours_ut_0.5},
the countercurrent flow regions of the IHU and PPU numerical fluxes, \textcolor{black}{defined respectively 
by $F^{\textit{PPU}}_w(\overline{u}_{T},S_L,S_R)
 F^{\textit{PPU}}_{n}(\overline{u}_{T},S_L,S_R) < 0$ and 
$F^{\textit{IHU}}_w(\overline{u}_{T},S_L,S_R)
 F^{\textit{IHU}}_{n}(\overline{u}_{T},S_L,S_R) <  0$,}
are in the top left and bottom right corners. We still consider the
case $\overline{u}_T = 1/2$.
}
\end{center}
\end{figure}

We now consider the viscous-capillary case with $\overline{u}_T = 1/2$ shown
in Figs.~\ref{fig:surface_plots_ut_0.5} and
\ref{fig:contours_slice_ut_0.5}. 
We see in Figs.~\ref{fig:surface_plots_ut_0.5} and
\ref{fig:slice_ut_0.5} that the PPU
approach still overestimates the capillary force, yielding
a large numerical flux for $S = 0$ and $S = 1$
while the numerical flux remains bounded
with the IHU scheme. But,
considering now Fig.~\ref{fig:contours_ut_0.5},
we observe that the two schemes produce
different numerical fluxes close to the center of the
saturation range (for $S_L \in [0.3, 0.7]$
and $S_R \in [0.3, 0.7]$). This is because
the split treatment of the mobilities in IHU
alters the viscous-capillary balance of forces 
and \textcolor{black}{produces a different cocurrent 
flow region as shown in 
Fig.~\ref{fig:contours_ut_0.5}. The PPU and IHU 
cocurrent flow regions contain the saturation 
pairs $(S_L,S_R)$ such that 
$F^{\textit{PPU}}_w(\overline{u}_{T},S_L,S_R)
 F^{\textit{PPU}}_{n}(\overline{u}_{T},S_L,S_R) > 0$ and 
$F^{\textit{IHU}}_w(\overline{u}_{T},S_L,S_R)
 F^{\textit{IHU}}_{n}(\overline{u}_{T},S_L,S_R) > 0$,
respectively.} We quantify this difference
with a truncation error analysis and show in
Section~\ref{subsection_forced_imbibition} that this leads to
different predictions of non-wetting phase trapping with PPU
and IHU.

Next, we derive the truncation error for the IHU and PPU schemes.
The calculation is done using the problem setup of
Section~\ref{subsection_forced_imbibition} since we will apply this
analysis to explain the respective behaviors of IHU and PPU for
cocurrent flow in a medium undergoing forced imbibition.
We assume one-dimensional cocurrent viscous-capillary flow with a (scalar) velocity
$u_T = 1/2$. We further assume that the saturation profile is
monotone, since this is the case in the homogeneous region
considered in Section~\ref{subsection_forced_imbibition}. Let the
wetting-phase saturation $x \mapsto S(x)$ be a twice
differentiable function of space, with 
$S_i = S(x_i)$. For cocurrent flow, the viscous truncation
error is by construction the same for the two upwindings: 
\begin{equation}
  \textcolor{black}{
  \mathcal{E}_{V}^{\textit{H}} = \mathcal{E}_{V}^{P} = - \frac{\Delta x}{2} \Big(\frac{\partial^2 }{\partial x^2} M_w(S) \Big)_{S_i} u_T + O(\Delta x^2),}
    \label{truncation_error_viscous_term}
\end{equation}
where $u_T$ is the scalar total velocity and $M_w$ is defined as
\begin{equation}
M_w(S) := \frac{\lambda_w(S)}{\lambda_w(S) + \lambda_{n}(S)}.
\end{equation}
Next, considering the case
$S_{max}< S_{i+1} \leq S_i \leq S_{i-1}$, where $S_{max}$ is the
saturation for which the capillary diffusion coefficient of the
matrix attains its maximum,
we use Taylor series expansions to obtain the following 
expression of the capillary truncation error:
\begin{equation}
  \textcolor{black}{
    \mathcal{E}^{\textit{H}}_C = \frac{k\Delta x}{2}\Bigg(\frac{\partial^2}{\partial x^2}\Big(D(S)\frac{\partial S}{\partial x}\Big)_{S_i}  - \frac{\partial}{\partial x}\Big(D(S)\frac{\partial^2 S}{\partial x^2}\Big)_{S_i}\Bigg) + O(\Delta x^2),}
\label{ihu_truncation_error_capillary_term}
\end{equation}
where $D$ is the capillary diffusion coefficient defined in
(\ref{nonlinear_capillary_diffusion_coefficient}) and 
$\Delta x := x_{i+1} - x_i := x_i - x_{i-1}$.
The same expression holds for the case
$S_{i-1} \leq S_i \leq S_{i+1} \leq S_{max}$.
The IHU capillary truncation error term  of
(\ref{ihu_truncation_error_capillary_term}) is 
always bounded which is in agreement
with Fig.~\ref{fig:surface_plot_ihu_ut_0.5}. 
With the PPU scheme, considering again the cocurrent
flow case, we write the capillary truncation error
term as
\begin{equation}
\textcolor{black}{
\mathcal{E}^{P}_C  =  -\mathcal{E}^{\textit{H}}_C + \frac{k\Delta x}{2}\Bigg[\frac{\partial}{\partial x}\Bigg(M_w(S) \lambda_n(S) \frac{\partial^2 P_c}{\partial S^2}(S)\Big(\frac{\partial S}{\partial x}\Big)^2\Bigg)_{S_i}\Bigg] + O(\Delta x^2).}
\label{ppu_truncation_error_capillary_term}
\end{equation}
\textcolor{black}{
We note that, in (\ref{ppu_truncation_error_capillary_term}),
the term between brackets in the right-hand
side contains the second derivative of capillary pressure and is
unbounded for small saturations. Therefore, for any $\epsilon_1 > 0$,
there exists a $\eta_1 > 0$ sufficiently small such that $S < \eta_1$
implies $|\mathcal{E}^{P}_C| > |\mathcal{E}^{H}_C| + \epsilon_1$,
leading
to large errors in the flux computation with PPU when the
wetting-phase saturation is close to zero. This unbounded term in
the leading capillary truncation error term is responsible for the 
large values observed in Figs.~\ref{fig:surface_plot_ppu_ut_0.5}
and \ref{fig:slice_ut_0.5}.}

\textcolor{black}{But, since the capillary pressure function
considered in this work has an inflection point, there is a
saturation, $S^{\text{inflec}}$, such that
$\frac{\partial^2 P_c}{\partial S^2}(S^{\text{inflec}}) = 0$.
Therefore, considering any small $\epsilon_2 > 0$, one can find a
$\Delta x_2 > 0$  and $\eta_2 > 0$ sufficiently small such that
$\Delta x < \Delta x_2$, $|S-S^{\text{inflec}}| < \epsilon_2$ implies
$| \mathcal{E}^P_C - (-\mathcal{E}^H_C)| < \epsilon_2$. This means
that in a neighborhood of $S^{\text{inflec}}$, the capillary
truncation errors produced by PPU and IHU have an opposite sign.
Since the two schemes produce the same viscous truncation error,
$\mathcal{E}^P_V = \mathcal{E}^H_V$, for the cocurrent case, the
interaction between the capillary and viscous errors is going to
affect the accuracy of the IHU and PPU schemes differently. 
}
  
Specifically, when we apply this analysis in
Section~\ref{subsection_forced_imbibition}, we will see that the
leading terms in $\mathcal{E}^{\textit{H}}_V$ and
$\mathcal{E}^{\textit{H}}_C$ have the same sign, which means that the
capillary truncation error of IHU tends to amplify the viscous
truncation error. Instead, the leading terms in
$\mathcal{E}^{P}_V$ and $\mathcal{E}^{P}_C$ have an opposite
sign, and therefore the two error terms of the PPU scheme tend to
cancel each other, yielding a smaller viscous-capillary
truncation error than with IHU.

\section{\label{section_numerical_examples}Numerical examples}

In this section, we refer to the standard Phase-Potential Upwinding
scheme as PPU. In PPU-C, the standard scheme is modified with the 
introduction of discrete interface conditions at the interface
between the matrix and the fracture. The fluxes involved in the 
discrete interface conditions are also discretized with PPU.
Finally, in IHU-C, Implicit Hybrid Upwinding is employed in 
the homogeneous regions as well as in the discretization of the
fluxes involved in interface conditions. The global
nonlinear systems are solved with Newton's method. Global convergence
is achieved when
\begin{equation}
\big| V_i\phi_i \frac{S_{\ell,i}^{n+1} - S_{\ell,i}^n}{\Delta t} +  \sum_{j \in \textit{adj}(i)} F_{\ell,ij}^{n+1} - V_iq_{\ell,i} \big| / ( \phi_i V_i )
 < 10^{-8} \quad \forall \ell \in \{ n, w \}.
\end{equation}

\subsection{\label{subsection_spontaneous_imbibition}Spontaneous imbibition} 

\begin{figure}[H]
\begin{center}
\tikzstyle{int}=[draw, minimum size=2em]
\tikzstyle{init} = [pin edge={to-,thick,black}]

\begin{tikzpicture}[node distance=3cm,auto,>=latex']

  \path (0,0) node (a) {};
  \path (9,2) node (b) {};
  \path [draw=black,thick] (a) rectangle (b); 
  \path (4.5,-0.05) node (a) {};
  \path (4.5,2.1) node (b) {};
  \path [draw=black,thick,dashed] (a) -- (b); 

  \path (2.25,1.3) node (e) {Matrix};
  \path (2.25,0.7) node (e) {$S^{\textit{init}}_{\textit{n}} = 1$};
  \path (6.75,1.3) node (e) {Fracture};
  \path (6.75,0.7) node (e) {$S^{\textit{init}}_{\textit{n}} = 0$};
  
  \path (3.85,1.25) node (a) {};
  \path (5.15,1.25) node (b) {};
  \path (4.8,1.425) node (c) {$\textit{n}$};
  \path [draw=gray,very thick,->] (a) -- (b); 

  \path (5.15,0.75) node (a) {};
  \path (3.85,0.75) node (b) {};
  \path (4.1,0.575) node (c) {$\textit{w}$};
  \path [draw=gray,very thick,->] (a) -- (b); 

\end{tikzpicture}
\vspace{-0.3cm}
\caption{\label{fig:schema_first_case}
Schematic of the one-dimensional matrix-fracture model considered
in this section. The low-permeability matrix is on the left and
the fracture is on the right. 
}
\end{center}
\end{figure}
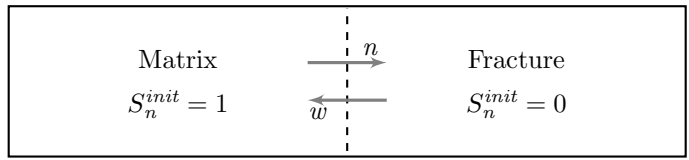

We first study the accuracy of the numerical schemes 
in predicting the hydrocarbon recovery rate driven by capillary
imbibition in a fractured porous medium to
illustrate the importance of the interface conditions. For
simplicity, we consider a one-dimensional horizontal
matrix-fracture system of length $L = 20 \, \text{m}$
sketched in Fig.~\ref{fig:schema_first_case}.
There is no source or sink term such that capillarity is
therefore the only driving force of the flow.
The left half of the domain is occupied by the matrix whose
uniform permeability is set to $k^{(m)} = 1 \, \text{mD}$. The
right half represents the fracture and has a permeability of
$k^{(f)} = 10^5 \, \text{mD}$.

The fracture pore volume is
artificially amplified by two orders of magnitude compared to
that of the matrix to simulate fast
wetting-phase injection through the fracture.
\textcolor{black}{Our assumption is that in a multidimensional
fracture network where counter-current imbibition occurs, the
non-wetting phase that enters the fracture will quickly move away due
to viscous or buoyancy forces. Hence, the wetting-phase saturation
in the fracture should remain close to one. A similar setup was
employed in the recent study of \cite{vo2019high}.}

\begin{figure}[H]
\begin{center}
\subfigure[$N^{(m)}  = N^{(f)} = 4$]{
\begin{tikzpicture}
\node[anchor=south west,inner sep=0] at (0,0){
\includegraphics[width=0.45\textwidth]{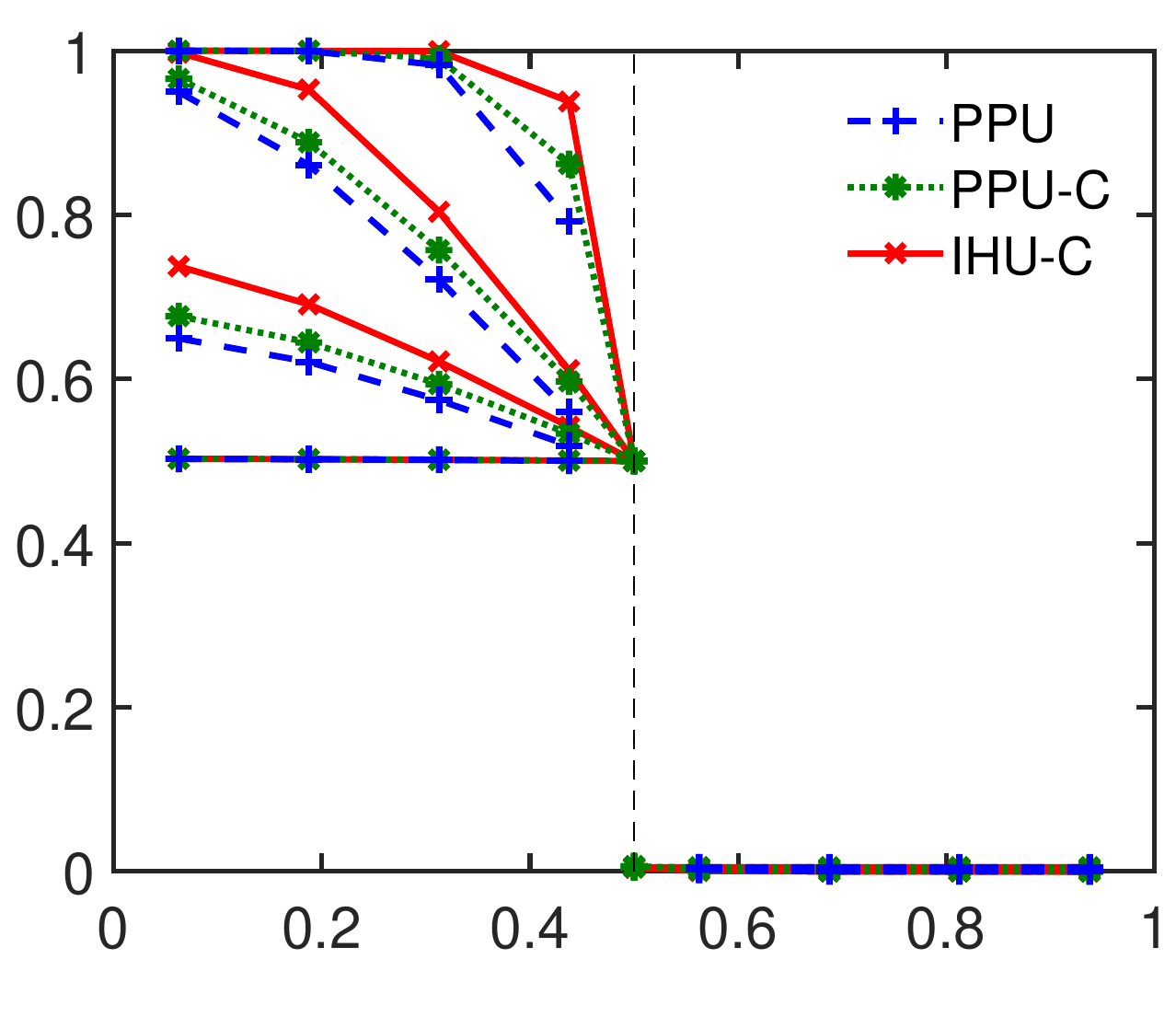}
};
\path (3.85,0.25) node (e) {$x_D$};
\node (ib_1) at (-0.15,3.35) {$S_{n}$};
\end{tikzpicture}
\label{fig:saturation_profiles_refinement_study_first_test_case_coarse_grid}}
\subfigure[$N^{(m)} = N^{(f)} = 128$]{
\begin{tikzpicture}
\node[anchor=south west,inner sep=0] at (0,0){
\includegraphics[width=0.45\textwidth]{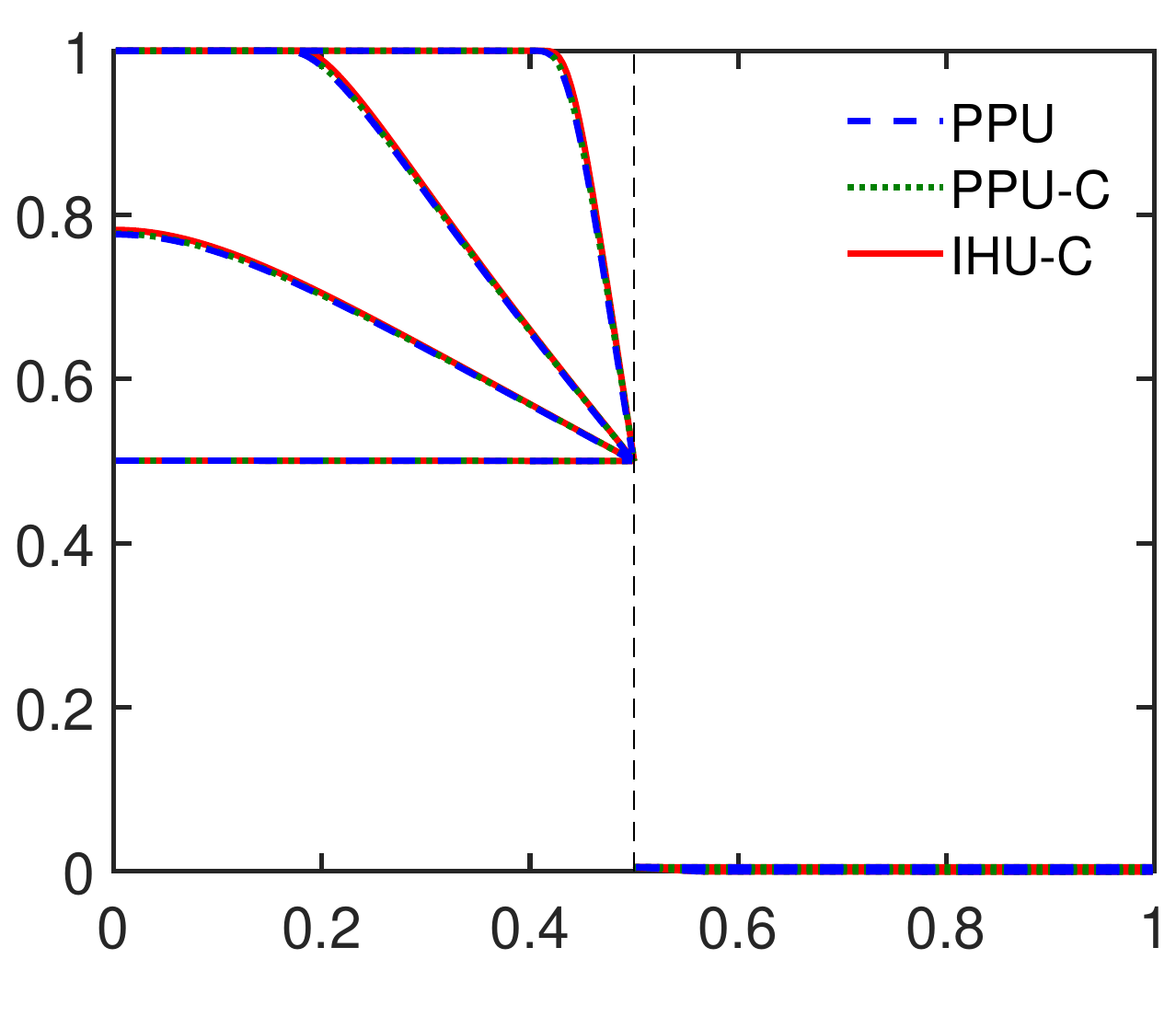}
};
\path (3.85,0.25) node (e) {$x_D$};
\node (ib_1) at (-0.15,3.35) {$S_{n}$};
\path (3.3,5.35) node (a) {};
\path (1.75,3.65) node (b) {};
\path [draw=black,very thick,->] (a) -- (b); 
\end{tikzpicture}
\label{fig:saturation_profiles_refinement_study_first_test_case_fine_grid}
}
\vspace{-0.4cm}
\caption{\label{fig:saturation_profiles_refinement_study_first_test_case}
Coarse-grid saturation profiles
\subref{fig:saturation_profiles_refinement_study_first_test_case_coarse_grid} and
fine-grid saturation profiles
\subref{fig:saturation_profiles_refinement_study_first_test_case_fine_grid} at $t_D = 0.0008$, $t_D = 0.008$, $t_D = 0.08$,
and $t_D = 0.8$ for spontaneous imbibition. We used quadratic
relative permeabilities in the matrix. The vertical dashed line
shows the
interface between the matrix on the left and the fracture on
the right. For IHU-C and PPU-C, the profiles include the
saturations obtained by solving (\ref{local_nonlinear_system})
at the interface between the matrix and the fracture.}
\end{center}
\end{figure}

We consider two sets of
saturation-dependent properties in the matrix.
In the first set, we use quadratic relative permeabilities,
leading to the diffusion coefficient of 
Fig.~\ref{fig:scal_data_diffusion_coefficient_quadratic}. In
the second set, we use cubic relative permeabilities and we 
obtain the diffusion coefficient of
Fig.~\ref{fig:scal_data_diffusion_coefficient_cubic}.
The matrix is initially fully saturated
with the non-wetting phase, and the fracture is fully saturated
with the wetting phase. The saturation profiles obtained with
the three numerical schemes until steady
state for quadratic relative permeabilities are in
Fig.~\ref{fig:saturation_profiles_refinement_study_first_test_case}. 

\begin{figure}[H]
\begin{center}
\subfigure[PPU]{
\begin{tikzpicture}
\node[anchor=south west,inner sep=0] at (0,0){
\includegraphics[width=0.45\textwidth]{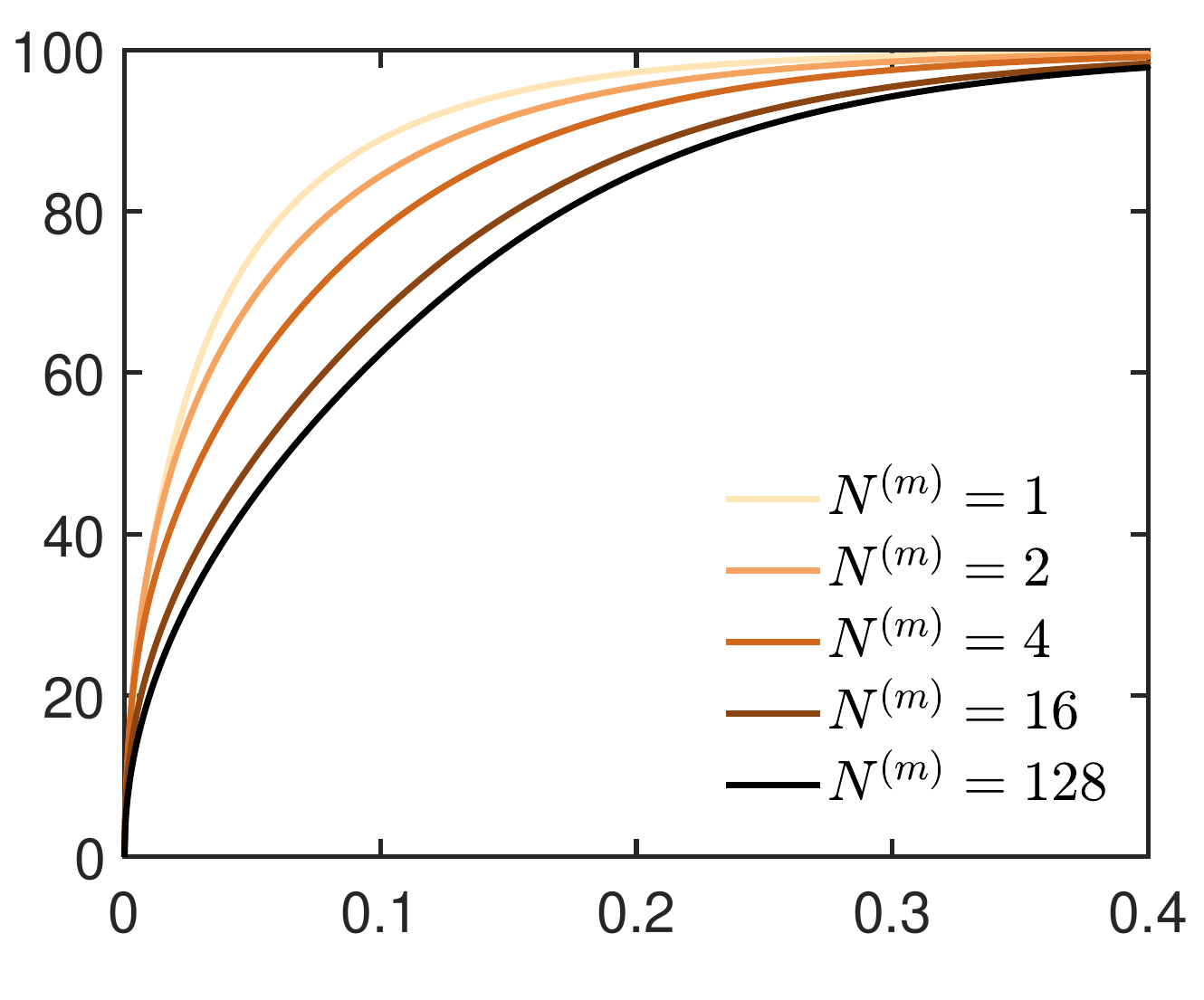}
};
\path (3.8,0.) node (e) {$t_D$};

\path (7.1,4.25) node (q) {\textcolor{white}{$\bigg\}$}};

\path (7,4.25) node (a) {};
\path (7.6,4.25) node (b) {};
\path [draw=white,very thick,->] (a) -- (b);

\node[rotate=90] (ib_1) at (-0.05,3.2) {Recovery percentage};
\end{tikzpicture}
\label{fig:recovery_rate_refinement_study_first_test_case_ppu}
}
\hspace{0.cm}
\subfigure[PPU-C]{
\begin{tikzpicture}
\node[anchor=south west,inner sep=0] at (0,0){
\includegraphics[width=0.45\textwidth]{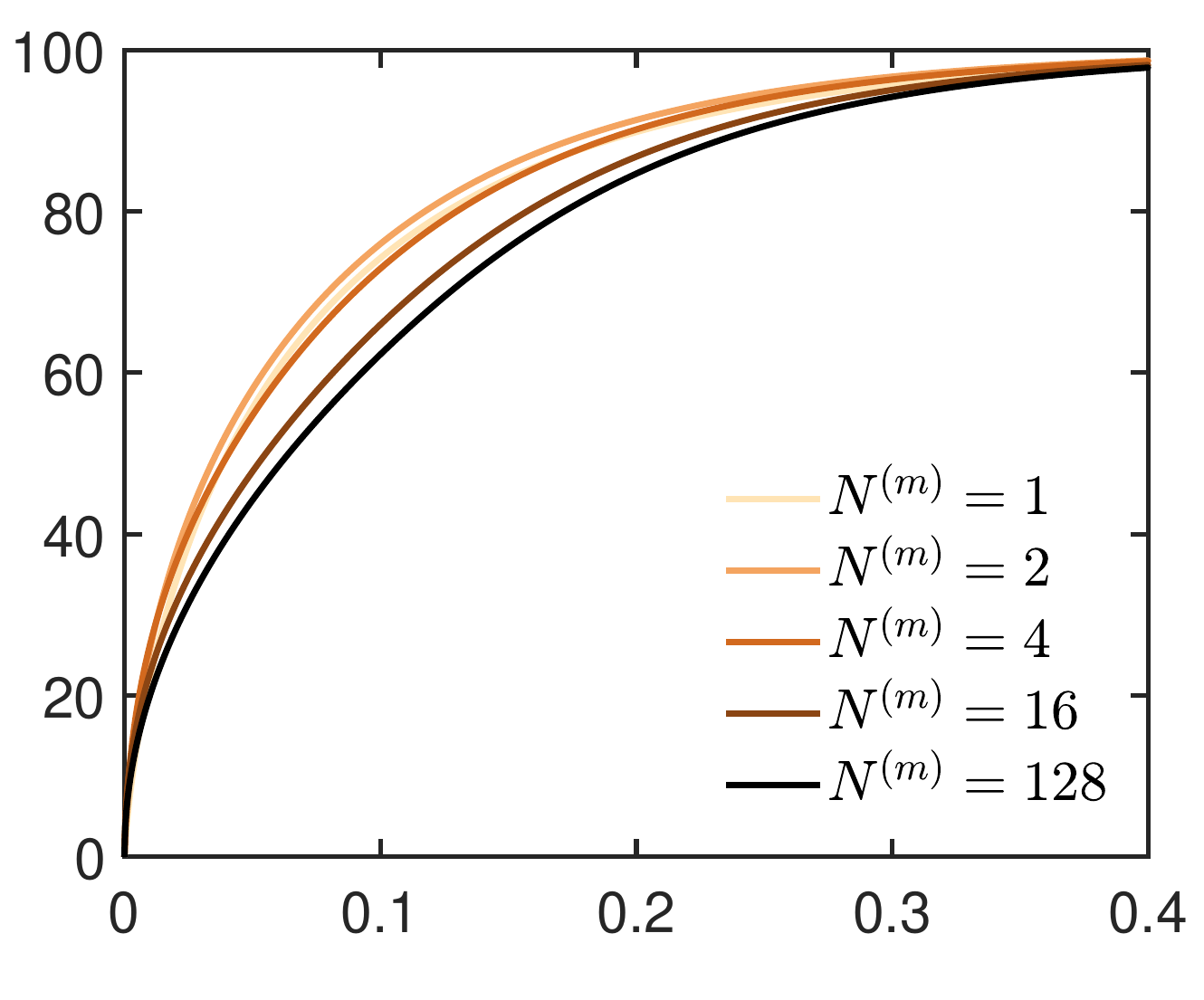}
};
\path (3.8,0.) node (e) {$t_D$};
\end{tikzpicture}
\label{fig:recovery_rate_refinement_study_first_test_case_ppu_c}
}
\subfigure[IHU-C]{
\begin{tikzpicture}
\node[anchor=south west,inner sep=0] at (0,0){
\includegraphics[width=0.45\textwidth]{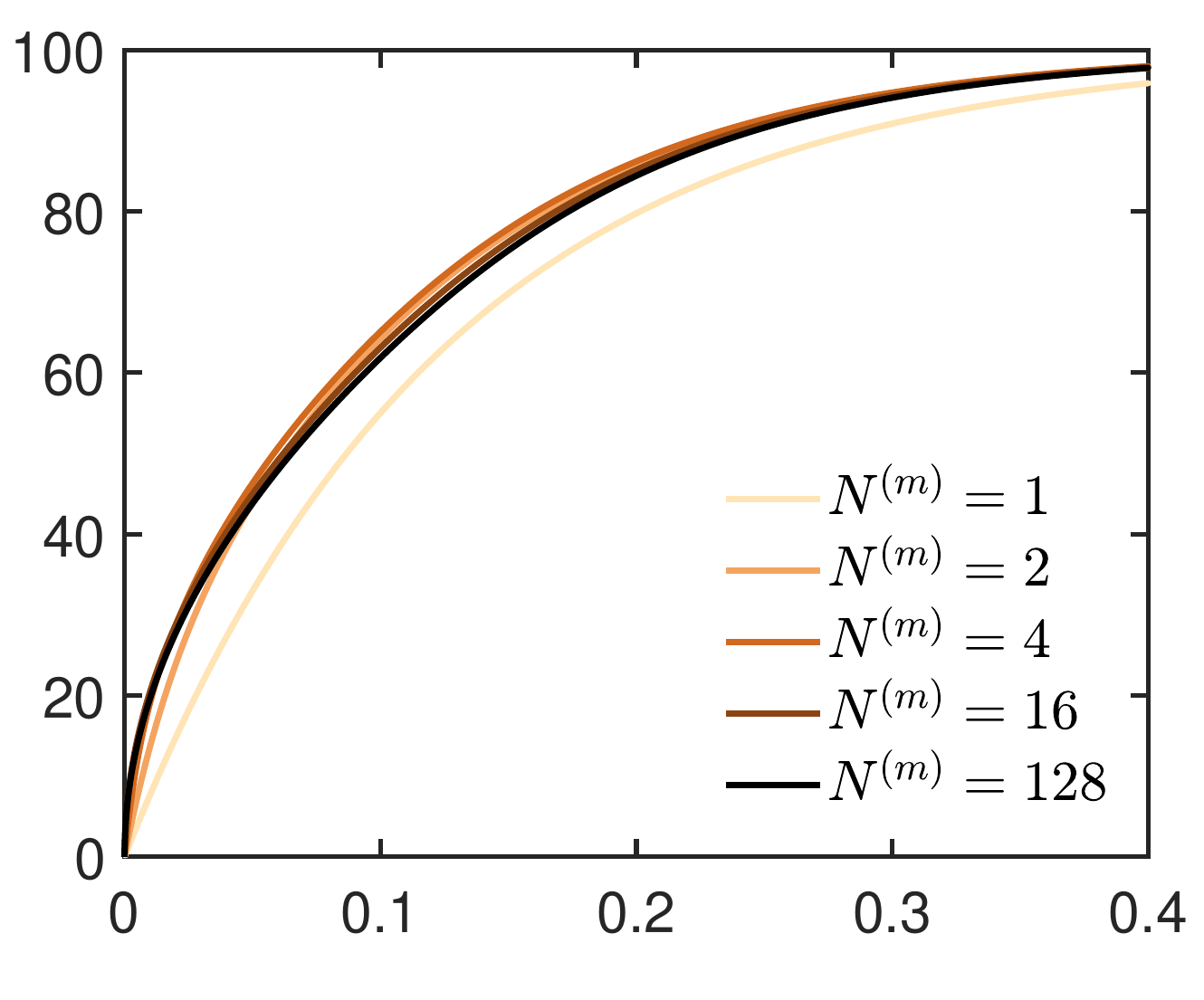}
};
\path (7.1,4.25) node (q) {$\bigg\}$};

\path (7,4.25) node (a) {};
\path (7.6,4.25) node (b) {};
\path [draw=black,very thick,->] (a) -- (b);

\path (3.8,0.) node (e) {$t_D$};
\node[rotate=90] (ib_1) at (-0.05,3.2) {Recovery percentage};
\end{tikzpicture}
\label{fig:recovery_rate_refinement_study_first_test_case_ihu_c}
}
\hspace{0.cm}
\subfigure[IHU-C (zoomed in)]{
\begin{tikzpicture}
\node[anchor=south west,inner sep=0] at (0,0){
\includegraphics[width=0.445\textwidth]{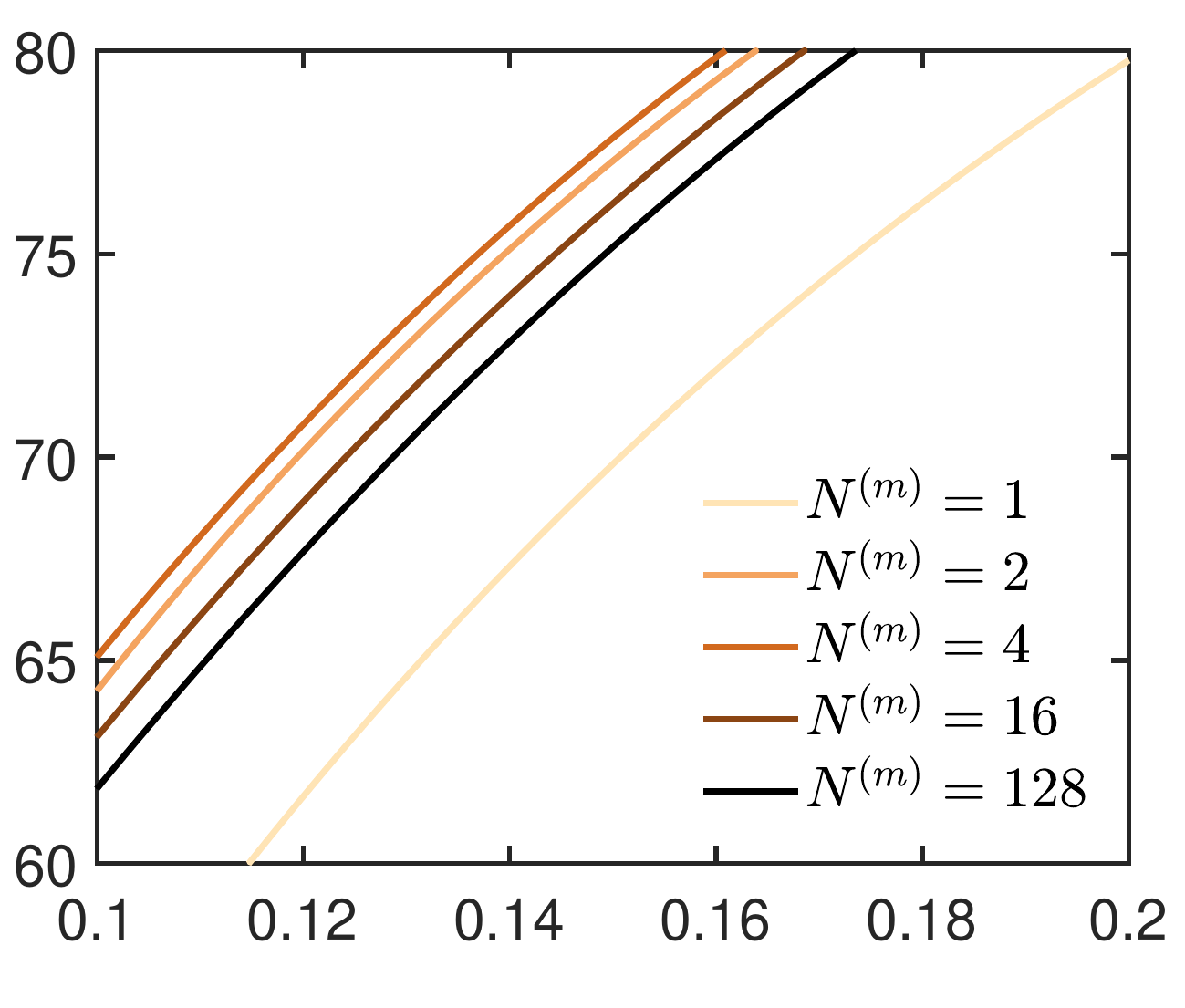}
};
\path (3.8,0.) node (e) {$t_D$};
\end{tikzpicture}
\label{fig:recovery_rate_refinement_study_first_test_case_ihu_c_zoomed}
}
\vspace{-0.4cm}
\caption{\label{fig:recovery_percentage_as_fct_of_time_refinement_study_first_test_case}
Recovery percentage as a function of time for different levels of
refinement in the matrix-fracture problem driven by spontaneous
imbibition for standard PPU
\subref{fig:recovery_rate_refinement_study_first_test_case_ppu},
PPU-C 
\subref{fig:recovery_rate_refinement_study_first_test_case_ppu_c},
and IHU-C 
\subref{fig:recovery_rate_refinement_study_first_test_case_ihu_c}, \subref{fig:recovery_rate_refinement_study_first_test_case_ihu_c_zoomed}. We used quadratic relative permeabilities. 
}
\end{center}
\end{figure}

We compute the recovery percentage at a given
dimensionless time as the ratio of the cumulative
amount of non-wetting phase that has left the matrix until this time 
over the cumulative amount at steady state. The recovery 
percentage is reported as a function of dimensionless diffusion
time defined as
\begin{equation}
 t_D :=\frac{k^{(m)} D^{(m)}_{\textit{max}} t}{\phi^{(m)} L^2}.
 \label{Eq26}
\end{equation}
The steady state is reached at $t_D \approx 0.8$.
In Fig.~\ref{fig:recovery_percentage_as_fct_of_time_refinement_study_first_test_case},
this numerical experiment is repeated for various
levels of spatial refinement in the matrix,
from a single cell in the matrix ($N^{(m)} = N^{(f)} = 1$)
to 128 cells in the matrix ($N^{(m)} = N^{(f)} = 128$).
In the refinement study,
the grid is refined at the same rate in the matrix and in the
fracture. The resolution given by $N^{(m)} = 128$ is too fine for
practical simulations but is used as a reference to compute the
error in the recovery percentage as a function of time. We see
that the standard PPU scheme significantly over-predicts the
recovery percentage for small $N^{(m)}$ -- i.e., for coarser levels
of spatial refinement. For instance, for $N^{(m)} = 1$, 
PPU predicts that the recovery reaches $80\%$
after $t_D = 0.063$), compared to $t_D = 0.171$ with the fine
reference. This over-prediction is reduced upon spatial 
refinement but with a slow convergence to the reference 
solution as $N^{(m)}$ is increased. 

The over-prediction of the recovery percentage is much less severe
for the schemes based on interface conditions.
For $N^{(m)} = 1$, PPU-C predicts the $80\%$ recovery time at
$t_D = 0.126$, as opposed to $t_D = 0.171$ in the reference solution. 
In addition, this over-prediction error remains lower than with
the standard PPU as the grid is refined. Finally, the IHU-C
scheme slightly under-predicts the recovery compared to PPU and
PPU-C. For $N^{(m)} = 1$, the IHU-C scheme predicts the $80\%$ 
recovery time at about $t_D = 0.202$ as compared to the reference
solution that predicts about $t_D = 0.171$. This under-prediction is
reduced with spatial refinement at a faster rate
than with PPU and PPU-C. Specifically, IHU-C with only two
cells in the matrix results in a very good match with the
reference solution. With $N^{(m)} = 2$ and $N^{(m)} = 4$, the
estimated times for the $80\%$-recovery are within $5\%$ of those
of the reference solution.

\begin{figure}[H]
\begin{center}
\subfigure[Quadratic relative permeabilities, $N^{(m)}=1$]{
\begin{tikzpicture}
\node[anchor=south west,inner sep=0] at (0,0){
\includegraphics[width=0.41\textwidth]{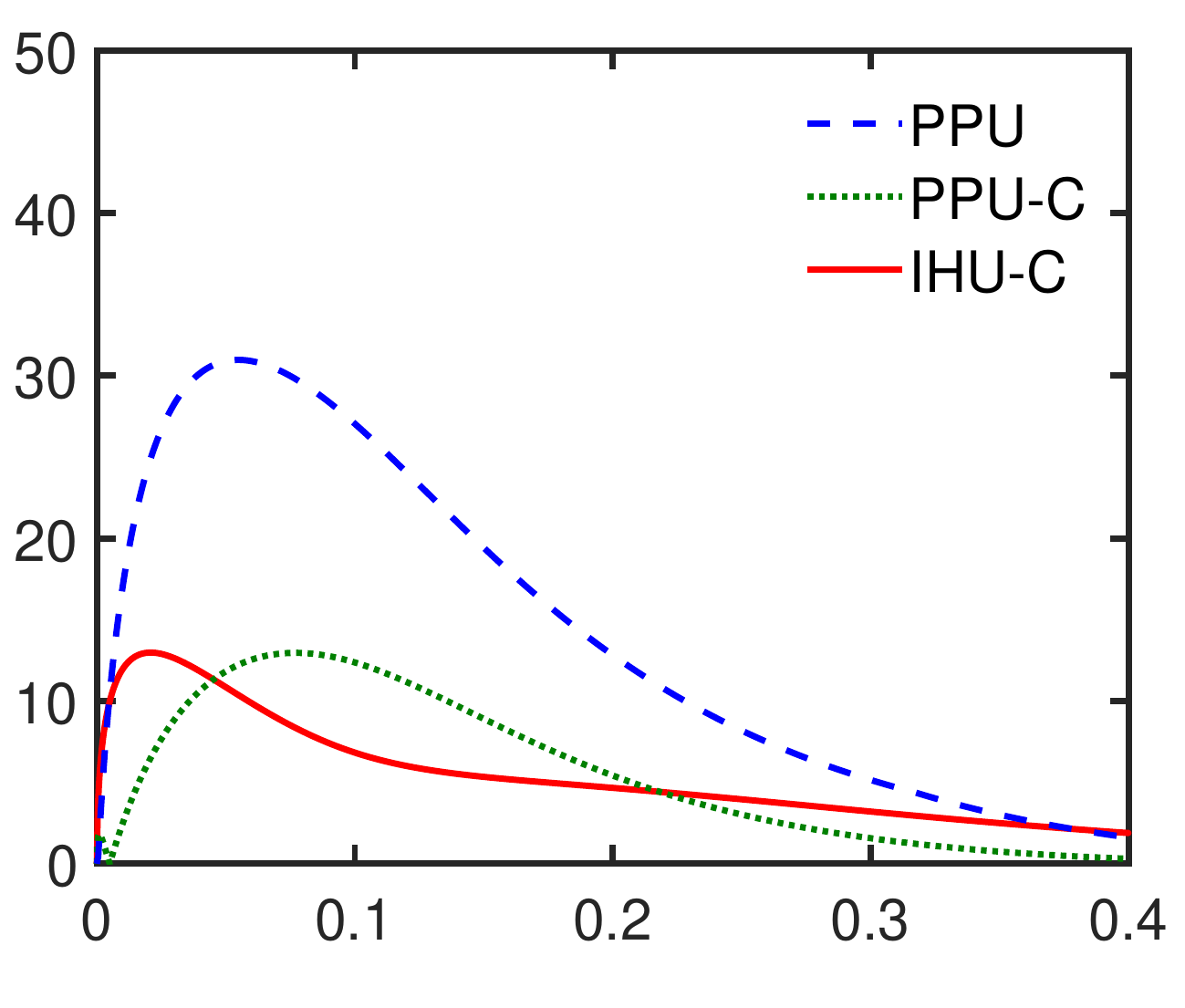}
};
\path (3.6,0.1) node (e) {$t_D$};
\node (ib_1) at (0,3.1) {$E_1$};
\end{tikzpicture}
\label{fig:recovery_rate_temporal_evolution_first_test_case_N_1_quadratic}
}
\hspace{0.4cm}
\subfigure[Cubic relative permeabilities, $N^{(m)}=1$]{
\begin{tikzpicture}
\node[anchor=south west,inner sep=0] at (0,0){
\includegraphics[width=0.41\textwidth]{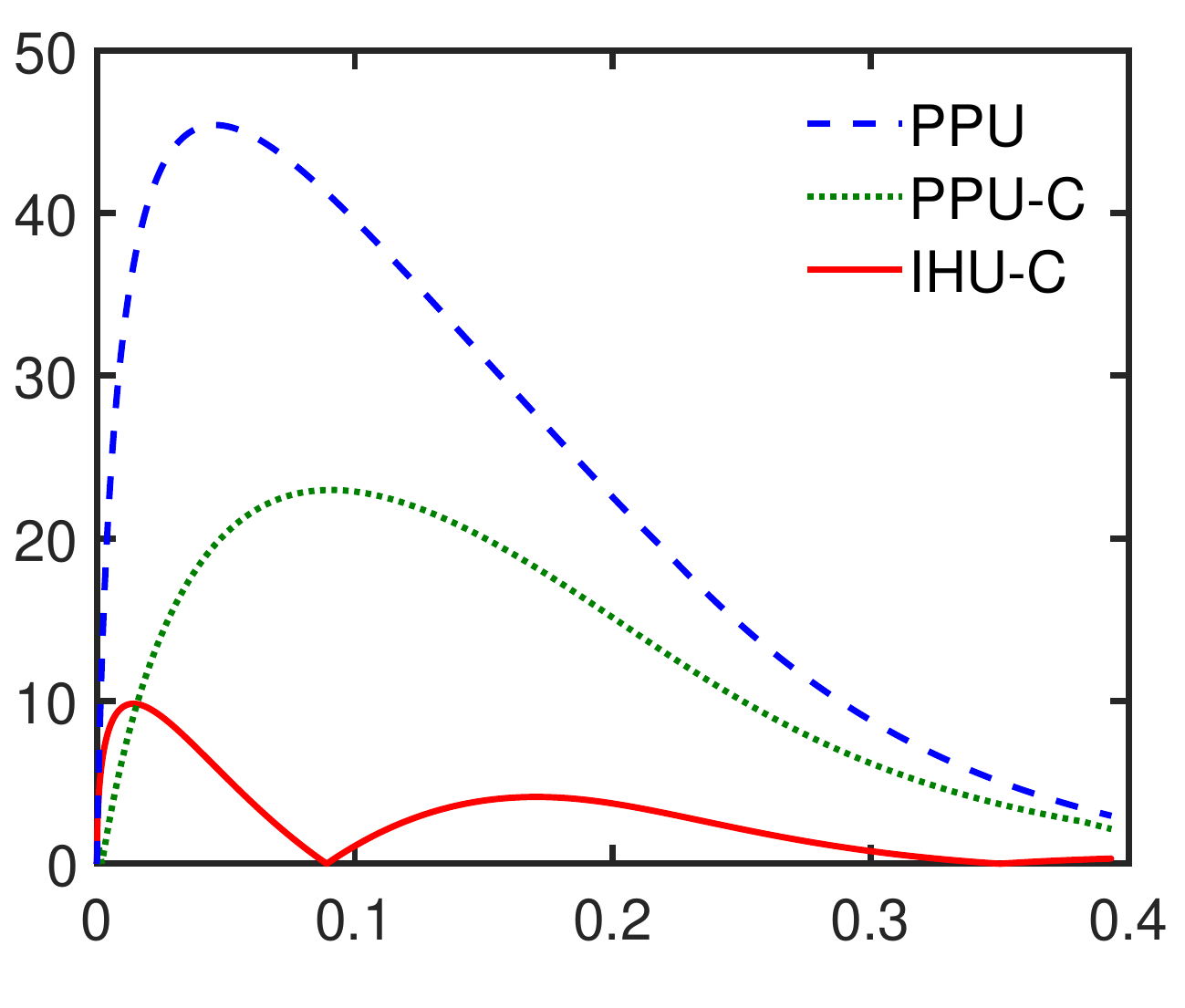}
};
\path (3.6,0.1) node (e) {$t_D$};
\end{tikzpicture}
\label{fig:recovery_rate_temporal_evolution_first_test_case_N_1_cubic}
} 

\subfigure[Quadratic relative permeabilities, $N^{(m)}=2$]{
\begin{tikzpicture}
\node[anchor=south west,inner sep=0] at (0,0){
\includegraphics[width=0.41\textwidth]{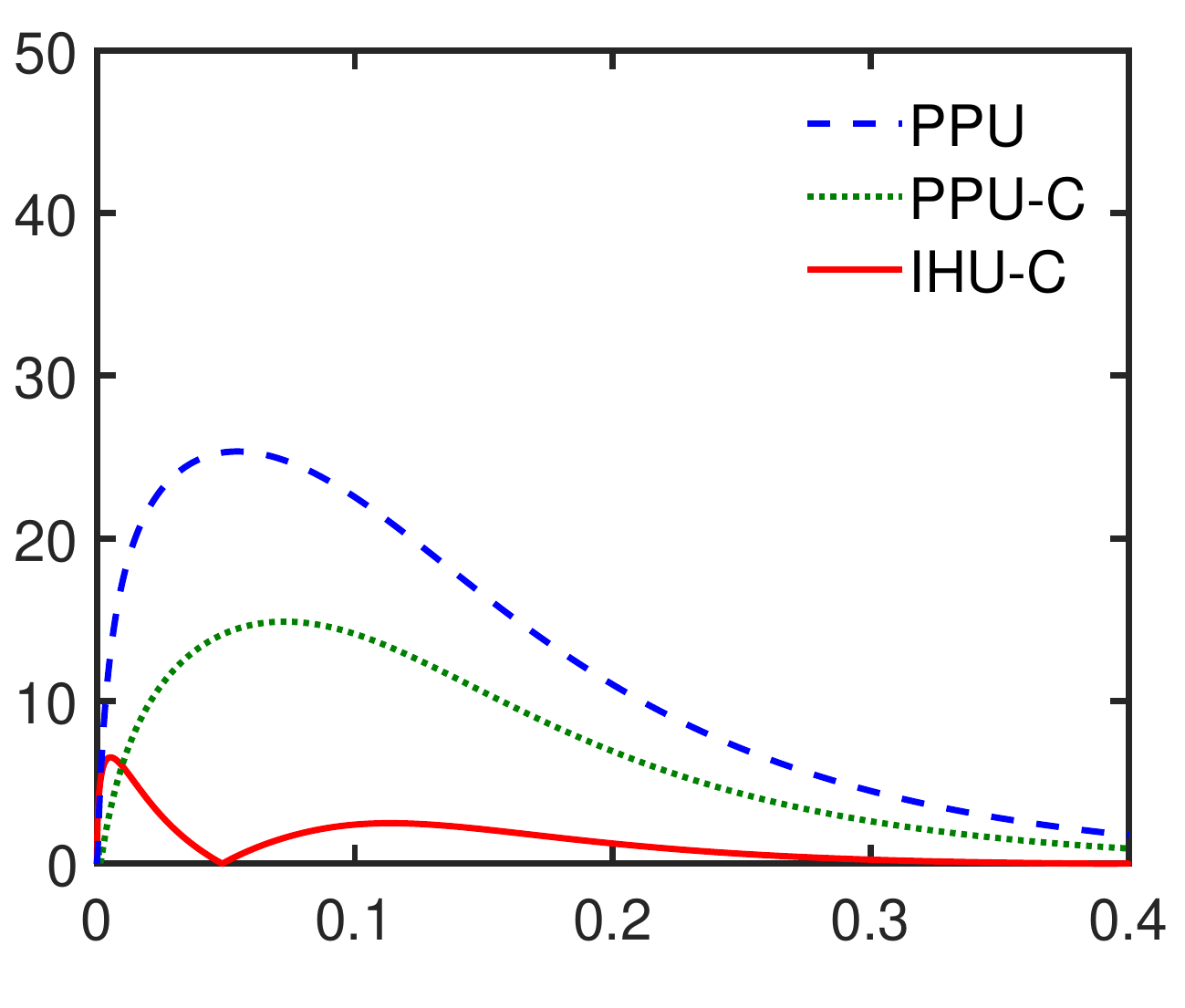}
};
\path (3.6,0.1) node (e) {$t_D$};
\node (ib_1) at (0,3.1) {$E_1$};
\end{tikzpicture}
\label{fig:recovery_rate_temporal_evolution_first_test_case_N_2_quadratic}
}
\hspace{0.4cm}
\subfigure[Cubic relative permeabilities, $N^{(m)}=2$]{
\begin{tikzpicture}
\node[anchor=south west,inner sep=0] at (0,0){
\includegraphics[width=0.41\textwidth]{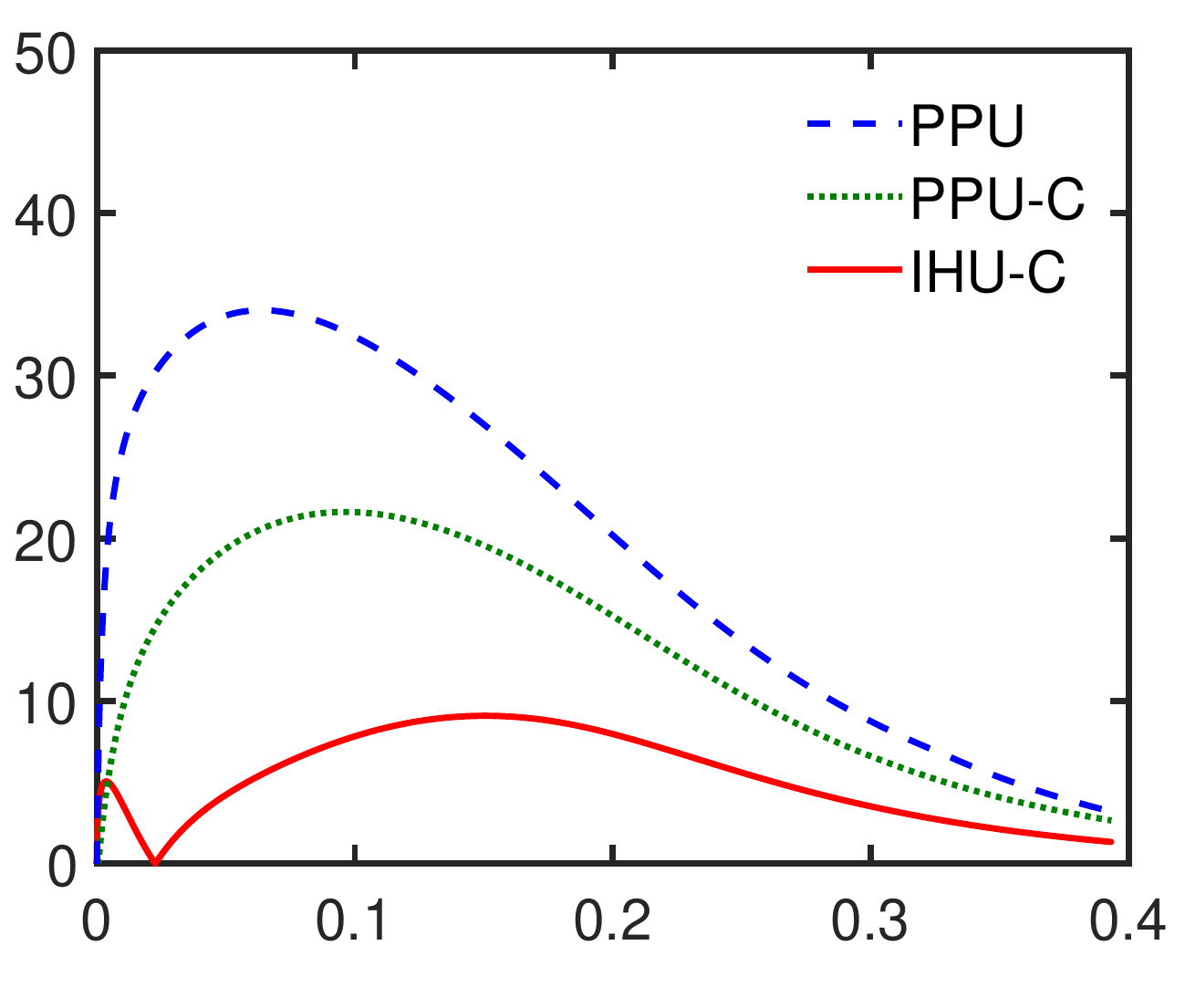}
};
\path (3.6,0.1) node (e) {$t_D$};
\end{tikzpicture}
\label{fig:recovery_rate_temporal_evolution_first_test_case_N_2_cubic}
}

\subfigure[Quadratic relative permeabilities, $N^{(m)}=8$]{
\begin{tikzpicture}
\node[anchor=south west,inner sep=0] at (0,0){
\includegraphics[width=0.41\textwidth]{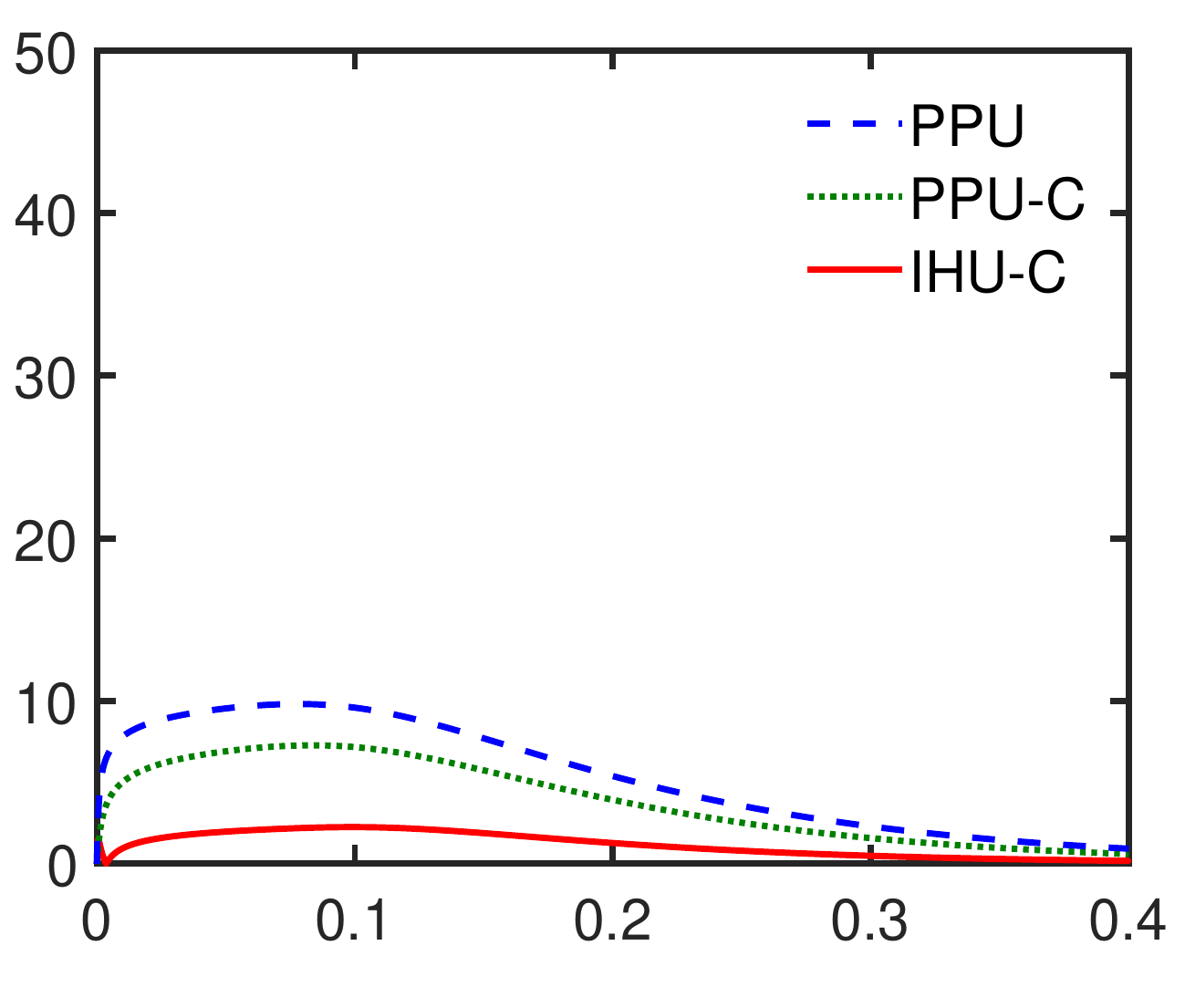}
};
\path (3.6,0.1) node (e) {$t_D$};
\node (ib_1) at (0,3.1) {$E_1$};
\end{tikzpicture}
\label{fig:recovery_rate_temporal_evolution_first_test_case_N_8_quadratic}
}
\hspace{0.4cm}
\subfigure[Cubic relative permeabilities, $N^{(m)}=8$]{
\begin{tikzpicture}
\node[anchor=south west,inner sep=0] at (0,0){
\includegraphics[width=0.41\textwidth]{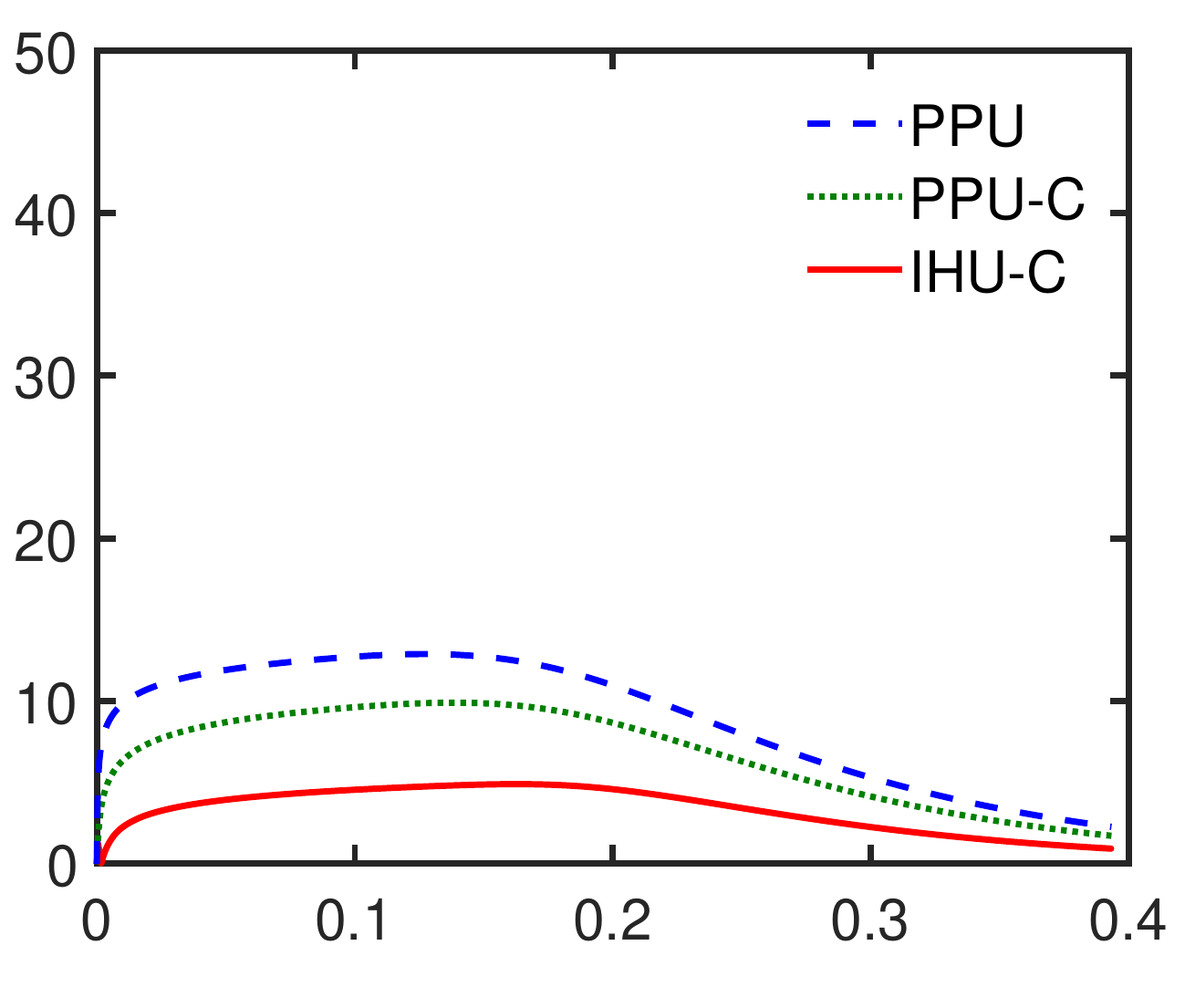}
};
\path (3.6,0.1) node (e) {$t_D$};
\end{tikzpicture}
\label{fig:recovery_rate_temporal_evolution_first_test_case_N_8_cubic}
}

\vspace{-0.4cm}
\caption[]{
\label{fig:recovery_rate_temporal_evolution_first_test_case}
Absolute error in recovery percentage as a function of dimensionless
time for different spatial resolutions until $t_D = 0.4$.
Quadratic relative
permeabilities are used in \subref{fig:recovery_rate_temporal_evolution_first_test_case_N_1_quadratic}, 
\subref{fig:recovery_rate_temporal_evolution_first_test_case_N_2_quadratic}, 
and 
\subref{fig:recovery_rate_temporal_evolution_first_test_case_N_8_quadratic}, whereas 
cubic relative
permeabilities are used in \subref{fig:recovery_rate_temporal_evolution_first_test_case_N_1_cubic}, 
\subref{fig:recovery_rate_temporal_evolution_first_test_case_N_2_cubic}, 
and 
\subref{fig:recovery_rate_temporal_evolution_first_test_case_N_8_cubic}.
}
\end{center}
\end{figure}

To distinguish early-time dynamics from the behavior at steady state,
Fig.~\ref{fig:recovery_rate_temporal_evolution_first_test_case}
shows the error in the recovery percentage as a function of
dimensionless time for different levels of spatial refinement.
The error in the recovery percentage is computed as
\begin{equation}
    E_1(N^{(m)}, \, t^n_D) := | R^n_{N^{(m)}} - R^n_{\textit{ref}} |,
\end{equation}
where $R^n_{N^{(m)}}$ (respectively, $R^n_{\textit{ref}}$) is the
recovery percentage at time $n$ for $N^{(m)}$ cells in 
the matrix (respectively, the recovery percentage at time $n$ 
for the reference solution). We consider both quadratic 
and cubic relative permeabilities. 
Figure~\ref{fig:recovery_rate_temporal_evolution_first_test_case}
illustrates that even though all the schemes predict 
the steady state solution correctly, the imbibition rate is
better captured by the schemes that rely on interface
conditions. In particular, we can see by comparing 
Figs.~\ref{fig:recovery_rate_temporal_evolution_first_test_case_N_1_quadratic}, 
\ref{fig:recovery_rate_temporal_evolution_first_test_case_N_2_quadratic}, 
and \ref{fig:recovery_rate_temporal_evolution_first_test_case_N_8_quadratic} that the
error obtained with IHU-C converges quickly 
as the number of cells is increased. 
\textcolor{black}{For cubic relperms 
(Figs.~\ref{fig:recovery_rate_temporal_evolution_first_test_case_N_1_cubic}, 
\ref{fig:recovery_rate_temporal_evolution_first_test_case_N_2_cubic}, 
and \ref{fig:recovery_rate_temporal_evolution_first_test_case_N_8_cubic}) the error decreases
at approximately the same rate for the three schemes but is 
maller with IHU-C.}

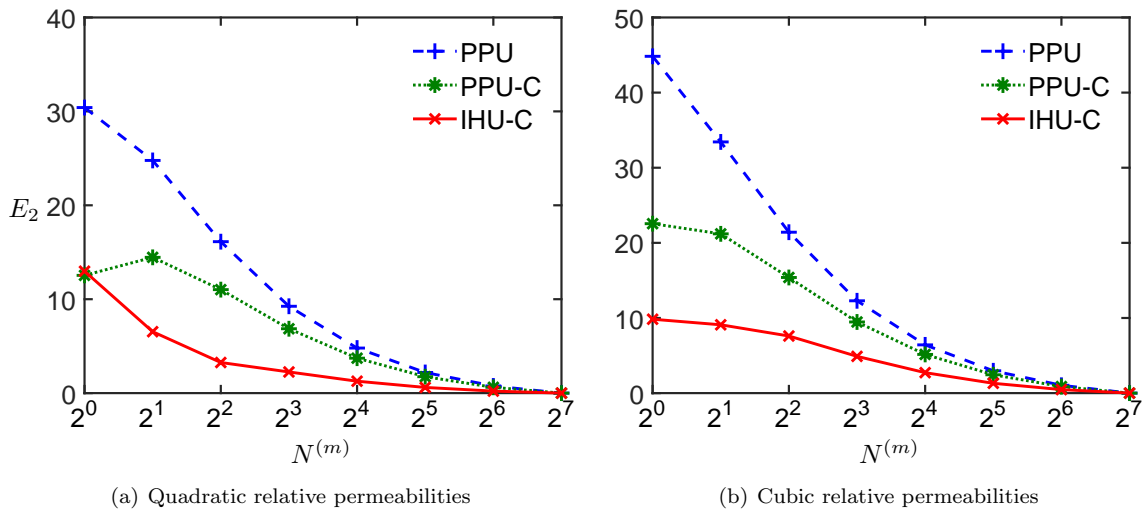
\begin{figure}[H]
\begin{center}
\subfigure[Quadratic relative permeabilities]{
\begin{tikzpicture}
\node[anchor=south west,inner sep=0] at (0,0){
\includegraphics[width=0.45\textwidth]{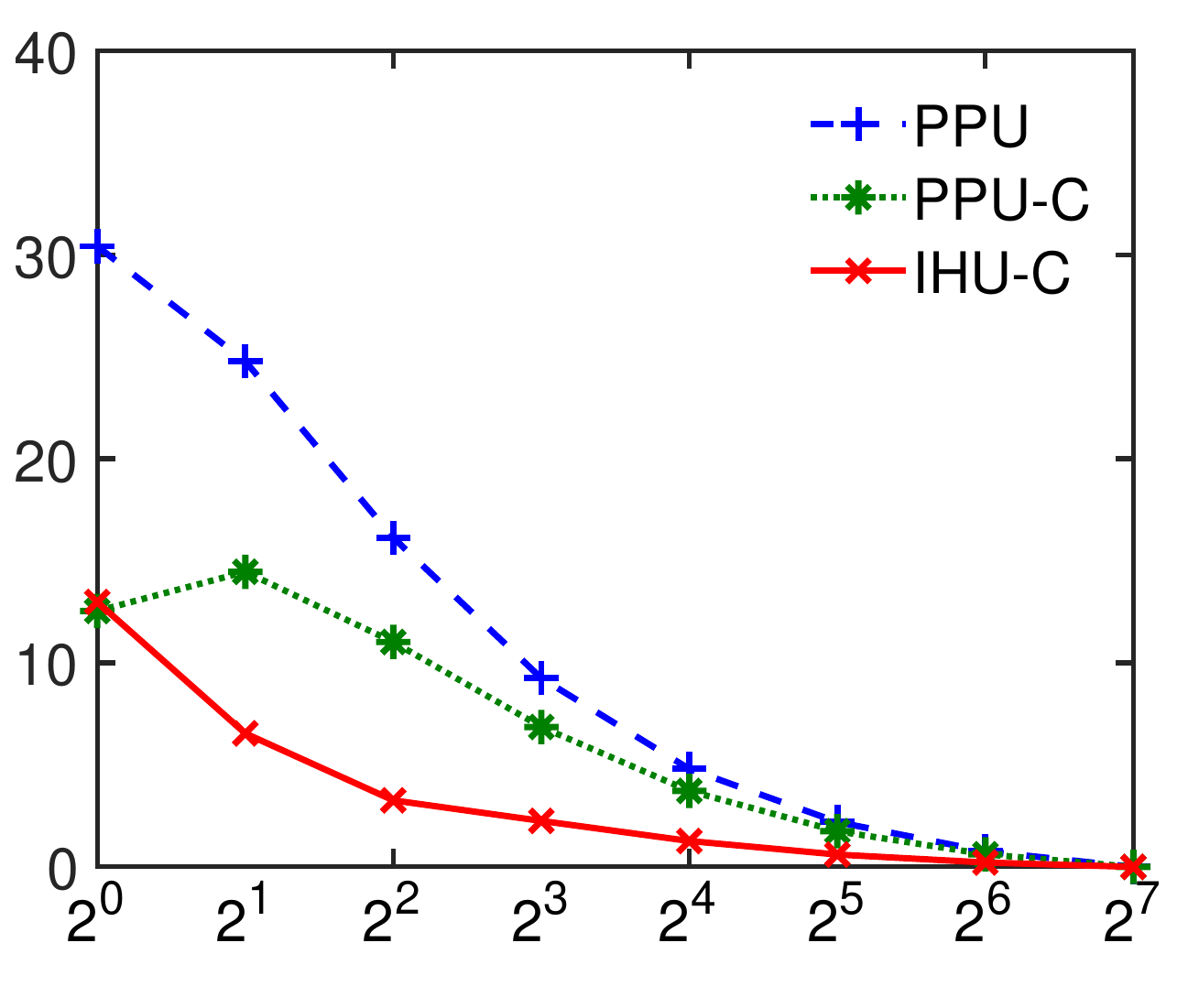}
};
\path (3.75,0.1) node (e) {$N^{(m)}$};
\node (ib_1) at (-0.2,3.3) {$E_2$};
\end{tikzpicture}
\label{fig:recovery_rate_refinement_study_first_test_case_quadratic_relperms}
}
\subfigure[Cubic relative permeabilities]{
\begin{tikzpicture}
\node[anchor=south west,inner sep=0] at (0,0){
\includegraphics[width=0.45\textwidth]{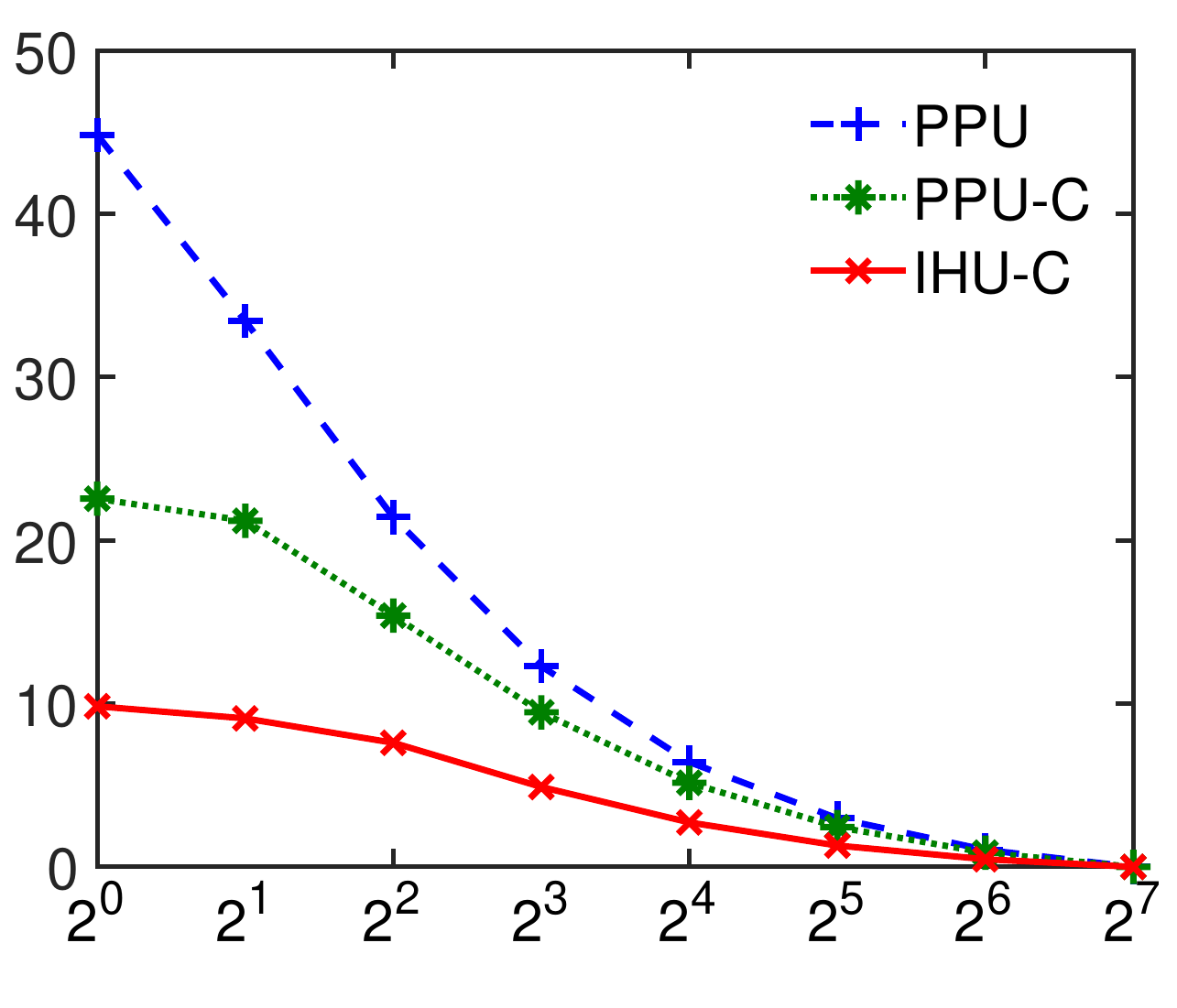}
};
\path (3.75,0.1) node (e) {$N^{(m)}$};
\end{tikzpicture}
\label{fig:recovery_rate_refinement_study_first_test_case_cubic_relperms}
}
\vspace{-0.4cm}
\caption{\label{fig:recovery_rate_refinement_study_first_test_case}
Error in the recovery percentage as a function of
the number of cells in the matrix, $N^{(m)}$. The error in 
the recovery percentage is integrated over time and is 
computed with
(\ref{norm_averaged_and_integrated_over_time}).  We show the
results 
for quadratic relative permeabilities
\subref{fig:recovery_rate_refinement_study_first_test_case_quadratic_relperms} 
and for cubic relative permeabilities
\subref{fig:recovery_rate_refinement_study_first_test_case_cubic_relperms}. 
}
\end{center}
\end{figure}

This is confirmed by
Fig.~\ref{fig:recovery_rate_refinement_study_first_test_case},
which compares the recovery error integrated over time obtained
with the three schemes for different levels of spatial
refinement. We compare the $L_{\infty}$-norm of the error in the
recovery percentage for the full simulation until steady state
as
\begin{equation}
 E_2(N^{(m)}) := \max_{n} E_1(N^{(m)}, \, t^n_D).
 \label{norm_averaged_and_integrated_over_time}
\end{equation} 
\textcolor{black}{
We highlight two key observations on
Fig. \ref{fig:recovery_rate_refinement_study_first_test_case}.
First, the three schemes produce the same 
solution upon refinement in space. Second, on 
coarser grids, large errors are observed with 
the standard PPU. These errors are significantly
reduced with the introduction of interface 
conditions in PPU-C and IHU-C.} For both PPU-C
and IHU-C, these results illustrate the 
importance of interface conditions to 
capture the flux at the matrix-fracture 
interface and accurately predict the rate of
imbibition in the matrix.

We underline that these results expressed in dimensionless
time can be used to extend our analysis to other matrix-fracture
configurations with different characteristic variables, such as 
absolute permeability, porosity, and length of penetration. In 
simulations not included here for brevity, we considered a modified 
version of the test case discussed here by doubling the
penetration length and quadrupling the absolute permeability in the matrix.
The error in recovery for this modified configuration, expressed as 
a function of dimensionless time until $t_D = 0.4$, is in very good
agreement with those of
Figs.~\ref{fig:recovery_percentage_as_fct_of_time_refinement_study_first_test_case} to
\ref{fig:recovery_rate_refinement_study_first_test_case}.

\textcolor{black}{
\begin{remark}
We have observed that reducing the volume of the fracture to let
the non-wetting phase saturation increase in the fracture does not
change the conclusions of this section. Therefore, the numerical
examples performed with smaller fracture volumes are not included
to the present paper for brevity.
\end{remark}
}

\subsection{\label{subsection_forced_imbibition}Forced imbibition}

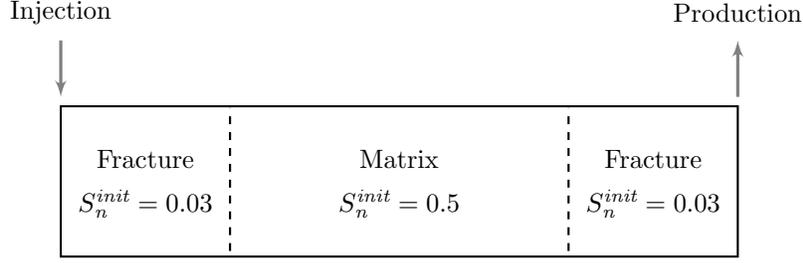
\begin{figure}[H]
\begin{center}
\tikzstyle{int}=[draw, minimum size=2em]
\tikzstyle{init} = [pin edge={to-,thick,black}]

\begin{tikzpicture}[node distance=3cm,auto,>=latex']

  \path (0,0) node (a) {};
  \path (9,2) node (b) {};
  \path [draw=black,thick] (a) rectangle (b); 
  \path (2.25,-0.05) node (a) {};
  \path (2.25,2.1) node (b) {};
  \path [draw=black,thick,dashed] (a) -- (b); 
  \path (6.75,-0.05) node (a) {};
  \path (6.75,2.1) node (b) {};
  \path [draw=black,thick,dashed] (a) -- (b); 

  \path (1.125,1.3) node (e) {Fracture};
  \path (1.125,0.7) node (e) {$S^{\textit{init}}_{\textit{n}} = 0.03$};
  \path (4.5,1.3) node (e) {Matrix};
  \path (4.5,0.7) node (e) {$S^{\textit{init}}_{\textit{n}} = 0.5$};
  \path (7.875,1.3) node (e) {Fracture};
  \path (7.875,0.7) node (e) {$S^{\textit{init}}_{\textit{n}} = 0.03$};
  
  \path (0,3) node (a) {};
  \path (0,2) node (b) {};
  \path (0,3.25) node (c) {Injection};
  \path [draw=gray,very thick,->] (a) -- (b); 
  
  \path (9,3) node (a) {};
  \path (9,2) node (b) {};
  \path (9,3.25) node (c) {Production};
  \path [draw=gray,very thick,->] (b) -- (a); 


\end{tikzpicture}
\vspace{-0.3cm}
\caption{\label{fig:schema_second_case}
  Schematic of the one-dimensional matrix-fracture model considered
  in this section. The low-permeability
  matrix is in between two high-permeability fractures.}
\end{center}
\end{figure}

We now extend the study of the accuracy and efficiency of the 
numerical schemes in a fractured porous medium by considering a 
forced imbibition process. We consider a one-dimensional domain
with a length $L= 20 \, \text{m}$. The domain,
sketched in
Fig.~\ref{fig:schema_second_case}, \textcolor{black}{is tilted
by an angle of 15}
degrees updip from left to right. The two fracture regions 
are located in the left quarter and in the right quarter of the
domain. The matrix occupies the middle of the domain. Unlike in
Section~\ref{subsection_spontaneous_imbibition}, the modified problem
has a source term and a sink term maintaining the wetting phase
saturation close to one in the fractures. Therefore, we do not
amplify the fracture pore volume and all the cells in the domain
have the same volume. 

\begin{figure}[H]
  \begin{center}
\subfigure[$N^{(m)} = 10$]{
\begin{tikzpicture}
\node[anchor=south west,inner sep=0] at (0,0){
\includegraphics[width=0.445\textwidth]{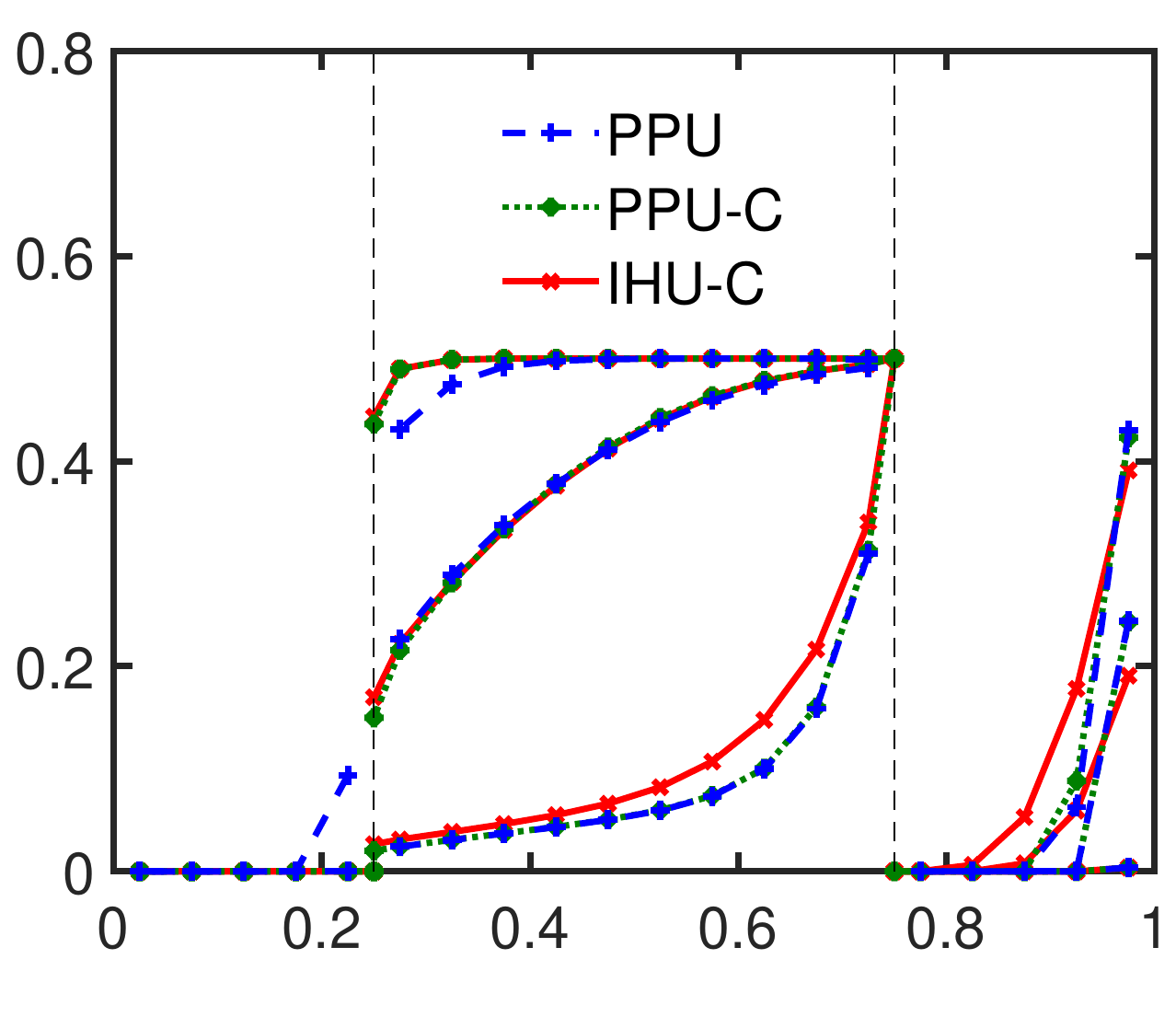}
};
\path (3.8,0.1) node (e) {$x_D$};
\node (ib_1) at (-0.225,3.375) {$S_{n}$};
\end{tikzpicture}
\label{fig:saturation_profiles_refinement_study_second_test_case_coarse_grid}
}
\subfigure[$N^{(m)} = 128$]{
\begin{tikzpicture}
\node[anchor=south west,inner sep=0] at (0,0){
\includegraphics[width=0.445\textwidth]{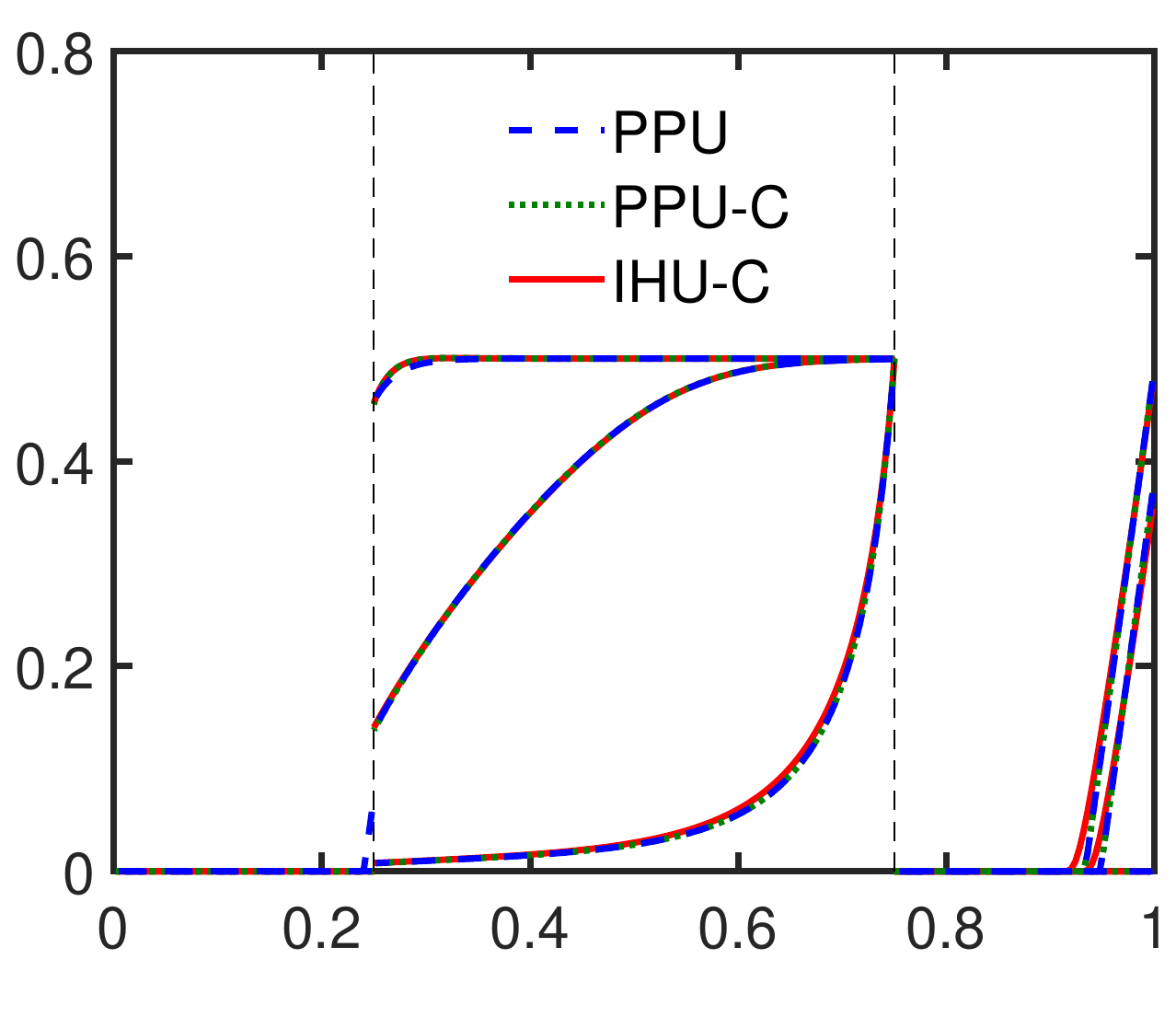}
};
\path (3.8,0.1) node (e) {$x_D$};
\node (ib_1) at (-0.225,3.375) {$S_{n}$};
\path (2.5,3.65) node (a) {};
\path (3.75,2.25) node (b) {};
\path [draw=black,very thick,->] (a) -- (b); 
\end{tikzpicture}
\label{fig:saturation_profiles_refinement_study_second_test_case_fine_grid}
}
\vspace{-0.4cm}
\caption{\label{fig:saturation_profiles_refinement_study_second_test_case}Coarse-grid
  saturation profiles ($N^{(m)} = 10$) in
  \subref{fig:saturation_profiles_refinement_study_second_test_case_coarse_grid}
  and fine-grid saturation profiles ($N^{(m)} = 128$) in
  \subref{fig:saturation_profiles_refinement_study_second_test_case_fine_grid} of
  the forced imbibition process. We show the saturation profiles
  at different times, respectively, 0.01 PVI, 0.1 PVI, and 1 PVI.
  The last reporting time is at steady state.}
\end{center}
\end{figure}

The values of the absolute permeabilities,
relative permeabilities,
and capillary pressures in the matrix and the
fractures are the same
as in the previous case. Initially, the matrix is
50\%-saturated with the non-wetting phase, and
the fracture is
97\%-saturated with the wetting phase. The
wetting phase is
injected from the left at a fixed rate, and a 
producer operates at a
fixed pressure of $3000 \, \text{psi}$ on the 
right. \textcolor{black}{We are interested
in the forced capillary imbibition process, so we 
make sure that
viscous and buoyancy forces are such that the 
saturation in the matrix
explores the negative section of the capillary 
pressure curve ($S \geq
0.5$).
In this configuration, buoyancy displaces the 
non-wetting
phase out of the matrix to the right of the 
domain, and the
wetting-phase injection creates a total flow from 
left to
right.} During this process, the local capillary
equilibrium at the interface
between the matrix and the fracture on the right imposes a non-zero
non-wetting phase saturation in the matrix. This capillary
end-effect \citep{richardson1952laboratory,huang1998capillary},
visible in the saturation profiles of
Fig.~\ref{fig:saturation_profiles_refinement_study_second_test_case},
leads to non-wetting phase trapping in the matrix at steady state. For
the two spatial resolutions of
Fig.~\ref{fig:saturation_profiles_refinement_study_second_test_case},
we observe that the non-wetting phase
remains relatively large in the matrix even though it vanishes
entirely in the fractures as a result of production. In particular,
the non-wetting phase saturation at the right boundary of the matrix
does not decrease over time and stays close to 0.5 due to the
capillary end-effect.

\begin{figure}[H]
  \begin{center}
\subfigure[]{
\begin{tikzpicture}
\node[anchor=south west,inner sep=0] at (0,0){
\includegraphics[width=0.44\textwidth]{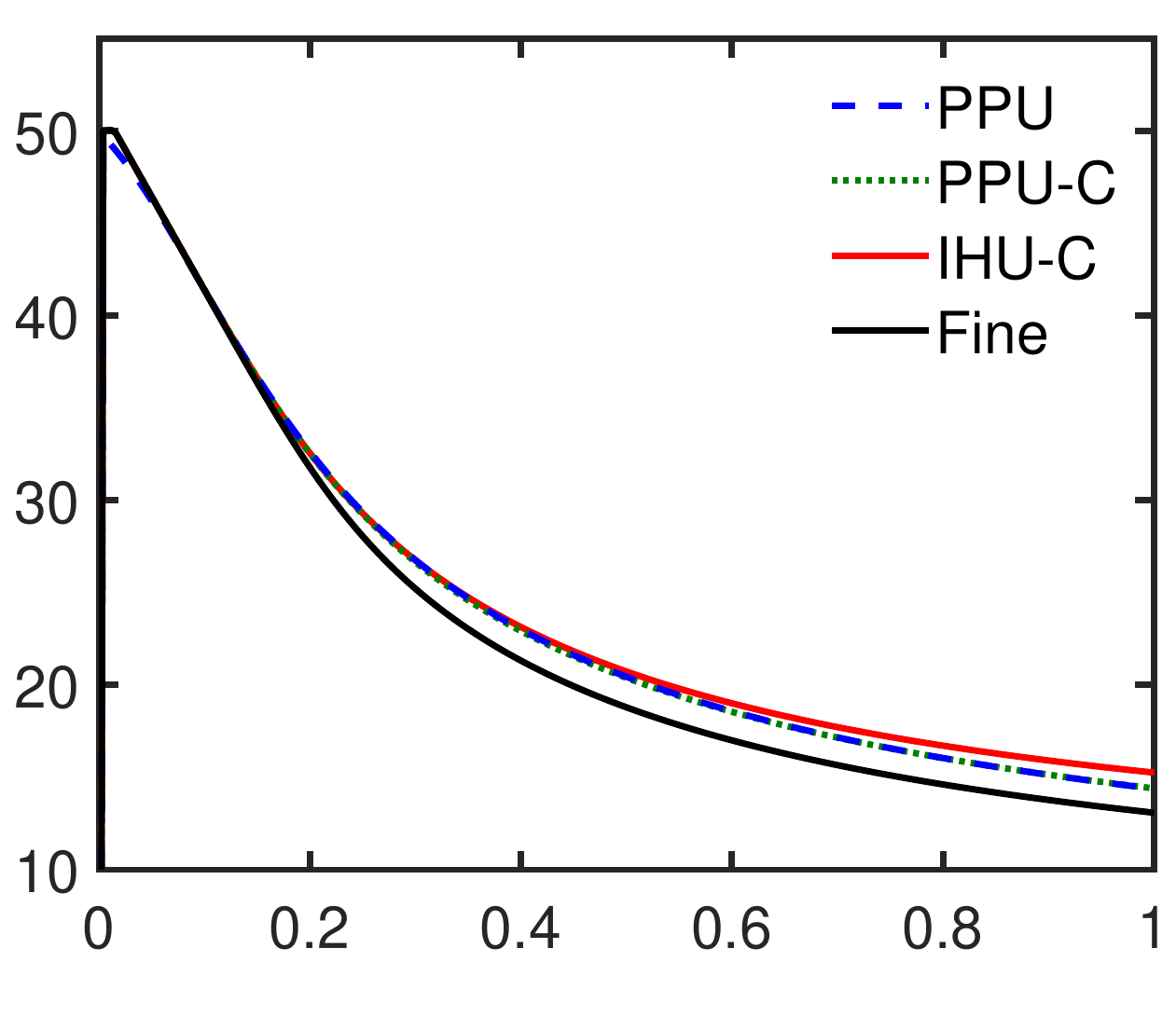}
};
\path (3.75,0.1) node (e) {PVI};
\node[rotate=90] (ib_1) at (-0.2,3.2) {Average $S_{n}$ in the matrix};
\end{tikzpicture}
\label{fig:trapping_temporal_evolution_second_test_case}
}
\subfigure[]{
\begin{tikzpicture}
\node[anchor=south west,inner sep=0] at (0,0){
\includegraphics[width=0.46\textwidth]{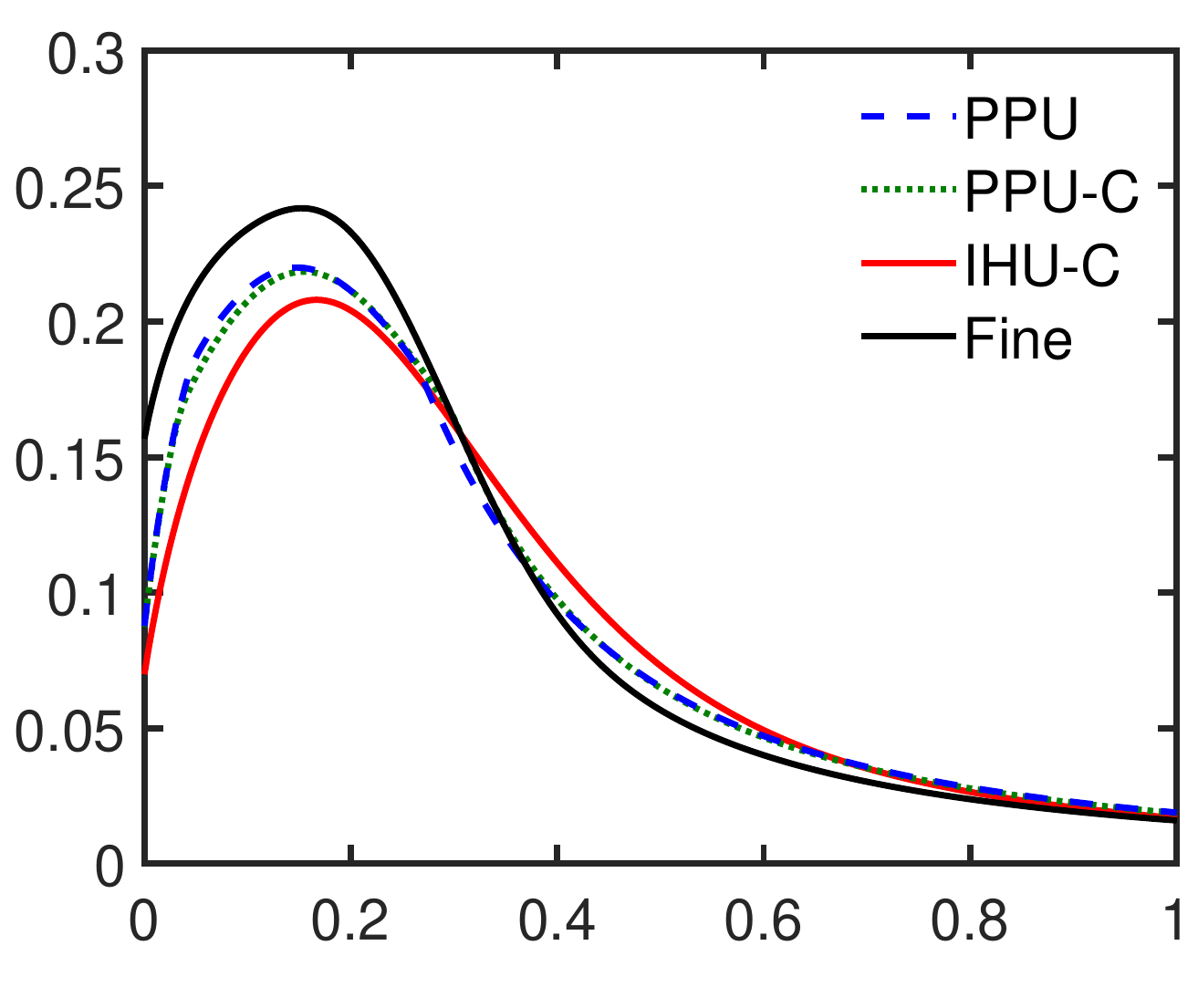}
};
\path (4,0.1) node (e) {PVI};
\node[rotate=90] (ib_1) at (-0.2,3.2) {Production rate};
\end{tikzpicture}
\label{fig:production_temporal_evolution_second_test_case}
}
\vspace{-0.4cm}
\caption{\label{fig:temporal_evolution_second_test_case}Comparison of
the coarse grid ($N^{(m)} = 10$) solution of the three schemes with the
fine-grid solution as a function of the pore volumes injected.
The average non-wetting phase saturation in
the matrix is shown in
\subref{fig:trapping_temporal_evolution_second_test_case} while
the non-wetting phase production
is in \subref{fig:production_temporal_evolution_second_test_case}.}
\end{center}
\end{figure}
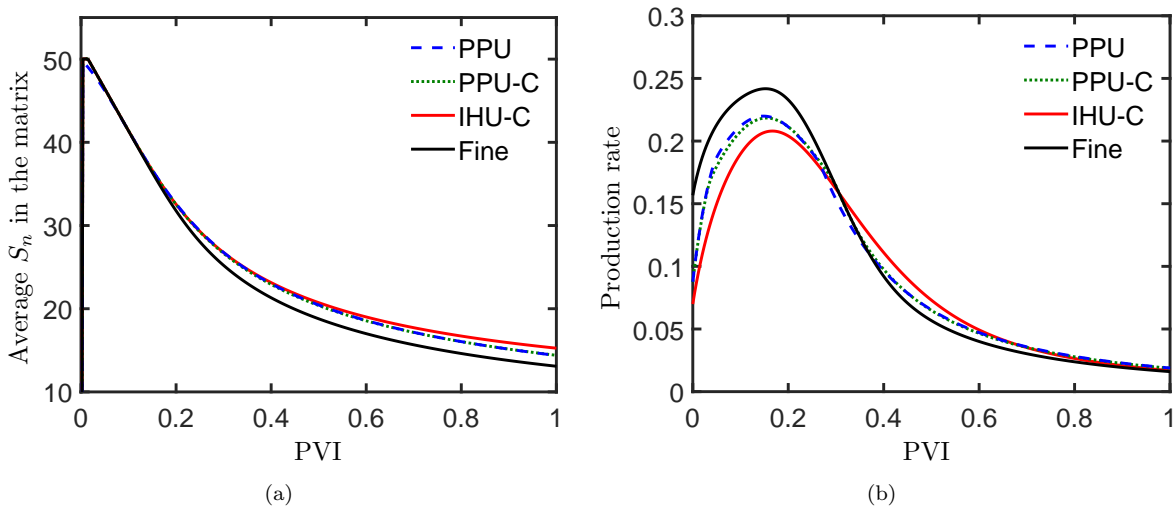

For this test case, we observe that this
interfacial saturation is relatively well
captured by standard PPU on the coarse grid.
We still see that the schemes based on interface
conditions are more accurate than standard PPU at the beginning
of the simulation (0.01 PVI), particularly at the boundary between
the left fracture and the matrix. But, at later times (0.1 PVI
and 1 PVI), the coarse saturation profiles yielded by PPU and PPU-C
are in agreement. At steady state, we note a mismatch at
$x_D \approx 0.6$ in the representation of the sharp gradient
between the coarse IHU-C results and the coarse profiles obtained
with the PPU-based results.
We see in Fig.~\ref{fig:trapping_temporal_evolution_second_test_case}
that the mismatch in the coarse saturation profiles results in an
over-estimation of the average non-wetting phase saturation remaining
in the matrix at steady state with IHU-C. In
Fig.~\ref{fig:production_temporal_evolution_second_test_case},
it also leads to an under-estimation of the amount of non-wetting
phase that is produced from the fracture on the right.

\begin{figure}[H]
  \begin{center}
    \begin{tikzpicture}
      \node[anchor=south west,inner sep=0] at (0,0){
        \includegraphics[width=0.53\textwidth]{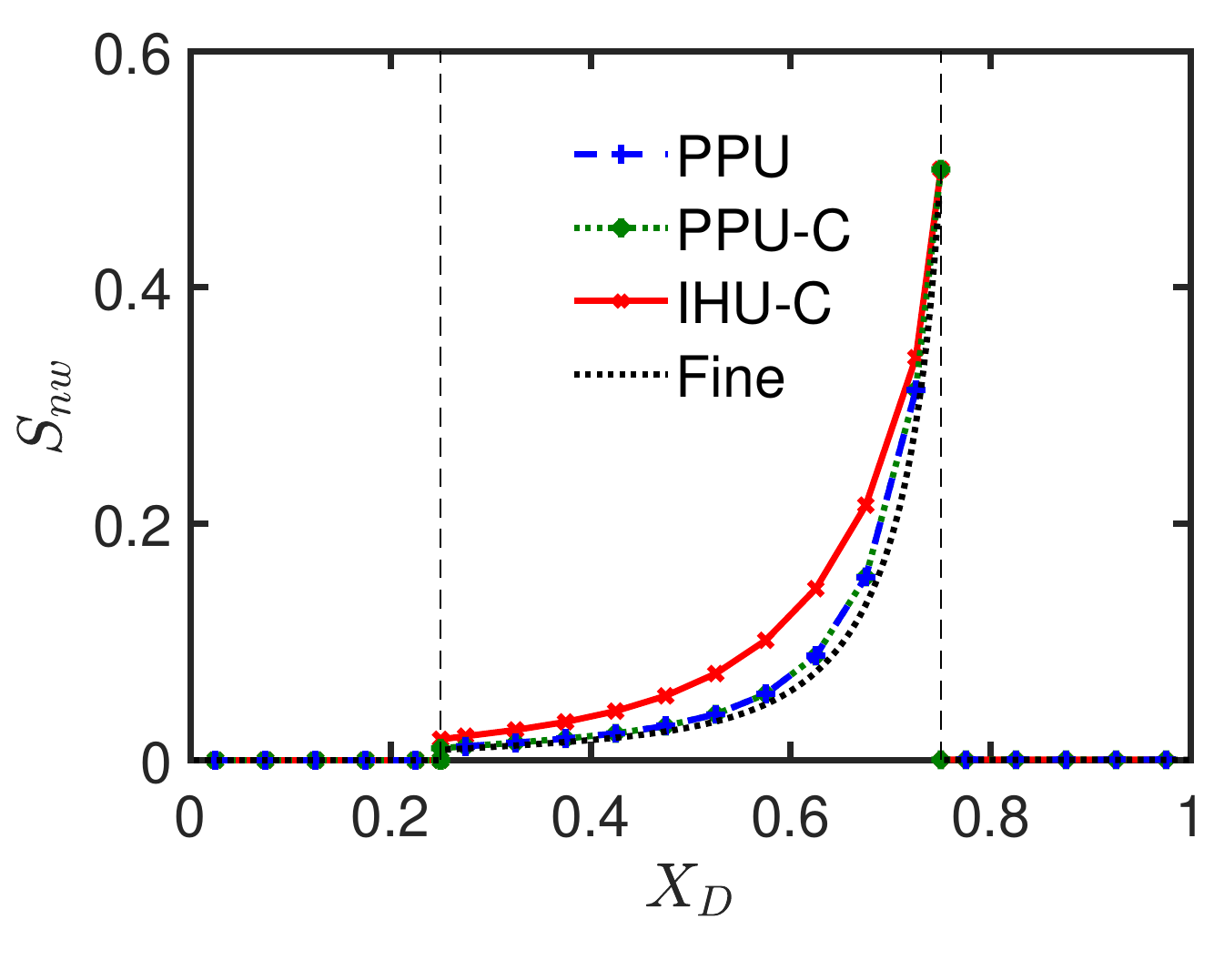}
      };
      \path (4.25,0.25) node (d) {};
      \path (5.65,0.9) node (e) {};
      \path [draw=white,fill=white] (d) rectangle (e);
      \path (4.85,0.6) node (e) {$x_D$};
      \path (0.1,2.8) node (d) {};
      \path (0.56,5) node (e) {};
      \path [draw=white,fill=white] (d) rectangle (e);
      \node (ib_1) at (0.3,4.) {$S_{n}$};
    \end{tikzpicture}
    \vspace{-0.4cm}
    \caption{\label{fig:saturation_profiles_refinement_study_second_test_case_coarse_grid_truncation_error}
      Coarse-grid saturation profiles ($N^{(m)} = 10$) at steady state
      (1 PVI) for the forced imbibition process in the absence of
      buoyancy forces. The other parameters are the same as in
      Fig.~\ref{fig:saturation_profiles_refinement_study_second_test_case_coarse_grid}.
    }
\end{center}
\end{figure}
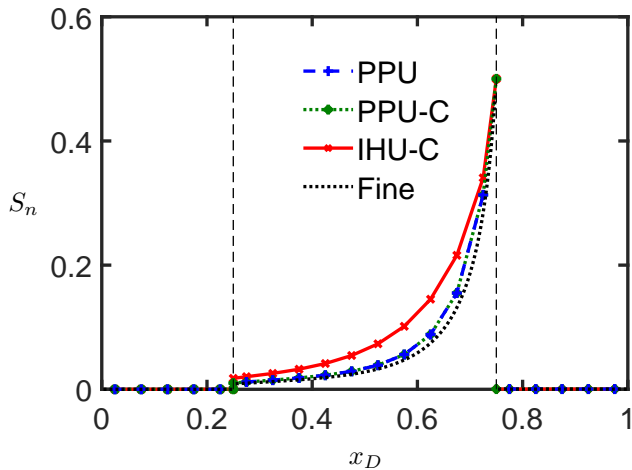

\begin{figure}[H]
  \begin{center}
    \subfigure[Viscous error]{
      \begin{tikzpicture}
        \node[anchor=south west,inner sep=0] at (0,0){
          \includegraphics[width=0.315\textwidth]{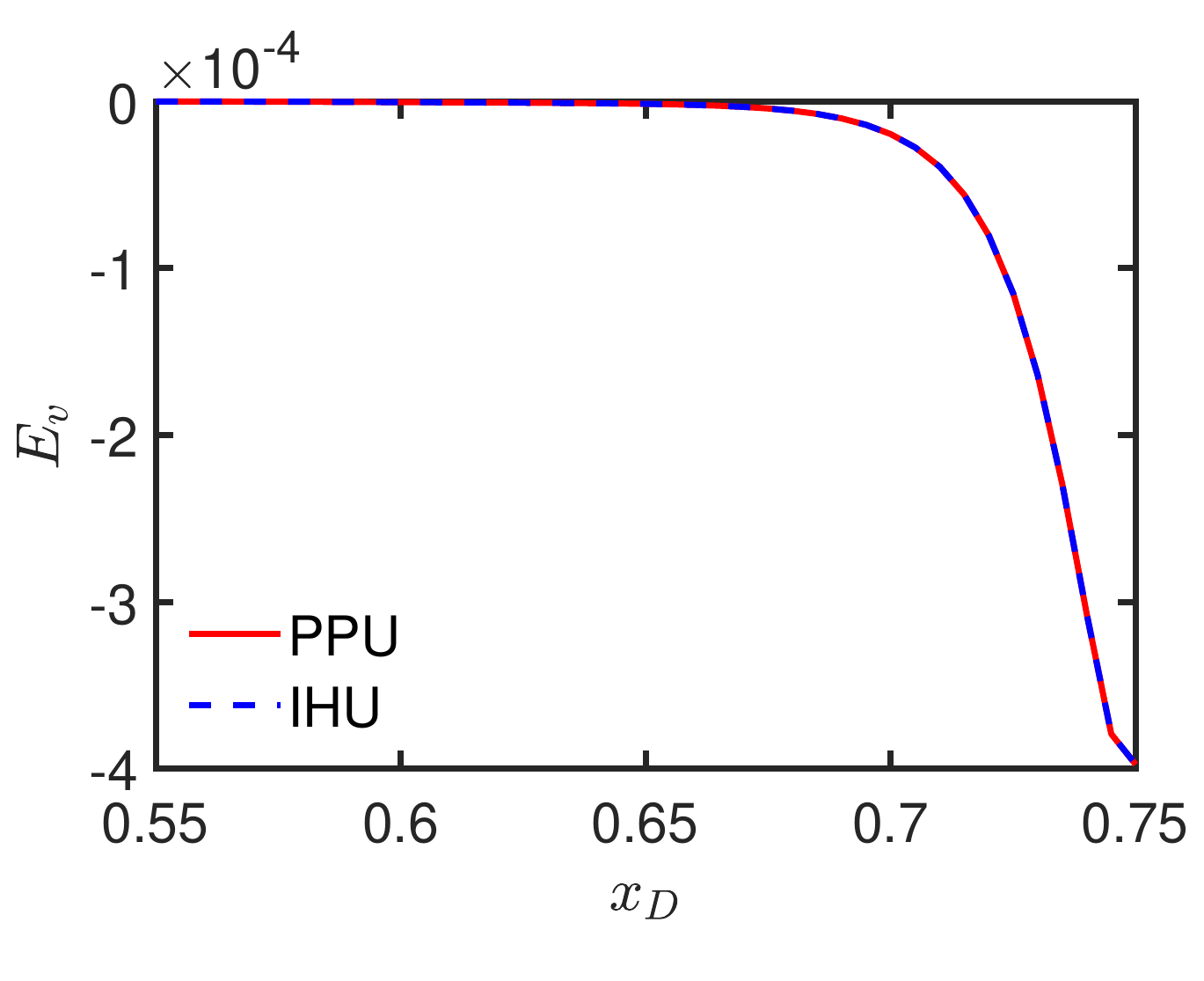}
        };
        \path (2.4,0.2) node (d) {};
        \path (3,0.5) node (e) {};
        \path [draw=white,fill=white] (d) rectangle (e);
        \path (2.75,0.3) node (e) {$x_D$};
        \path (0.,2.1) node (d) {};
        \path (0.4,2.8) node (e) {};
        \path [draw=white,fill=white] (d) rectangle (e);
        \node[rotate=90] (ib_1) at (0.3,2.3) {$\mathcal{E}_V$};
      \end{tikzpicture}
       \label{fig:truncation_error_viscous_term}
    }
    \hspace{-0.35cm}
    \subfigure[Capillary error]{
      \begin{tikzpicture}
        \node[anchor=south west,inner sep=0] at (0,0){
          \includegraphics[width=0.315\textwidth]{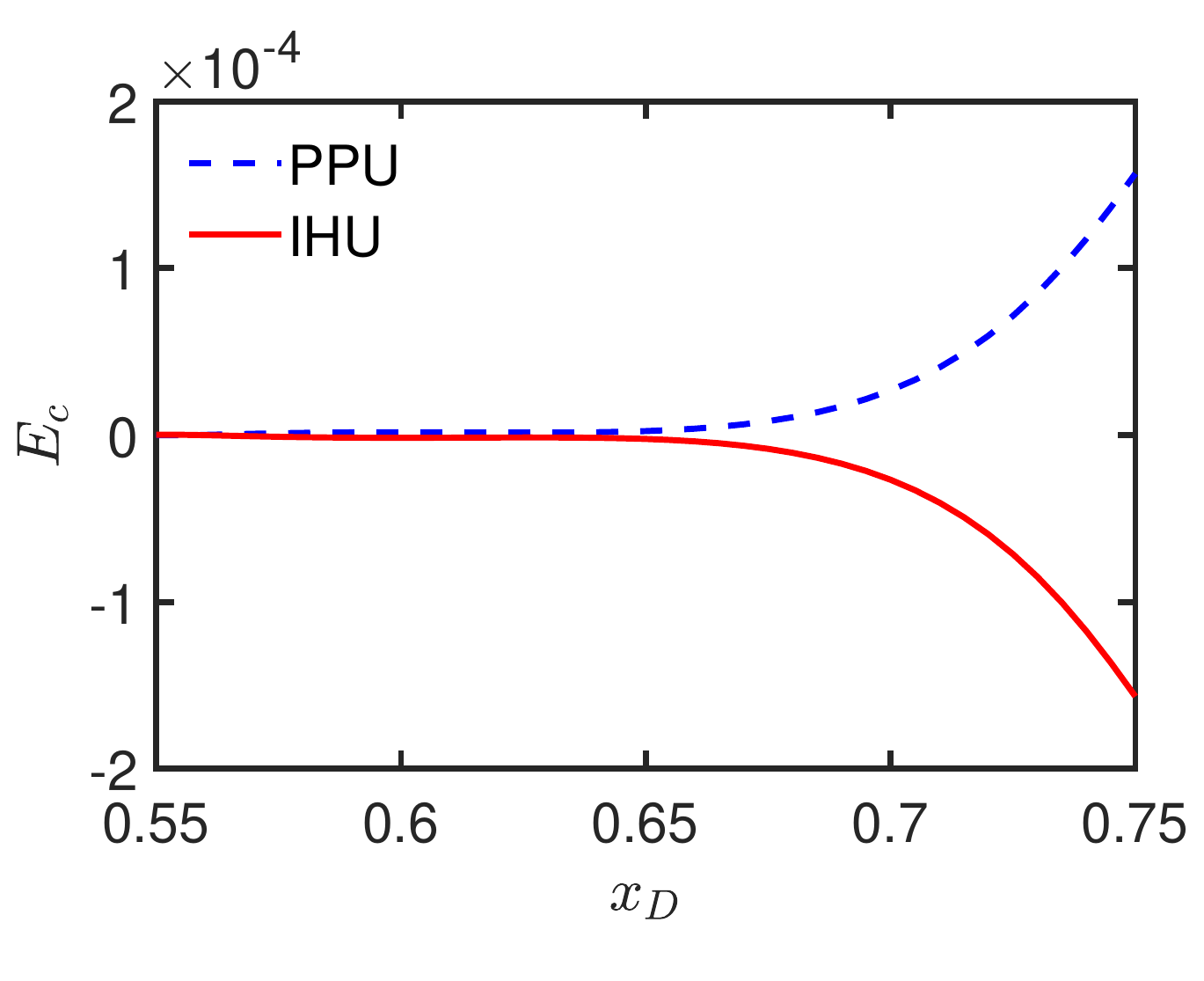}
        };
        \path (2.4,0.2) node (d) {};
        \path (3,0.5) node (e) {};
        \path [draw=white,fill=white] (d) rectangle (e);
        \path (2.85,0.3) node (e) {$x_D$};
        \path (0.,2.1) node (d) {};
        \path (0.37,2.8) node (e) {};
        \path [draw=white,fill=white] (d) rectangle (e);
        \node[rotate=90] (ib_1) at (0.35,2.3) {$\mathcal{E}_C$};
      \end{tikzpicture}
      \label{fig:truncation_error_capillary_term}
    }
    \hspace{-0.35cm}
    \subfigure[Viscous-capillary error]{
      \begin{tikzpicture}
        \node[anchor=south west,inner sep=0] at (0,0){
          \includegraphics[width=0.315\textwidth]{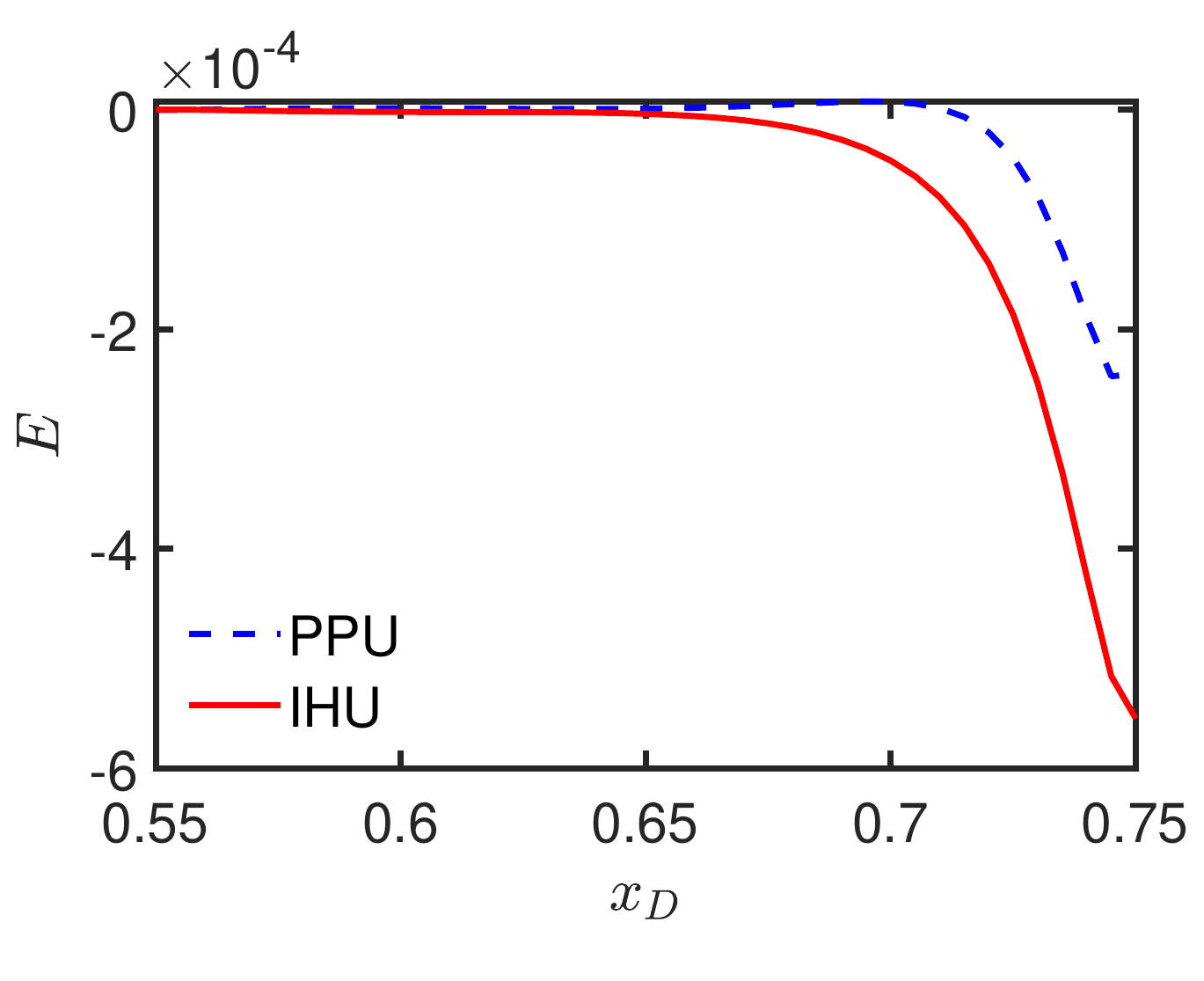}
        };
        \path (2.4,0.2) node (d) {};
        \path (3,0.5) node (e) {};
        \path [draw=white,fill=white] (d) rectangle (e);
        \path (2.75,0.3) node (e) {$x_D$};
        \path (0.,2.1) node (d) {};
        \path (0.37,2.8) node (e) {};
        \path [draw=white,fill=white] (d) rectangle (e);
        \node[rotate=90] (ib_1) at (0.3,2.3) {$\mathcal{E}_{\textit{VC}}$};
      \end{tikzpicture}
      \label{fig:truncation_error_viscous_capillary_term}
    }
    \vspace{-0.4cm}
    \caption{\label{fig:truncation_error_terms}
      Leading truncation error terms as a function of space for the viscous
      flux in \subref{fig:truncation_error_viscous_term}, the capillary flux
      in \subref{fig:truncation_error_capillary_term}, and the combined
      viscous-capillary flux in \subref{fig:truncation_error_viscous_capillary_term}.
      The computations are based on the expressions of the truncation error terms
      given in (\ref{truncation_error_viscous_term})-(\ref{ppu_truncation_error_capillary_term}) and applied
      to the saturation profiles of Fig.~\ref{fig:saturation_profiles_refinement_study_second_test_case_coarse_grid_truncation_error}.}
  \end{center}
\end{figure}
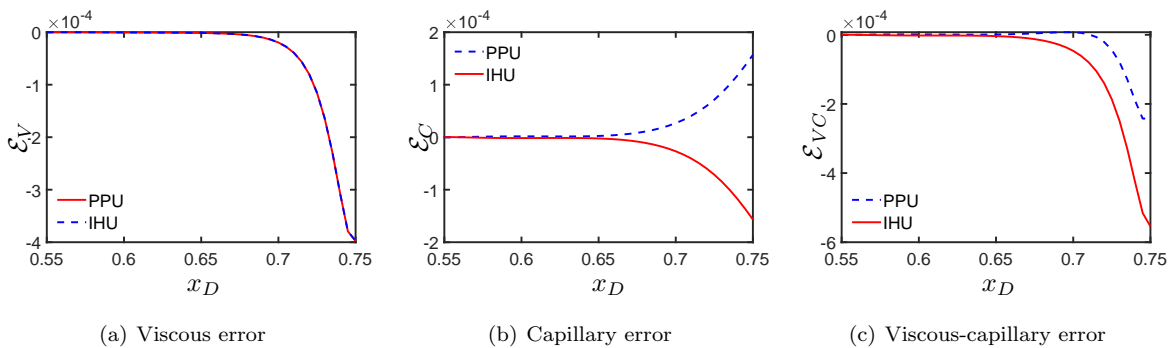

Next, we use the truncation error analysis of
Section~\ref{section_truncation_error_analysis} to explain the
discrepancy between the PPU-based schemes and IHU-C for the forced
imbibition process. To simplify the analysis, we omit buoyancy
forces and only consider the viscous-capillary equilibrium. The
other parameters remain unchanged, resulting in the coarse-grid
saturation profiles of
Fig.~\ref{fig:saturation_profiles_refinement_study_second_test_case_coarse_grid_truncation_error}.
We apply the truncation error analysis of
Section~\ref{section_truncation_error_analysis} to the saturation
profiles in the matrix ($x \in [0.55,0.75]$) where the difference
between the PPU-C and IHU-C results are relatively large. The flow
in this region is cocurrent from left to right, which corresponds
to the configuration studied in
Section~\ref{section_truncation_error_analysis}. The leading terms
of the viscous, capillary, and total truncation errors are shown
in Fig.~\ref{fig:truncation_error_terms}.

For the viscous term in
Fig.~\ref{fig:truncation_error_viscous_term}, we see that the two
schemes lead to the same truncation error since they are based on
the same discretization in the presence of cocurrent flow. In
Fig.~\ref{fig:truncation_error_capillary_term}, the PPU and IHU
capillary truncation error terms computed in
(\ref{ihu_truncation_error_capillary_term})-(\ref{ppu_truncation_error_capillary_term})
have a similar absolute magnitude
but an opposite sign. This is key to understand the behavior
of the total error, obtained by summing the viscous and capillary
error terms and shown in
Fig.~\ref{fig:truncation_error_viscous_capillary_term}. We observe
that the capillary term tends to reduce the viscous error with PPU,
but instead amplifies the viscous error with IHU, yielding
a larger total error with IHU than with PPU. This is particularly 
the case when the saturation gradient is very sharp in the
neighborhood of the interface with the fracture on the right
($x \in [0.65, 0.75]$).
In other words, the discrepancy does not
result from the treatment of the interfacial saturations, but
instead arises because of the split treatment of the
viscous-buoyancy-capillary flux in the homogeneous regions.
This methodology introduces splitting errors that affect the accuracy
of IHU-C in the presence of sharp saturation gradients. 
We plan to address this limitation with a high-resolution of IHU-C
along the lines of \cite{mykkeltvedt2017fully}.

\section{Conclusion}

In this work, we have studied a fully implicit finite-volume scheme
for capillary-dominated multiphase flow in matrix-fracture systems.
In particular, we have studied the impact of discrete interface
conditions enforcing a local capillary equilibrium between the
matrix and the fracture on the accuracy of the predicted imbibition
rate into the matrix. We have also considered the interaction of the
discrete interface conditions with the upwinding schemes -- namely,
PPU and IHU -- used to evaluate the interfacial saturation-dependent
coefficients.

\textcolor{black}{We demonstrate that both the PPU and IHU schemes
based on discrete interface conditions (and referred to in the paper
as PPU-C and IHU-C, respectively) yield an interfacial flux that
is more accurate than with the standard scheme. 
Our results show that using interface conditions to compute the
numerical flux between the matrix and the fracture
improves the accuracy of the prediction of the imbibition rate
compared to the standard scheme without interface conditions.
Achieving the same accuracy with the standard scheme would require
significant mesh refinement, which would dramatically increase the
cost of large-scale, three-dimensional simulations.} Future
directions include extending these results for the three-phase
case to further demonstrate the applicability of the scheme to
practical simulation problems.

\appendix

\section{\label{appendix_capillary_pressure_interpolation}Capillary pressure interpolation}

Here, we explain the procedure used to enforce an upper bound,
$P_{c,max}$, and a lower bound, $P_{c,min}$, on the capillary
pressure function and therefore avoid infinite gradients when
capillary pressure is accounted for. This is done by 
replacing the expression given in (\ref{brooks_corey_pc}) with
a polynomial function when the saturation is close to one of the
endpoints, as follows:
\begin{equation}
\bar{P}_c(S) = \left\{
\begin{array}{ll}
a S^2 + b S + P_{c,\textit{max}} & \text{if} \, S \leq S^- \\[2pt]
c (1-S)^2 + d (1-S) + P_{c,\textit{min}} & \text{if} \, S \geq S^+ \\[2pt]
P_c(S) & \text{otherwise,} 
\end{array}
\label{capillary_pressure_interpolation}
\right. 
\end{equation}
where $P_c$ is the function defined in (\ref{brooks_corey_pc}).
In a preprocessing step, we compute the triplets $(a,b,S^-)$ and $(c,d,S^+)$
such that the capillary pressure function defined in 
(\ref{capillary_pressure_interpolation}) is continuous and twice 
differentiable. Computing the first triplet involves solving the
following system 
\begin{equation}
 \left\{
\begin{array}{ll}
      P_{c} (S^-) = a (S^-)^2 + b S^- + P_{c,\textit{max}} \\[4pt]
     \displaystyle \frac{dP_{c}(S^-)}{dS} = 2a S^- + b \\[4pt]
     \displaystyle \frac{d^2P_{c}(S^-)}{dS^2} = 2a.
    \label{pc_preprocessing_procedure}\\
\end{array} 
\right. 
\end{equation}
\textcolor{black}{The nonlinear system} solved to compute $(c,d,S^+)$ is analogous.
All the figures and numerical examples presented
in this paper rely on (matrix) capillary pressures computed with
(\ref{capillary_pressure_interpolation}). We set
$P_{c,max} = 15 \, \text{psi}$ and $P_{c,min} = -15 \, \text{psi}$.

\section{\label{appendix_local_solver_derivatives}Derivatives of $S^{(\alpha)}_{ij}$ with respect to cell-centered variables}

\textcolor{black}{
The last lines of Algorithm~\ref{alg:local_nonlinear_solver} require
the computation of the derivatives of the saturation at interface
$(ij)$, denoted by $S^{(\alpha)}_{ij}$, with respect to the
cell-centered primary variables, $p_i$, $p_j$, $S_i$, and $S_j$.
We remind the reader that the total flux,
$\overline{u}_{T,ij}(\Delta p_{ij},S_i,S_j)$,
is a function of $\Delta p_{ij}$, $S_i$, and $S_j$. Consider the
function
}
\begin{equation}
  R_{ij}: (a,b,c,d)
  \longmapsto
  F_{\ell,ij}(\overline{u}_{T,ij}(a,b,c), b, d)
  +
  F_{\ell,ji}(-\overline{u}_{T,ij}(a,b,c), c, h(d)).
\end{equation}
\textcolor{black}{
Given $\Delta p_{ij}$, $S_i$, and $S_j$, the local nonlinear solver
finds the interfacial saturation, $S^{(\alpha)}_{ij}$, such that
}
\begin{equation}
R_{ij}(\Delta p_{ij},S_i,S_j,S^{(\alpha)}_{ij}) = 0.
\end{equation}
\textcolor{black}{
Noting that the monotonicity properties of the flux imply that
}
\begin{equation}
\frac{\partial R_{ij}}{\partial d}(\Delta p_{ij},S_i, S_j, S^{(\alpha)}_{ij}) \neq 0,
\end{equation}
\textcolor{black}{
we apply the implicit function theorem to compute the derivatives of
$S^{(\alpha)}_{ij} = f_{ij}(\Delta p_{ij},S_i,S_j)$ with respect to
$\Delta p_{ij}$, $S_i$, and $S_j$. For $\tau = \{ a, b, c \}$, we
obtain
}
\begin{equation}
  \frac{\partial f_{ij}}{\partial \tau}(\Delta p_{ij},S_i,S_j) =
  \bigg( \frac{\partial R_{ij}}{\partial d}(\Delta p_{ij},S_i, S_j, S^{(\alpha)}_{ij}) \bigg)^{-1}
  \frac{\partial R_{ij}}{\partial \tau}(\Delta p_{ij}, S_i, S_j, S^{(\alpha)}_{ij}).
\end{equation}

\bibliography{biblio}

\end{document}